\documentclass[10pt,prd,aps,onecolumn,notitlepage,nofootinbib,nobalancelastpage,superscriptaddress]{revtex4-1}

\usepackage{amsmath}
\usepackage{amssymb}
\usepackage{graphicx}
\usepackage[colorlinks=true,
            urlcolor=blue,
            citecolor=blue,
            linkcolor=red,
            menucolor=blue,
            linktocpage=true]{hyperref}
\usepackage{amsfonts} 
\usepackage{mathrsfs}
\usepackage{bm}
\usepackage{multirow}
\usepackage{color}
\usepackage{caption}
\usepackage{subcaption}
\usepackage{ragged2e}
\usepackage{enumerate}
\DeclareCaptionJustification{justified}{\justifying}
\DeclareMathAlphabet{\mathpzc}{OT1}{pzc}{m}{it}

\def\eprint#1{\href{http://www.arxiv.org/abs/#1}{{\tt arXiv:#1}}}
\def\eprintt#1{\href{http://www.arxiv.org/abs/#1}{{\tt #1}}}

\def\be{\begin{equation}}\def\ee{\end{equation}}
\def\bea#1\eea{\begin{align}#1\end{align}}
\def\mbf#1{\mathbf{#1}}
\def\mrm#1{\mathrm{#1}}

\def\x{\mathbf{x}}
\def\r{\mathbf{r}}

\def\k{\mathbf{k}}
\def\q{\mathbf{q}}
\def\s{\mathbf{s}}
\def\z{\mathbf{z}}
\def\p{\mathbf{p}}
\def\u{\mathbf{u}}
\def\n{\hat{\mathbf{n}}}
\def\v{\mathbf{v}}

\newcommand\rme{\mathrm{e}}

\newcommand\delD{\delta_\mathrm{D}}
\newcommand\dif{\mrm{d}}

\newcommand\ii{\mathrm{i}}

\newcommand\dk{\frac{\dif^3\k}{(2\pi)^3}}

\def\fint#1{\int_{#1}}

\newcommand\calH{\mathcal{H}}
\newcommand\calL{\mathcal{L}}
\newcommand\Ks{K_{\mathrm{S}}}
\newcommand\Gs{G_{\mathrm{S}}}
\newcommand\PL{P_\mathrm{L}}
\newcommand\KA{\mathcal{K}^{(A)}}
\newcommand\KB{\mathcal{K}^{(B)}}

\newcommand\KsA{\KA_\mrm{S}}
\newcommand\KsB{\KB_\mrm{S}}

\newcommand\Da{D^{\:2}_\text{1pt}}
\newcommand\Dab{D^{\:2}_\text{2pt}}

\newcommand\vz{v_z}
\newcommand\uz{u_z}
\newcommand\qz{q_z}
\newcommand\kz{k_z}
\newcommand\nablaz{\nabla_z}

\newcommand\Pu{P_{uu}}
\newcommand\Pv{P_{vv}}
\newcommand\xiv{\xi_{vv}}
\newcommand\Mpc{\,\mathrm{Mpc}}

\begin{document}

\title{Exploring the redshift-space peculiar velocity field and its power spectrum}

\author{Lawrence~Dam}
\email[Corresponding author\\]{ldam4036@uni.sydney.edu.au}
\affiliation{Sydney Institute for Astronomy, School of Physics, A28, The University of Sydney, NSW 2006, Australia}
\author{Krzysztof~Bolejko}
\affiliation{School of Natural Sciences, College of Sciences and Engineering, University of Tasmania, Private Bag 37, Hobart TAS 7001, Australia}
\author{Geraint~F.~Lewis}
\affiliation{Sydney Institute for Astronomy, School of Physics, A28, The University of Sydney, NSW 2006, Australia}

\begin{abstract}
Redshift-space distortions (RSD) generically affect any spatially-dependent observable that is mapped
using redshift information. The effect on the observed clustering of galaxies is the primary example of
this. This paper is devoted to another example: the effect of RSD on the apparent peculiar motions of
tracers as inferred from their positions in redshift space (i.e.\ the observed distance). Our theoretical
study is motivated by practical considerations, mainly, the direct estimation of the velocity power
spectrum, which is preferably carried out using the tracer's redshift-space position (so as to avoid
uncertainties in distance measurements).
We formulate the redshift-space velocity field and show that RSD enters as a higher-order effect.
Physically, this effect may be interpreted as a dissipative correction to the usual perfect-fluid
description of dark matter. We show that the effect on the power spectrum is a damping on relatively
large, quasilinear scales ($k\gtrsim0.01\,h\Mpc^{-1}$), as was observed, though unexplained, in
$N$-body simulations elsewhere. This paper presents two power spectrum models for the peculiar
velocity field in redshift space, both of which can be considered velocity analogues of existing
clustering models. In particular, we show that the ``Finger-of-God'' effect, while also present in the
velocity field, cannot be entirely blamed for the observed damping in simulations. Our work provides
some of the missing modelling ingredients required for a density\textendash velocity multi-tracer analysis,
which has been proposed for upcoming redshift surveys.
\end{abstract}

\date{\today}
\maketitle

\newpage
\tableofcontents

\section{Introduction\label{sec:intro}}
Galaxy redshift surveys provide a useful way to probe cosmology through the large-scale structure of the Universe.
The process begins with the basic task of mapping the distribution of galaxies (or other mass tracers). In principle,
this is quite simple: place each galaxy in three-dimensional space by measuring both their angular position
and radial distance. In practice, distances are rarely available, and one resorts to the redshift (and a model prior)
to infer the galaxy's distance. Redshifts, however, are imperfect indicators; they incur a Doppler shift from the
galaxy's own peculiar velocity, which translates to a shift---from the ``real-space'' position $\x$ to the observed
``redshift-space'' position $\s$---according to
\be\label{eq:mapping}
\x\to\s=\x+\frac{\v(\x)\cdot\n}{aH}\,\n,
\ee
where $a$ is the scale factor, $H$ is the Hubble parameter, $\n$ is the direction of observation, and
$\v(\x)\cdot\n$ is the galaxy's peculiar velocity along the line of sight (LOS). This mapping modifies
only the radial positions of galaxies, leading to \emph{redshift-space distortions} (RSD)
in the galaxy clustering \cite{Kaiser:1987qv,Peebles:1980,Jackson:1972,Sargent:1977}.

Far from being a nuisance, though, the measurement of RSD now forms a key science goal of redshift
surveys, having been observed in numerous surveys, among various tracers
\cite{Davis:1983,Hamilton:1993,Fisher:1994,Peacock:2001,Blake:2012,Beutler:2012,delaTorre:2013rpa,Alam:2017,Gil-Marin:2021}.
The interest in RSD is that it can be used to probe $f$, the growth rate of cosmic structure
\cite{Sargent:1977,Kaiser:1987qv,Hamilton:1992}.
This is through the fact that the distortions induce a systematic anisotropic
signal on the clustering pattern, the size of which being determined by the growth rate
through the amplitude of the displacement in eq.~\eqref{eq:mapping}.
The growth rate is of particular interest as different values are predicted by
different dark-energy scenarios and modified-gravity theories \cite{Linder:2005in,Ishak:2006,Song:2008qt,Ishak:2019,Baker:2021}.
Constraining it thereby provides a convenient way to distinguish the $\Lambda$CDM model from other
alternatives. The RSD measurements to date are largely consistent with that predicted by $\Lambda$CDM
(and so general relativity). A more demanding test of $\Lambda$CDM~\cite{Alam:2020}
will come with the next generation of redshift surveys, such as DESI~\cite{Aghamousa:2016}, Euclid~\cite{Amendola:2016},
and SPHEREx~\cite{Dore:2014}.

As this paper illustrates, RSD is not only limited to the galaxy clustering: any tracer sampled in
three-dimensional redshift space will present a distorted view of the large-scale structure. This
includes the velocity field underlying the large-scale streaming motions of galaxies---the focus
of this paper. Just as galaxies \emph{cluster} differently in redshift space, so too they \emph{move}
differently. In the former case the galaxy overdensity field becomes intertwined with the
velocity field; in the latter case the velocity field becomes intertwined with itself.
In both cases, there is additional cosmological information not present when galaxies
are viewed at their true, real-space positions.
Usefully, since the particular dependence on the growth rate changes, this can allow
parameter degeneracies to be broken and tighter constraints to be obtained.
Extracting this information, though, is more challenging given the
difficulties of modelling in redshift space.

Besides the well-known cosmological value, there is a practical reason as to why a redshift-space-based analysis
is preferable over a real-space one. As can be seen from eq.~\eqref{eq:mapping}, when a tracer's peculiar
velocity is known (when distances are available) it becomes possible to shift the tracer back to its actual
position. The shift, however, introduces a sizeable uncertainty through the peculiar velocity, which
propagates to the inferred real-space position. A typical $20\%$ uncertainty in the peculiar velocity
(which largely derives from the distance estimate) then implies a position uncertainty of at least the
same size. In order to estimate the velocity power it thus becomes necessary to take into account the
uncertainty in the discrete sampling of the velocity field. This complication can be avoided if the positions
are instead taken to be that determined by the spectroscopic redshifts.
(Though note that the large velocity uncertainties still affect the power spectrum estimates through the shot
noise term, which is weighted by these errors, unlike the density power spectrum.)
So though the velocity field can in principle be analyzed in either real space or redshift space,
the positions are simply better determined in the latter, being more closely connected to
what is observed, rather than inferred.

\subsection{Motivation}
This paper is motivated by the present need for a theoretical model of the
\emph{redshift-space peculiar velocity field} and particularly its two-point statistics.
This need first arose in the work of Koda~{\it et al.}~\cite{Koda:2014}, who showed using a Fisher forecast
the ability to substantially improve growth rate constraints with the ``multi-tracer'' method~\cite{McDonald:2009}
applied to the galaxy density and peculiar velocities; i.e.\ with a joint analysis of the auto- and
cross-power spectra of the redshift-space galaxy density and velocity fields
(see also refs.~\cite{Burkey:2003rk,Kim:2020}).
Without a physical model for the redshift-space velocity power spectrum, a simple empirical model
calibrated on $N$-body simulations~\cite{Poole:2014} was adopted.
The simulations showed that the monopole moment of the measured power spectrum was damped, much as the
``Fingers-of-God'' (FoG) effect damps the galaxy power spectrum. This behaviour was found to be well
described by a simple fitting function in the form of a FoG-like damping factor, which was parametrized
by an empirical velocity dispersion parameter.
Unfortunately, this nuisance parameter was shown to degrade growth rate constraints by $30$--$50\%$,
depending on the set of parameters chosen (as well as likely introducing a systematic bias).
A number of other works have also relied on this model
(e.g.~\cite{Howlett:2017a,Howlett:2017b,Adams:2020,Amendola:2021}).

To minimize the loss of statistical power a physical model is required which can explain
the damping ideally from first principles. The aim of this paper is to provide such a model.
It is also timely, coming ahead of the imminent peculiar velocity surveys (e.g.~\cite{Taipan,Wallaby}),
which are set to probe a larger volume with a denser sampling (and so improve the shot-noise
properties of power spectrum measurements).
Another opportunity~\cite{Gordon:2007,Howlett:2017b,Mukherjee:2018,Kim:2019,Amendola:2021} will be
provided by the large sample of up to $10^6$ type Ia supernovae from LSST~\cite{LSST_science}.

To our knowledge a detailed treatment on the redshift-space velocity field has not
previously been carried out. However, a couple of works have considered some related aspects.

Kaiser and Hudson \cite{Kaiser:2014jca} described two effects arising from
random motion on the measurement of a galaxy's peculiar velocity from its redshift-space position.
The first effect is a consequence of the radial number density of galaxies increasing
with distance: 
Given that a galaxy will be displaced from its actual location due to random motion, a galaxy with
observed distance $r$ is thus more likely to have ``scattered'' down from a larger distance (say, $r+\dif r$),
than up from a shorter one ($r-\dif r$). This selection bias affects the observed sample,
but it is a separate consideration that does not bear on our theory-based study.

More closely related to our work is the second effect. This is of a similar nature to the first effect,
but arises specifically for an inhomogeneous
mass distribution (whereas the first exists even if the galaxies were to be homogeneously distributed).
In a spatial region centered, for example, on the nearside of a galaxy cluster (where the number
density of galaxies is increasing as a function of distance) the number of galaxies scattered into
the region from the farside will be larger than on the nearside, where there are fewer galaxies.
When conditioned on the cluster---modelled simply as a plane-wave density
perturbation $\delta_g$---the mean displacement is no longer zero-centered.
The measured LOS peculiar velocities $v_\|\equiv\v\cdot\n$ in this region then changes by an amount
$\Delta v_\|\simeq-(\partial\delta_g/\partial r)\sigma_v^2/H_0$, with $\sigma_v$ the velocity dispersion,
and $H_0$ the Hubble constant.
A linear-theory calculation then showed that the peculiar velocity is biased low by
$\Delta v_\|/v_\|\simeq-{k^2\sigma_v^2}/{(\beta H_0^2)}$, with $\beta\equiv f/b$ the distortion
parameter, $b$ the linear galaxy bias, and $k$ the wavenumber.
As a result, the velocity power spectrum is damped (isotropically) by a factor
$D^2(k)=(1-k^2\sigma_v^2/(\beta H_0^2))^2$.
(In fact, this damping factor holds more generally than the plane-wave mass distribution
considered by Kaiser and Hudson.%
\footnote{ We find that the same damping factor $D(k)$ applies even for an arbitrary mass distribution
$\delta=\delta_g/b$. This follows from recalling that $\v$ is a potential flow, and adopting the
plane-parallel limit (in which the LOS is fixed). Given these considerations we can write
$\partial(\nabla\cdot\v)/\partial r
    =\nabla^2 v_\|$.
Then by the linearized continuity equation
$\partial\delta_g/\partial r
=-\nabla^2 v_\|/(\beta H_0)$,
and so $\Delta v_\| \simeq {\sigma_v^2 \nabla^2 v_\|}/({\beta H_0^2})$.
The result is obtained upon taking the Fourier transform of $v_\|+\Delta v_\|$.
})

Though this simple analytic model can provide the right amount damping observed in the simulations of
Koda~{\it et al.}~\cite{Koda:2014}, it also predicts problematically that (\emph{i}) for a given wavenumber,
damping affects all Fourier modes equally, not just the LOS modes; and that (\emph{ii}) the amount of
damping depends on galaxy bias. Both (\emph{i}) and (\emph{ii}) do not follow from the mapping~\eqref{eq:mapping},
for if the transverse modes were damped it would imply that the angular positions also shift under
the mapping. Moreover, because only the velocity field participates in the mapping, galaxy bias is not
expected to appear. As we will show in this work, any field prescribed over redshift space can be formulated
in terms of its real-space version convolved with eq.~\eqref{eq:mapping}, independent of the dynamical
relation between the galaxy (or matter) density and velocity fields.

The linear analysis by Kaiser and Hudson was not aimed at producing a fully-working model, but
it does nevertheless identify a mechanism that can explain the damping. Here we will go further.
With the systematic approach taken in this work, we will show (among other things) that there is also a
long-range FoG-like effect, and because of which the random motions cannot be solely blamed for the
damping observed in simulations.

In another work, Okumura~{\it et al.}~\cite{Okumura:2014} performed a study similar to our own but for the
\emph{density-weighted} redshift-space velocity field (i.e.\ the ``momentum'' field).
Such a field is closely related~\cite{Sugiyama:2016} to the kinetic Sunyaev--Zel'dovich effect~\cite{kSZ};
and, in the context of a direct measurement of the velocity power, it is perhaps a more practical
summary statistic (in terms of having a mass weighting; see refs.~\cite{Howlett:2019,Zhang:2015}).
However, in this work we specifically have in mind the aforementioned multi-tracer analysis of which
the scientific return from upcoming surveys has been well studied~\cite{Koda:2014,Howlett:2017a,Howlett:2017b}.
(Though, note that our results also have relevance to a redshift-space-based analysis of velocity
two-point correlations~\cite{Jaffe:1995,Abate:2008,Johnson:2014kaa,Howlett:2017,Adams:2017,Turner:2021}.)

Like this earlier work~\cite{Okumura:2014}, we will also make use of the
``distribution-function approach''~\cite{Seljak:2011tx,McDonald:2009hs,DF2,DF3,DF4} of phase space. This approach
allows us to systematically compute from a derivative expansion the effects of RSD on density-weighted fields;
the velocity field, however, is not density weighted but we will show how it still admits a similar formulation.
We will go beyond this approach, though, by showing that another treatment is possible which does not
make any approximations related to the mapping~\eqref{eq:mapping}. (We will show in passing that the same
treatment applies equally well to the momentum field.)

This paper is organized as follows. In Section~\ref{sec:basic} we briefly review the theory of RSD
as it applies to the matter density field, and fix some of our notation. In Section~\ref{sec:vel-RS} we
introduce a more general formalism for RSD in terms of the phase-space distribution function \cite{Seljak:2011tx};
we will give a different derivation of the so-called velocity-moment expansion, and show in particular how
density-weighted fields in redshift space follow from simple convolution integrals.
In Section~\ref{sec:v} we derive the redshift-space velocity field using the results of the previous 
section; we show that the (exact) velocity field, not being density weighted, follows from a modified integral
formula in terms of a Green's function.
Section~\ref{sec:vel-Pk} presents the first model of the redshift-space velocity power spectrum using
the velocity-moment expansion and shows how the effect of RSD modifies a set of mode-coupling kernels.
Section \ref{sec:cum-exp-model} presents the second model, which is obtained using the integral formulae;
in this model we show that the power spectrum can alternatively be written in terms of the
statistics of pairwise velocities. To complete the study of the two-point statistics, in Section~\ref{sec:config} we
derive the redshift-space correlation function of the LOS velocities.
Conclusions and discussion follow in Section \ref{sec:discussion}.
Our main results are summarized in Section \ref{sec:conclusions}.
Several appendices collect supplementary material; of note is Appendix~\ref{app:FFT} describing
our fast numerical implementation of the power spectrum models which exploits the
Fast Fourier Transform (the ``FFTLog'' method~\cite{Simonovic:2017mhp}).\footnote{Our code is
publicly available and can be found at \url{https://github.com/lhd23/RSDPT-FFTLog/}.}
\\
\\
\noindent
{\it Notation and conventions.}
Our Fourier convention is
\be\label{eq:ft}
f(\x) =\fint{\k}\, \widetilde f(\k) \rme^{-\ii\k\cdot\x}\,,
\quad\quad
\widetilde f(\k) =\int\! \dif^3\x\, f(\x) \rme^{\ii\k\cdot\x}\,,
\ee
where $\fint{\k}$ is a shorthand for $\int\!{\dif^3\k}/{(2\pi)^3}$, which we will often
use throughout this paper. We work with the dimensionless velocity divergence defined as 
$\theta\equiv-\nabla\cdot\v/(aHf)$,
where $\v$ is the peculiar velocity field (or ``velocity''), $a$
is the scale factor, $H$ is the Hubble parameter, and $f$ is the linear growth rate.
We will largely use the conformal Hubble parameter $\calH\equiv (\dif a/\dif\tau)/a=aH$, with
$\tau$ the conformal time.
For numerical work we adopt a spatially-flat $\Lambda$CDM cosmology with
$\Omega_{m0}=0.315$, $\Omega_{b0}=0.04904$, $\sigma_{8}=0.829$, $n_s=0.966$; the
growth rate is parametrized in the usual way, i.e.\ $f(z)=\Omega_m(z)^{0.55}$~\cite{Wang:1998gt}.

\section{Background}\label{sec:basic}
This paper is about the peculiar velocity field in redshift space. It will however be useful
to recall some well-known results relating to RSD in the context of clustering
(see ref.~\cite{Hamilton:1997zq} for a review), as it will help us to understand some generic
features in a simpler setting.

The starting point to study clustering in redshift space is to use the fact that the number of
galaxies does not change in going from real space to redshift space; that is, the mapping \eqref{eq:mapping}
is number conserving. Formally, for an infinitesimal volume element $\dif^3\s$ in redshift space
centered at $\s$, and $\dif^3\x$ in real space centered at $\x$, we have the relation
\be\label{eq:mass-cons}
\big[1+\delta^s(\s)\big] \dif^3\s=\big[1+\delta(\x)\big] \dif^3\x
\ee
between the galaxy overdensity $\delta^s(\s)$ in redshift space and $\delta(\x)$
in real space. (Here we are assuming no galaxy bias for simplicity.)
From this relation we have that
\be\label{eq:dels-jac}
\delta^s(\s)=\frac{1}{J(\x)}\big[1+\delta(\x)\big]-1,
\qquad
J(\x)
\equiv \bigg|\frac{\dif^3\s}{\dif^3\x}\bigg| = 1+\frac{1}{\calH}\,\n\cdot\nabla(\v\cdot\n),
\ee
where $J$ is the Jacobian of the mapping \eqref{eq:mapping}.
By linearizing eq.~\eqref{eq:dels-jac} we obtain Kaiser's formula~\cite{Kaiser:1987qv},
which reads in Fourier space
\be\label{eq:kaiser-limit}
\tilde\delta^s(\k)=\big(1+f\mu^2\big)\tilde\delta(\k),
\ee
where $f$ is the linear growth rate, and $\mu$ is the cosine of the angle of separation between the wavevector $\k$
and the LOS.
According to this relation, structures viewed in redshift space will appear squashed along the LOS as the
coherent infall of galaxies causes a LOS-directed displacement towards higher-density regions. Kaiser's formula
is valid in the ``plane-parallel limit'' in which the LOS is fixed (at $\n=\hat\z$, say). In practice, this is a
good approximation when the pairwise separations of the galaxy sample are small compared to their distances.
Note that this formula assumes no velocity bias between galaxy velocity $\v_g$ and
the underlying matter flow, i.e.\ $\v_g=\v$.%
\footnote{It can be argued~\cite{bias_review} based on the equivalence principle that
galaxies and matter must respond identically to the gravitational field (on suitably large scales);
any deviations from $\v_g=\v$ enter as terms that are higher-order derivatives in $\v$,
which are suppressed on large scales. Quantitatively, this is also supported by $N$-body simulations that
show negligible halo velocity bias on scales $k\lesssim 0.2\,h\Mpc^{-1}$~\cite{Chen:2018,Zheng:2015}.}
At least on perturbative scales this appears to be a quite valid assumption~\cite{Chen:2018}, and so we will
assume no velocity bias throughout this work.

From the Kaiser formula~\eqref{eq:kaiser-limit} we have the power spectrum
\be\label{eq:Pk-kaiser}
P^s(k,\mu)=\big(1+f\mu^2\big)^2 \PL(k),
\ee
where $\PL(k)$ is the (real-space) linear power spectrum.
This shows that the power in redshift space is enhanced along the LOS. It also shows that it contains
information not present in real space: Distortions induce an anisotropy in the clustering statistics,
and the degree of anisotropy observed can be used to constrain the growth rate $f$.

Equation~\eqref{eq:Pk-kaiser} does not account for the fact that galaxies (typically residing in clusters)
also possess virial motion, in addition to their large-scale streaming motions. In redshift space,
virial motion gives rise to an elongation of structures along the LOS.
This is the FoG effect and, in contrast to the Kaiser effect, arises from the nonlinear regime. As such,
in the past a phenomenological approach was common~\cite{Ballinger:1996cd}, whereby eq.~\eqref{eq:Pk-kaiser} is simply
multiplied by a damping function, which preserves the Kaiser effect on large scales but suppresses the power on
small scales.

Recent modelling efforts are largely based on a framework developed by Scoccimarro~\cite{Scoccimarro:2004tg} which
is able to provide a consistent treatment of clustering on a wide range of scales.
In particular, it was shown that (in the plane-parallel limit) an exact formula for the redshift-space power spectrum
can be obtained:
\be\label{eq:P-sco04}
P^s(\k)
=\int\dif^3\r\: \rme^{\ii\k\cdot\r}
    \Big\langle\rme^{-\ii f\kz\Delta\uz}\big[1+\delta(\x)\big] \big[1+\delta(\x')\big]\Big\rangle,
\ee
where $\r=\x-\x'$ is the separation, $\kz\equiv\hat\k\cdot\hat\z$ is the LOS component of the wavevector, 
$\Delta\uz\equiv\uz(\x)-\uz(\x')$ is the pairwise LOS velocity, and $\uz\equiv\v\cdot\hat\z/(-\calH f)$.
Here the power spectrum is described in terms of moments of the pairwise LOS velocities; in particular,
we have that $\langle\cdots\rangle$ is the pairwise LOS velocity generating function.
This general relation is the starting point for a number of power spectrum models (e.g.~\cite{Taruya:2010,Vlah:2019}).
In configuration space, eq.~\eqref{eq:P-sco04} implies the anisotropic two-point correlation function,
\be\label{eq:xi-stream}
1+\xi^s(s_\perp,s_\|)=\int^\infty_{-\infty} \dif r_\|\,\big[1+\xi(r)\big] \, p(s_\|-r_\|\mid\r),
\ee
where $s_\|=s\mu$ and $r_\|=r\mu$ are real- and redshift-space separations along the LOS, respectively;
$s_\perp=r_\perp$ is the transverse separation;  $r=(r_\|^2+r_\perp^2)^{1/2}$; $\xi(r)$ is the (real-space)
two-point function; and $p$ is the probability distribution function of the pairwise LOS velocities at separation $\r$.
We see again that the distortions are characterized by the statistics of the pairwise velocities,
but this time recalling earlier ``streaming models''~\cite{Peebles:1980,Fisher:1995}. In the case of a (scale-dependent)
Gaussian distribution, we obtain the aptly named ``Gaussian streaming model''~\cite{Reid:2011,Wang:2014}, which is
widely-used today~\cite{Reid:2012,Bautista:2020}. Skew corrections to the Gaussian assumption have also
been considered~\cite{Cuesta-Lazaro:2020ihk}.

It is important to note that eqs.~\eqref{eq:P-sco04} and \eqref{eq:xi-stream} are derived by appealing
to number conservation~\eqref{eq:mass-cons}, which is specific to tracers of the density field.
Nevertheless, we will show, beginning with the distribution-function approach, that expressions similar
to eqs.~\eqref{eq:P-sco04} and \eqref{eq:xi-stream} also exist for the velocity field.

\section{Distribution-function approach to RSD\label{sec:vel-RS}}
A general procedure for prescribing fields in redshift space can be realized with the phase-space
distribution function of dark matter particles \cite{Seljak:2011tx,McDonald:2009hs,DF2,DF3,DF4}.
While this will require introducing a certain amount of formalism, the advantage gained is a
systematic way to treat RSD, extending the notion of ``density field in redshift space'' to
any other bulk quantity, such as the velocity field.
Indeed, by taking moments of the distribution function we can obtain the fluid description of dark
matter, given either in real space or redshift space. This section briefly reviews the
distribution-function approach to RSD, focussing on the ``velocity-moment expansion''
first given in \cite{Seljak:2011tx}. We will further show that the velocity-moment expansion follows
from more intuitive convolution formulae which permits another treatment
of RSD with its own advantages.

\subsection{Moments in real space}
\sloppy We begin with the one-particle phase-space distribution function,
$f_1(\x,\p,\tau)$, which gives the probability $f_1(\x,\p,\tau)\,\dif^3\x\,\dif^3\p$
of finding one particle in an infinitesimal phase-space volume $\dif^3\x\,\dif^3\p$
centered on comoving position $\x$ and momentum $\p$, at conformal time $\tau$.%
\footnote{We retain the subscript in $f_1$ that is often omitted so as to distinguish it from
the growth rate (denoted $f$).} The number-conservation argument may then be understood as a
consequence of conservation of  phase-space volume.
In the following, as we do not yet need to consider dynamics, we will
suppress the time dependence in $f_1(\x,\p,\tau)$ for brevity, and simply write $f_1(\x,\p)$
(much as for any quantities derived from it). 

The zeroth and first moments of the distribution function are of primary interest.
The zeroth moment corresponds to the mass density and the first moment corresponds to the
momentum density; the first cumulant is the bulk velocity and is given by the first moment
divided by the zeroth moment. That is, we have
\bea
1+\delta(\x)&= \int\!\dif^3\p\:f_1(\x,\p), \label{eq:1pdelta-x} \\
\v(\x)&
=\int \dif^3\p\,\Big(\frac{\p}{ma}\Big)\, f_1(\x,\p)\,
    {\Big/} \!\int \dif^3\p\:f_1(\x,\p), \label{eq:v-x}
\eea
where $m$ is the particle mass and $a$ is the scale factor. (The bulk velocity of particles
in the cosmological context is the peculiar velocity.) The second moment is the stress tensor,
$T_{ij}(\x)=[1+\delta(\x)]\sigma_{ij}(\x)$, where $\sigma_{ij}$ is the velocity dispersion given by
\be
\sigma_{ij}(\x)
=\bigg[\int\dif^3\p\:f_1(\x,\p) \,\Big(\frac{p_i}{ma}\Big)\Big(\frac{p_j}{ma}\Big)\, 
    {\Big/} \int\dif^3\p\:f_1(\x,\p)\bigg] - v_i(\x) v_j(\x).
\ee
We also have the mass density $\rho(\x)= ma^{-3}[1+\delta(\x)]$, and the momentum
density $\bm{\pi}(\x)=[1+\delta(\x)]\v(\x)$.

On cosmological scales the action of gravity is dominant and it is customary to treat
the matter particles as a pressureless perfect fluid (PPF), i.e.\ with vanishing stress
tensor $T_{ij}=0$ and pressure $p=0$. This simplifies the dynamics considerably: all higher
moments vanish and the ``Boltzmann hierarchy'' is closed for the zeroth and first moments.
This yields a system of equations in the density and velocity which can be consistently solved.
In this case the particle velocities are single-valued at each $\x$ (``single-streaming''), and
implies the distribution function takes the form
\be\label{eq:f-PPF}
f_1(\x,\p)=f_1^\mrm{PPF}[\delta,\v]=\big[1+\delta(\x)\big]\,\delD\big(\p-am\v(\x)\big),
\ee
i.e.\ in terms of the zeroth and first moments only.

\subsection{Moments in redshift space\label{sec:df-rsd}}
Redshift-space distortions cause a radial shift in the positions of tracers, with
the apparent position depending on the tracer's own state of motion. At the level of
phase space the apparent positions---the positions in redshift space---of point
particles become functions of their momenta. The observed configuration space then results
from a LOS ``projection'' of phase-space dynamics onto the actual configuration space.
This may be realized by convolving the real-space distribution function with the
real-to-redshift-space mapping~\eqref{eq:mapping} to obtain the \emph{redshift-space distribution function},
\be\label{eq:fs}
f_1^s(\s,\p)
\equiv\int \dif^3\x\,f_1(\x,\p)\,
    \delD\Big(\s-\x-\calH^{-1}\frac{\n\cdot\p}{ma}\n\Big)
=f_1\Big(\s-\calH^{-1}\frac{\n\cdot\p}{ma}\n,\p\Big)
\ee
where $\n=\x/|\x|=\s/|\s|$ is the LOS unit vector, and $\calH\equiv aH$.
(Throughout this paper a superscript $s$ will be used to indicate a quantity defined
over redshift space.) Progress can be made if we expand $f_1^s$ in powers of $\n\cdot\p/(ma)$, then
integrate out momenta $\p$ to, e.g.\ obtain the redshift-space density field, as done in ref.~\cite{Seljak:2011tx}.
As we will show, there is another approach that can be taken with its own advantages.

First, we note that the usual number-conservation argument provides a useful relation \eqref{eq:mass-cons}
between real- and redshift-space density fields, but that it cannot be applied more generally to the wanted
velocity field. We can however see that eq.~\eqref{eq:fs} is consistent with number conservation (as it should be).
For if the phase-space volume is conserved, 
$\int\dif^3\x\int\dif^3\p\,f_1(\x,\p)=\int\dif^3\s\int\dif^3\p\,f_1^s(\s,\p)$, we have locally
that $[1+\delta^s(\s)]\dif^3\s=[1+\delta(\x)]\dif^3\x$.

Next, making use of the specific PPF form given by eq.~\eqref{eq:f-PPF}, we can derive a convolution
formula from which the velocity-moment expansion is obtained.
To demonstrate, observe for the density field [cf.~eq.~\eqref{eq:1pdelta-x}]
\bea
1+\delta^s(\s)
&=\int\dif^3\p\: f_1^s(\s,\p) \nonumber\\
&=\int\dif^3\x\int\dif^3\p\:
    \delD\Big(\s-\x-\calH^{-1}\frac{\n\cdot\p}{ma}\n\Big)\, \big[1+\delta(\x)\big]\,\delD\big(\p-am{\v}(\x)\big) ,
\eea
where in the second line we substituted $f^s_1$ for eq.~\eqref{eq:fs}, in which $f_1$
is given by its PPF form \eqref{eq:f-PPF}. Upon integrating out momenta we have
\be\label{eq:1pdelta-s}
1+\delta^s(\s)
=\int \dif^3\x\: \big[1+\delta(\x)\big] \,\delD\big(\s-\x-\calH^{-1}v_\|(\x)\n\big).
\ee
This simple formula makes intuitive sense: to construct the redshift-space version of $1+\delta$,
take all real-space mass elements $1+\delta$ at $\x$ and reassign them to $\s$, according
to the real-to-redshift-space mapping (realized through the delta function).
Of course different $\x$ may give rise to the same $\s$, so naturally we are to
integrate over real space.
(Specifically, since the transverse components are unaffected by the mapping, we only need
to integrate over the LOS component $x_\|\equiv\x\cdot\n$.)

It is not hard to see that the formula \eqref{eq:1pdelta-s} generalizes straightforwardly to any
other moment we care to compute (in the PPF approximation), since we have that the $n$th moment is just
$M_{i_1\ldots i_n}\equiv(1+\delta)v_{i_1}\cdots v_{i_n}$. For example, the first moment is the
momentum field and reads
\be\label{eq:m-s}
\big[1+\delta^s(\s)\big]\v^s(\s)
=\int\dif^3\x\:
    \big[1+\delta(\x)\big]\v(\x)\, \delD\big(\s-\x-\calH^{-1} v_\|(\x)\n\big),
\ee
with $\v^s$ the redshift-space velocity field. Here we
see that, unlike the momentum field, the velocity field $\v^s$
is not density weighted so cannot apparently be expressed purely in terms of $\v$; rather it
is the first \emph{cumulant} of the distribution function.
However, as we will soon see, we can still write $\v^s$ in a way that resembles the
intuitive convolution form given by eqs.~\eqref{eq:1pdelta-s} and \eqref{eq:m-s} but
with slight modification.

\subsubsection{Relation to velocity-moment expansion}
We now show that the velocity-moment expansion of the distribution-function approach
follows from the integral representation \eqref{eq:1pdelta-s}. 
We will go through the steps in some detail as they will turn out to be instructive
for a subsequent calculation on the velocity field. 

Beginning with eq.~\eqref{eq:1pdelta-s}, we replace the Dirac delta function with its
plane-wave expansion to obtain
\bea\label{eq:del-s-plane-waves}
1+\delta^s(\s)
&=\int \dif^3\x\:
    \big[1+\delta(\x)\big]\,
    \int\!\dk\, \rme^{-\ii\k\cdot(\s-\x)} \rme^{\ii\calH^{-1}v_\|(\x)\k\cdot\n}.
\eea
This form gives another way to view eq.~\eqref{eq:1pdelta-s}, namely, as the
convolution of its real-space density field with plane waves of different velocity-induced phases.
Next, we define $k_\|\equiv\k\cdot\n$ and Taylor expand the second exponential to get
\be\label{eq:plane-wave-exp}
\rme^{\ii\calH^{-1}v_\|k_\|}
=\sum_{n=0}^\infty \frac{1}{n!} \big(\ii \calH^{-1}v_\|k_\|\big)^n
= 1 + \frac{1}{\calH}\ii v_\|k_\| - \frac{1}{2\calH^2}(v_\|k_\|)^2 + \cdots.
\ee
Since $(\nabla_\|)^n \rme^{-\ii\k\cdot(\s-\x)}=(-\ii k_\|)^n \rme^{-\ii\k\cdot(\s-\x)}$, each term in
the foregoing expansion can be generated by the action of a basis of differential operators acting on plane
waves. We can thus exchange powers of $k_\|$ for powers of $\nabla_\|=\n\cdot\partial/\partial\s$
(acting on the inner integral), and write eq.~\eqref{eq:del-s-plane-waves} as
\begin{align*}
1+\delta^s(\s)
&=\int \dif^3 \x\:
    \big[1+\delta(\x)\big]\,
    \sum_{n=0}^\infty\frac{1}{n!}\Big(-\frac{1}{\calH}v_\|(\x)\nabla_\|\Big)^n
    \int\dk\, \rme^{\ii\k\cdot(\s-\x)} \\[5pt]
&=\int \dif^3\x\: \big[1+\delta(\x)\big]\delD(\s-\x)
    - \frac{1}{\calH}\nabla_\|\int\dif^3\x\:\big[1+\delta(\x)\big]v_\|(\x)\delD(\s-\x)
    + \cdots. 
\end{align*}
Owing to the delta functions the integrals are easily done, and altogether yield the
\emph{velocity-moment expansion} in position space,
\be\label{eq:1pdel-s-sum}
1+\delta^s(\s)
=\sum_{n=0}^\infty \frac{1}{n!}\Big(\frac{-1}{\calH}\Big)^n {\nabla_\|}^n \, T_\|^{(n)}(\s),
\ee
where the density-weighted LOS velocity moments are defined as
\be\label{eq:T}
T_\|^{(n)}(\s)\equiv \big[1+\delta(\s)\big] v_\|(\s)^n.
\ee
Taking the Fourier transform of eq.~\eqref{eq:1pdel-s-sum} we then recover
[see equation (2.6) in ref.~\cite{Seljak:2011tx}]
\be
\tilde\delta^s(\k)
=\sum_{n=0}^\infty\frac{1}{n!}\bigg(\frac{\ii k_\|}{\calH}\bigg)^n \widetilde{T}_\|^{(n)}(\k),
\ee
with $\widetilde{T}_\|^{(n)}$ the Fourier transform of $T_\|^{(n)}$, and
the background ($k=0$) mode omitted.
Note that in ref.~\cite{Seljak:2011tx} $T_\|^{(n)}$ is defined in terms of a general
distribution function, whereas here we have adopted its PPF form \eqref{eq:f-PPF}. In the end this
makes no difference; we are invoking the simplifying PPF approximation at the
level of phase space, which will be assumed anyway for predictions using perturbation theory.

The above calculation can be repeated for any other moment of $f^s$, always resulting in the same form 
\eqref{eq:1pdel-s-sum}. More precisely, we also have expressions for the $n$th moment $M^s_{i_1\ldots i_n}$
by simply replacing $1+\delta^s$ with $M^s_{i_1\ldots i_n}$, and $1+\delta$ with $M_{i_1\ldots i_n}$.
For example, the momentum field $\bm\pi^s\equiv(1+\delta^s)\v^s$ is
\be
\bm\pi^s(\s)
=\sum_{n=0}^\infty \frac{1}{n!}\Big(\frac{-1}{\calH}\Big)^n {\nabla_\|}^n\Big[\big(1+\delta(\s)\big)v_\|(\s)^n\v(\s)\Big],
\ee
and the LOS component $\pi^s_\|\equiv\bm\pi^s\cdot\n$ can be written in terms of the LOS velocity
moments $T_\|^{(n)}$ as [cf.~eq.~\eqref{eq:1pdel-s-sum}]
\be\label{eq:pi-s-sum}
\pi^s_\|(\s)
=\sum_{n=0}^\infty \frac{1}{n!}\Big(\frac{-1}{\calH}\Big)^n {\nabla_\|}^n\, T_\|^{(n+1)}(\s).
\ee
We note that the Fourier transform of this expression was given in ref.~\cite{Okumura:2014}.

It is easy to check that Kaiser's formula \eqref{eq:kaiser-limit} is recovered
by truncating eq.~\eqref{eq:1pdel-s-sum} at $n=1$ and using linear theory.
It can also be shown that the two-point function $\langle(1+\delta^s)(1+\delta^s)\rangle$
implied by eq.~\eqref{eq:1pdelta-s} recovers Scoccimarro's streaming model~\eqref{eq:xi-stream}.
(In contrast to our previous calculations one instead leaves unexpanded the second
exponential in eq.~\eqref{eq:del-s-plane-waves}. Such an approach will be preferable
to the velocity-moment expansion for reasons we will discuss later in Section \ref{sec:virial}.)

\section{Redshift-space velocity field}\label{sec:v}
\subsection{Position space}
The velocity field in redshift space is defined in terms of the phase-space distribution
function by [cf.~eq.~\eqref{eq:v-x}]
\be
\v^s(\s)
\equiv\int \dif^3\p\,\Big(\frac{\p}{ma}\Big)\,f_1^s(\s,\p)
	{\,\Big/} \int \dif^3\p\,f_1^s(\s,\p),
\ee
i.e.\ the first cumulant of $f^s_1$ (first moment divided by the zeroth moment).
Specializing to the case of a pressureless perfect fluid (appropriate for dark matter), the moments
are thus given by eqs.~\eqref{eq:1pdelta-s} and \eqref{eq:m-s}, and we have the following exact
expression in terms of spatial integrals over real-space $\delta$ and $\v$:
\be\label{eq:v-s-frac}
\v^s(\s)
=\frac{\raisebox{0.04in}{$\int\dif^3\x\: [1+\delta(\x)]\v(\x)\,\delD\big(\s-\x-\calH^{-1}v_\|(\x)\n\big)$}}
    {\raisebox{-0.04in}{$\int\dif^3\x\: [1+\delta(\x)]\,\delD\big(\s-\x-\calH^{-1}v_\|(\x)\n\big)$}}.
\ee
According to this expression, the velocity field in redshift space appears to be density weighted.
Of course, in real space the velocity field is volume weighted, and we will show that upon formal
expansion of the right-hand side of eq.~\eqref{eq:v-s-frac}, it remains volume weighted in redshift
space, as well; i.e.\ the apparent density weighting vanishes through cancellation.
To obtain the expansion we introduce the bookkeeping parameter
$\epsilon$, then Taylor expand about $\epsilon=0$.
Thus, starting with eq.~\eqref{eq:v-s-frac}, performing a plane-wave expansion of the Dirac delta
functions contained therein, as in eq.~\eqref{eq:del-s-plane-waves}; taking $\v\to\epsilon\v$;
performing a formal expansion, $\v^s(\s) = \sum_n\epsilon^n\v^s_n(\s)$;%
\footnote{One can also expand in powers of $\delta$, though this turns out to be
unnecessary. The velocity is not a density-weighted field and the apparent dependence
on $\delta$ in eq.~\eqref{eq:v-s-frac} drops out in redshift space, as (trivially) occurs
in real space. As eq.~\eqref{eq:v-s-G1} shows, $\v^s$ can be written in terms of $\v$ only.}
setting $\epsilon=1$; we find
\be\label{eq:v-s-series}
\v^s(\s)
= \v(\s) + \sum_{n=1}^\infty \frac{1}{n!}\Big(\frac{-1}{\calH}\Big)^n{\nabla_\|}^{n-1}
        \Big[v_\|(\s)^n\nabla_\|\v(\s)\Big],
\ee
where $\nabla_\|=\n\cdot\nabla$.
Clearly we have that at leading order $\v^s\simeq\v$. (Contrast this with the overdensity
$\delta^s$ where, working to the same precision, we have eq.~\eqref{eq:kaiser-limit},
the Kaiser formula.)
Equation \eqref{eq:v-s-series} shows that RSD in the velocity field (given by the sum)
enters at next-to-leading order in the real-to-redshift-space mapping. As mentioned, notice
also that $\delta$ does not appear; this is to be expected as only $\v$ appears in the mapping~\eqref{eq:mapping}.

As a consistency check, we have verified that the product of the
expansions, eqs.~\eqref{eq:v-s-series} and \eqref{eq:1pdel-s-sum}, recovers the expansion \eqref{eq:pi-s-sum},
i.e.\ that we have $\bm\pi^s=(1+\delta^s)\v^s$. This is easy to check order-by-order;
the general case requires more work, which we give details on in Appendix~\ref{app:v-details}.

The physical meaning of the higher-order terms in eq.~\eqref{eq:v-s-series} can be understood
by observing that the series is organized as a derivative expansion,\footnote{Note that the density
and momentum expansions, eqs.~\eqref{eq:1pdel-s-sum} and \eqref{eq:pi-s-sum}, can also be described
as derivative expansions.}
\be\label{eq:grad-exp}
\v^s[\v]=\v^s_{(0)}+\v^s_{(1)}+\v^s_{(2)}+\cdots,
\ee
in that the $(n+1)$th term $\v^s_{(n)}=\v^s_{(n)}[\v]$ depends linearly on terms containing $n$th-order
derivatives, which includes products of lower-order derivatives.
We can thus view eq.~\eqref{eq:v-s-series} as a \emph{hydrodynamic gradient expansion}~\cite{Kovtun:2012},
where derivative terms $\v^s_{(1)},\v^s_{(2)},\ldots$ correspond to corrections to the perfect-fluid
approximation (zeroth-order terms). In particular, first derivatives give rise to dissipative effects in
the fluid when viewed in redshift space: it is no longer the perfect fluid of real space.
(Recall the continuity equation and Euler's equation describe the motion of a perfect fluid. When there
are dissipative effects Euler's equation is modified but the continuity equation is not \cite{LandauLifshitzFluid}.)
Since we will compute the leading nontrivial-order effect of RSD on the power spectrum we will need
to keep terms in the expansion to third order in $\v$:
\begin{subequations}\label{eq:vi-s}
\bea
\v^s_{(0)}&=\v, \\[6pt]
\v^s_{(1)}&=-\frac{1}{\calH}v_\|\nabla_\|\v, \\[3pt]
\v^s_{(2)}&=\frac{1}{\calH^2}v_\|\nabla_\|v_\|\nabla_\|\v +\frac{1}{2\calH^2}v_\|^2\nabla_\|^2\v.
\eea
\end{subequations}

The derivative expansion in eq.~\eqref{eq:v-s-series} should of course be understood
perturbatively so that---with the velocity field generated by a small matter density
perturbation $\delta$---the size of successive terms are of decreasing relevance.
This is easy to see if we consider for concreteness a plane-wave perturbation with wavenumber $k$,
for then derivative terms in eq.~\eqref{eq:v-s-series} are order $\nabla_\|v_\|/\calH\sim v_\|k/\calH$.
The $n$th term in the series \eqref{eq:v-s-series} then gives a correction to
real-space $\v$ of order
\be
(v_\|k/\calH)^{n-1}\sim \theta^{n-1}\sim\delta^{n-1},
\ee
where $\theta$ is the velocity divergence [defined below by eq.~\eqref{eq:theta}]. We therefore see
that higher-derivative terms correspond to higher powers of $\delta$, and are accordingly suppressed.

\subsubsection{A Green's function formula}
We now show that the derivative expansion of $\v^s$ results from a convolution integral of
the form given by eq.~\eqref{eq:1pdelta-s} for $1+\delta^s$. The formula is equivalent to the
series expansion \eqref{eq:vi-s} when considered to all orders, but has certain other advantages
when we come to compute the two-point statistics (Section \ref{sec:cum-exp-model}).

We have mentioned that the velocity $\v^s$ is the first cumulant of the distribution function
and so cannot apparently be written as a simple convolution.
This is partly true; it cannot be done for $\v^s$ but can be done for the \emph{gradient}
of $\v^s$. The trick is in observing that if we act with $\nabla_\|$ on
eq.~\eqref{eq:v-s-series} the resulting expansion is precisely of the same form as
the moment expansion \eqref{eq:1pdel-s-sum}; that is, we have
\be\label{eq:nablav-s-sum}
\nabla_\|\v^s(\s)
=\sum_{n=0}^\infty \frac{1}{n!}\Big(\frac{-1}{\calH}\Big)^n{\nabla_\|}^n
        \Big[v_\|(\s)^n\nabla_\|\v(\s)\Big].
\ee
By comparing this series to that of eq.~\eqref{eq:1pdel-s-sum}, we see that here
$\nabla_\|\v^s$ takes the place of $1+\delta^s$, while $\nabla_\|\v$ takes the place of
$1+\delta$ (in $T^{(L)}_\|$).
Because eq.~\eqref{eq:1pdel-s-sum} is the series expansion of eq.~\eqref{eq:1pdelta-s}
we deduce the relation
\be\label{eq:v-s-delta}
\nabla_\|\v^s(\s)
=\int\dif^3\x\:\nabla_\|\v(\x)\,
    \delD\big(\s-\x-\calH^{-1}v_\|(\x)\n\big),
\ee
i.e.\ the formal expansion of this integral formula is eq.~\eqref{eq:nablav-s-sum}.

One can view eq.~\eqref{eq:v-s-delta} as an equation of the form $\mathcal{L}\v^s=\bm\Delta$,
with the linear differential operator $\mathcal{L}=\nabla_\|$
and $\bm\Delta(\s)$ regarded as a source term. To solve this equation we will specialize
to the LOS component and adopt the plane-parallel limit (treating the LOS $\n$ as constant).
Thus projecting the differential equation onto $\n$ we have $\mathcal{L}v^s_\|=\Delta_\|(\s)$,
which can be solved by means of a Green's function $G$ to give
\bea
v_\|^s(\s)
&=\int\dif^3\s'\, G(\s-\s')
    \int\dif^3\x\:\nabla_\| v_\|(\x)\,
    \delD\big(\s'-\x-\calH^{-1}v_\|(\x)\n\big) \nonumber\\
&=\int\dif^3\x\:\nabla_\| v_\|(\x)\,
    G\big(\s-\x-\calH^{-1}v_\|(\x)\n\big). \label{eq:v-s-G1}
\eea
Comparing the second line with eq.~\eqref{eq:1pdelta-s} shows that the Green's function
takes the place of the delta function (as might be guessed from the nonlocal nature of $v^s_\|$).
To construct the Green's function we recall that $G$ is such that $\mathcal{L}G=\delD$.
In Fourier space this equation is solved by $\widetilde{G}(\k)=1/(-\ii\k\cdot\n)$, so
we have that%
\footnote{In the second line we have used the spectral representation
\be
\frac{1}{-\ii\k\cdot\n}
=\frac12\int^\infty_{-\infty}\dif\lambda\: \rme^{\ii\lambda \k\cdot\n}\, \mrm{sgn}(\lambda),
\ee
where $\mrm{sgn}(x)$ is the sign function. In other words, the Fourier transform of
$\mrm{sgn}(\lambda)/2$ is $1/(-\ii\k\cdot\n)$, for $\lambda$ dual to $\k\cdot\n$.}
\bea
G(\s)
&=\int\!\dk\: \frac{\rme^{-\ii\k\cdot\s}}{-\ii\k\cdot\n} \label{eq:G-Fourier}\\[3pt]
&=\frac12\int^\infty_0\dif\lambda\:\delD(\s-\lambda\n)
    -\frac12\int^\infty_0\dif\lambda\:\delD(\s+\lambda\n). \label{eq:G-config}
\eea
Note that this reduces to one-dimensional integrals because the LOS in the mapping~\eqref{eq:mapping}
only affects the radial coordinate.
We can also recognize the first and second term in the second line as the advanced and retarded
Green's function, respectively. We can then write $v_\|^s$ as (after making the change of
variable $\lambda=\calH^{-1}w$)
\be\label{eq:v-s-soln}
\begin{split}
v_\|^s(\s)
&=\frac12\int\dif^3\x\: \nabla_\| v_\|(\x)
    \int^\infty_0\calH^{-1}{\dif w}\:\delD\Big[\s-\x-\calH^{-1}\big(v_\|(\x)+w\big)\n\Big] \\[2pt]
&\quad-\frac12\int\dif^3\x\: \nabla_\| v_\|(\x) 
    \int^\infty_0\calH^{-1}{\dif w}\:\delD\Big[\s-\x-\calH^{-1}\big(v_\|(\x)-w\big)\n\Big].
\end{split}
\ee
This position-space formula is exact, and can be understood as an integral representation of the
expansion \eqref{eq:v-s-series} (in that we recover the LOS component of eq.~\eqref{eq:v-s-series}
if we expand the Green's function in plane waves).%
\footnote{It might be noticed that eq.~\eqref{eq:v-s-soln} is not the most general solution, for
if we return to $\mathcal{L}v^s_\|=\Delta_\|$ we see that we are free to add to
eq.~\eqref{eq:v-s-soln} a vector field $\v_h$ that satisfies the homogeneous equation
$\mathcal{L}\v_h=0$. This means $\v_h$ must have constant radial component.
Furthermore, because $\v_h$ is irrotational (because $\v^s$ is) we conclude that
it is a constant radial vector field (emanating from the observer).
This homogeneous solution should be discarded if are to recover the correct solution
\eqref{eq:v-s-series};
by not doing so the LOS component (the observable part) can
be specified arbitrarily, and therefore physically meaningless.
Of course, choosing suitable boundary conditions, such as that $\int\dif^3\x\,\v(\x)=0$
(no bulk flow), eliminates this freedom.
}
We will return to these expressions in Section \ref{sec:virial}.

\subsection{Fourier space\label{sec:Fourier-space}}
The power spectrum of eq.~\eqref{eq:vi-s} (and related two-point statistics) will be presented
in the following sections. We will thus need expressions for eq.~\eqref{eq:vi-s} in
Fourier space, which we will present in the remainder of this section.
It will be convenient, however, to work with the \emph{scaled} velocity field
$\u\equiv-\v/(\calH f)$, for $\u$ then has units of length, $\nabla\cdot\u$ is dimensionless,
and (in linear theory) $\delta=\theta$,
where
\be\label{eq:theta}
\theta(\x)\equiv\nabla\cdot\u = -\nabla\cdot\v/(\calH f)
\ee
is the (scaled) velocity divergence. Analogous relations also hold for the redshift-space versions.

We will now turn attention to the velocity divergence. In Fourier space eq.~\eqref{eq:theta}
reads
\be\label{eq:tilde-theta}
\tilde\theta(\k) = -\ii\k\cdot\widetilde\u(\k)=\ii\k\cdot\widetilde\v(\k)/(\calH f),
\ee
where tildes denote Fourier-space fields. 
In general, the velocity field can be decomposed into a curl-free (irrotational) part and
divergence-free (rotational) part. In our case the cosmic velocity field is to a good
approximation a potential flow (being sourced by the gradient of the potential $\Phi$) and
is determined by a scalar, the velocity divergence.%
\footnote{Vorticity does not arise in standard gravitational instability theory.
Although this does not preclude vorticity that was generated in the early Universe, any initial
vorticity will decay by the Hubble expansion as $1/a$ in the absence of anisotropic forces.}
Thus, knowledge of $\theta$ is sufficient to fully recover the velocity field, and in Fourier
space eq.~\eqref{eq:tilde-theta} yields the simple relation
\be\label{eq:v-theta}
\widetilde\v(\k)
=-\calH f\frac{\ii\k}{k^2}\,\tilde\theta(\k).
\ee
Passing to redshift space, the velocity field $\v^s$ (or $\u^s$) remains irrotational
so that it suffices to work with $\theta^s$. In other words, no curl modes are induced
by going into redshift space; the redshift mapping only modifies the \emph{radial} positions.
(It is easily checked that $\nabla\times\u^s=0$, starting from eq.~\eqref{eq:v-s-series}
and writing $\u$ as the gradient of a potential.)

The remainder of this section is devoted to the evaluation of
$\theta^s\simeq\theta^s_1+\theta^s_2+\theta^s_3$ [cf.~eq.~\eqref{eq:vi-s}] in terms
of the Fourier transform of the real-space field $\theta$ (for which predictions of its
$n$-point statistics are well known \cite{Bernardeau:2001qr}). To ease notation we will drop
the tildes on Fourier-space quantities and write, e.g. $\theta^s(\k)$, instead of $\tilde\theta^s(\k)$.
We compute the Fourier transform of each term $\theta^s_n$ in turn, using eq.~\eqref{eq:tilde-theta},
and noting that $\theta^s_1$ is leading order, $\theta^s_2$ is next-to-leading order, and $\theta^s_3$ is
next-to-next-to-leading order.
For the first term, clearly $\theta^s_1=\theta$ since $\v^s_{(0)}=\v$. The second
term is a composite of two fields so, using the convolution theorem and noting that the
Fourier transform of $\nabla_\|\v\equiv(\n\cdot\nabla)\v$ is $(-{\ii}\n\cdot\k)\v(\k)$,
we have
\be\label{eq:v2}
\v^s_{(1)}(\k)
=\frac{\ii}{\calH}\fint{\q}\,
	(\n\cdot\q)\,\v(\q)\,{v}_\|(\k-\q)
\ee
with our Fourier shorthand. This implies
\bea
\theta^s_2(\k)
&=
	f\!\fint{\q}\:
	\frac{(\n\cdot\q)(\q\cdot\k)\big[(\k-\q)\cdot\n\big]}{q^2|\k-\q|^2}\,
	\theta(\q) \, \theta(\k-\q) \nonumber \\[3pt]
&\equiv
	f\!\fint{\q}\:
	K^{(2)}(\k-\q,\q;\n)\, \theta(\q) \, \theta(\k-\q). \label{eq:theta-u-s}
\eea
In the second line we have defined the (dimensionless) mode-coupling kernel $K^{(2)}$, but
it will turn out to be more convenient to work with the symmetrized kernel defined as%
\footnote{Equation \eqref{eq:theta-u-s} can be equivalently written as
$\fint{\q_1}\fint{\q_2}K^{(2)}(\q_1,\q_2)\, \theta(\q_1)\theta(\q_2)(2\pi)^3\delD(\k-\q_1-\q_2)$.
Since all terms in eq.~\eqref{eq:v-s-series}
are products purely of velocity, or derivatives thereof, they become convolutions in Fourier space
(filtered by some kernel).
We then use the standard trick that, because the product of the fields is symmetric in the
variables $\q_1$ and $\q_2$, the integral against the anti-symmetric part of $K$ is zero. Thus,
only the symmetric part of $K$ needs to be considered.}
\bea
\Ks^{(2)}(\q_1,\q_2;\n)
&\equiv\frac12\Big(K^{(2)}(\q_1,\q_2;\n)+K^{(2)}(\q_2,\q_1;\n)\Big) \nonumber \\[2pt]
&=\frac12|\q_1+\q_2|^2\frac{(\q_1\cdot\n)(\q_2\cdot\n)}{(q_1\,q_2)^2}. \label{eq:K2}
\eea
(Note that there is no quantitative difference whether the integral \eqref{eq:theta-u-s} is
evaluated using $\Ks^{(2)}$ or $K^{(2)}$.)

Finally, the third-order term ${\v}^s_{(2)}$ corresponds to a scalar divergence
\be
\theta^s_3(\k)
= f^2 \fint{\q_1} \fint{\q_2} \fint{\q_3}
    \Ks^{(3)}(\q_1,\q_2,\q_3) \,
    \theta(\q_1) \, \theta(\q_2) \, \theta(\q_3) \,
    (2\pi)^3 \, \delD(\k-\q_1-\q_2-\q_3),
\ee
where we find for the symmetrized kernel after a straightforward (if lengthy) calculation
\bea
\Ks^{(3)}(\q_1,\q_2,\q_3;\n)
&\equiv\frac{1}{3!}\Big(K^{(3)}(\q_1,\q_2,\q_3;\n)+\text{5 perm.}\Big) \nonumber \\[2pt]
&=\frac16
    |\q_1+\q_2+\q_3|^2
    \big[(\q_1+\q_2+\q_3)\cdot\n\big] 
    \frac{(\q_1\cdot\n)(\q_2\cdot\n)(\q_3\cdot\n)}{(q_1\,q_2\,q_3)^2}. \label{eq:K3}
\eea
In particular, the only configuration we will need collapses this to the simple form
\be
\Ks^{(3)}(\k,\q,-\q;\n)
=-\frac{(\k\cdot\n)^2(\q\cdot\n)^2}{6\,q^4}.
\ee

Because all terms in eq.~\eqref{eq:v-s-series} are products of $\v$ (and derivatives thereof) we
find that the $n$th term in the series expansion of $\theta^s$ is given by the general formula
\be\label{eq:theta-s-n}
\theta^s_n(\k,z)
=f^{n-1} \bigg[\,\prod_{i=1}^n  \int\!\frac{\dif^3\q_i}{(2\pi)^3}\,\theta(\q_i,z)\bigg]
\Ks^{(n)}(\q_1,\ldots,\q_n)\,(2\pi)^3\delD\bigg(\k-\sum_{j=1}^n\q_j\bigg).
\ee
This is similar to expressions found in perturbation theory (Section \ref{sec:beyond}).
One difference here is each term has an overall factor of a certain power of the growth rate $f$;
higher-order terms have greater sensitivity to the growth rate (though understood perturbatively,
and being higher-derivative, these terms are of decreasing significance).
Here we have $\Ks^{(1)}=1$ (no distortions at leading order since $\theta^s=\theta$), while higher-order kernels $\Ks^{(n)}$
will depend on $\n$ via the LOS momenta $\hat\q_1\cdot\n$, $\hat\q_2\cdot\n$, etc.
We may also notice that $\theta^s$, unlike $\delta^s$, is determined only by its real-space
equivalent $\theta$. This can be traced to the real-to-redshift-space mapping being a function
of the velocity field; the velocity field is volume weighted, not density weighted (as are all
other cumulants of the distribution function).

To summarize, $\theta$ (and therefore $\theta^s$) is here taken to be a fully-nonlinear field. The
resulting formal expansion is of course to be treated perturbatively (for the series to be convergent),
i.e.\ $\theta\sim vk/\calH\sim\delta$ is a small fluctuation about zero, as is the working assumption
in the distribution-function approach. We have hence exhibited in
eq.~\eqref{eq:vi-s} the expansion up to third order, and presented the associated (symmetrized)
kernels in eqs.~\eqref{eq:K2} and \eqref{eq:K3}, which we will use in the following section.

\section{Power spectrum model I: moment-expansion approach\label{sec:vel-Pk}}
Now that we have an expression \eqref{eq:theta-s-n} for the redshift-space velocity field
we will compute its power spectrum.
So far though we have only treated (rather formally) the mapping from real space to
redshift space \eqref{eq:mapping} without dynamical considerations;
we did not make explicit use of the Boltzmann (or fluid) equations when considering
the distribution function (or its moments). In order to obtain a model of the power spectrum
we will thus need to specify the dynamical evolution.
Since eq.~\eqref{eq:theta-s-n} describes the
redshift-space velocity divergence $\theta^s(\k,z)$ in terms of its real-space equivalent
$\theta(\k,z)$, it will suffice to fix the dynamics in real space.
The velocity power spectrum is easily recovered from the velocity divergence power spectrum
using eq.~\eqref{eq:v-theta}.

We will require a nonlinear description of $\theta(\k,z)$. As we
have mentioned, linear theory is inadequate to see the effect of RSD;
there is no analogue to the well-known Kaiser effect in the large-scale limit, for
there is no mean streaming velocity (in the way that there is a mean density).
Nonlinearities must be considered and for this we will use Eulerian
standard perturbation theory (SPT) at one-loop precision \cite{Bernardeau:2001qr}.
This will allow us to compute the leading-order contribution to the nonlinear power spectrum
(the one-loop power spectrum).
As we will see the RSD imprint on quasilinear and
nonlinear scales, in contrast to the clustering picture, which is apparent
on all scales.

\subsection{Real-space velocity divergence power spectrum\label{sec:beyond}}
The large-scale structure exhibited in the distribution of matter is the result of
gravitational instability. Regions slightly overdense initially are further enhanced
over time as more matter accumulates. 
The velocity field then arises from potential gradients set up by matter density fluctuations.
In the regime where the fluctuations may be considered small the matter density field
(or velocity field) is well described by linear dynamics, which has the property that
(large-scale) Fourier modes grow at the same rate.
However, over time linear theory breaks down as gravity drives initially small density
fluctuations towards nonlinearity.

This subsection reviews how to systematically compute nonlinear corrections
to linear theory using one-loop PT. We will keep the discussion brief by presenting only the results
needed to specify our power spectrum model (given by eq.~\eqref{eq:Ps-NL} below).

In SPT the dynamics are governed by the continuity, Euler, and Poisson
equations, describing the time evolution of a self-gravitating, pressureless perfect fluid.
Provided that the amplitude of the overdensity field $\delta$ and velocity
divergence field $\theta$ is small these equations can be solved perturbatively with the
power-series ansatz
\bea
\delta(\k,z)&=\sum_{n=1}^\infty D(z)^n \delta^{(n)}(\k), \label{eq:delta-power}\\[2pt] 
\qquad
\theta(\k,z)&=\sum_{n=1}^\infty D(z)^n \theta^{(n)}(\k), \label{eq:theta-power}
\eea
where $D(z)$ is the linear growth factor (here normalized to unity at $z=0$), and the velocity
divergence is defined in eq.~\eqref{eq:theta}.\footnote{Our definition of $\theta$ absorbs an overall factor
of $-\calH f$ that often appears on the right-hand side of eq.~\eqref{eq:theta-power}.}
In the following we work at $z=0$ to suppress unimportant factors of $D^n$ in calculations; they
are of course trivial to carry through and can be restored in the end result. (Note that
peculiar velocity surveys probe low redshifts, $z\lesssim0.01$, so setting $D=1$ is appropriate.)

The first-order solution shows that $\delta^{(1)}(\k)=\theta^{(1)}(\k)$ and recovers
the usual linear theory predictions---namely,
$\delta(\k,z)=D(z)\delta^{(1)}(\k)=\theta(\k,z)$ and $\v(\k,z)=-(\ii\k/k^2)\calH f\delta(\k,z)$.
If we consider the second-order solution, we will also have a term quadratic in the linear
field $\delta^{(1)}$; in general, the $n$th-order term (e.g.\ $\theta^{(n)}$)
contains a product of $n$ linear fields.
Working at the same level of precision as before, nonlinearities in the divergence
field need to be considered up to third order, $\theta\simeq\theta^{(1)}+\theta^{(2)}+\theta^{(3)}$.
Solving the fluid equations with the ansatz \eqref{eq:theta-power} we have the
standard mode-coupling formula,
\be\label{eq:theta-n}
\theta^{(n)}(\k)
=\bigg[\,\prod_{i=1}^n \int\!\frac{\dif^3\q_i}{(2\pi)^3}\,\delta^{(1)}(\q_i)\bigg]
    \Gs^{(n)}(\q_1,\ldots,\q_n)\,(2\pi)^3\delD\bigg(\k-\sum_{j=1}^n\q_j\bigg),
\ee
where the mode-coupling kernels $\Gs^{(n)}$ are obtained from a set of
recursion relations (that can be found in ref.~\cite{Bernardeau:2001qr}).\footnote{A similar expression
exists for $\delta^{(n)}$ in terms of kernels $F_\mrm{S}^{(n)}$, but we will not need it.}
In general, the $n$th-order kernel $\Gs^{(n)}$ depends on a combination of lower-order kernels,
up to $F_\mrm{S}^{(n-1)}$ and $\Gs^{(n-1)}$. At one loop we need only consider the second- and
third-order kernels. The kernel $\Gs^{(3)}$ is not widely used, so for
convenience we present it here (together with $\Gs^{(2)}$):
\bea
\Gs^{(2)}(\q_1,\q_2)
&=\frac37 + \frac47 \frac{(\q_1\cdot\q_2)^2}{q_1^2q_2^2}
    + \frac12\frac{\q_1\cdot\q_2}{q_1q_2}\left(\frac{q_1}{q_2}+\frac{q_2}{q_1}\right), \label{eq:G2} \\[10pt]
\Gs^{(3)}(\k,\q,-\q)
&=\frac1{|\k-\q|^2}\bigg[
-\frac{2 (\k\cdot\q)^3}{63 k^2 q^2}+\frac{4 (\k\cdot\q)^3}{63 q^4}+\frac{(\k\cdot\q)^2}{36 k^2}-\frac{5 (\k\cdot\q)^2}{84 q^2} \nonumber\\[2pt]
&\phantom{=\frac1{|\k-\q|^2}\!}\quad
-\frac{k^2 (\k\cdot\q)^2}{18 q^4}+\frac{\k\cdot\q}{252}+\frac{19 k^2 \k\cdot\q}{252 q^2}-\frac{k^2}{42}
\bigg] \nonumber\\[3pt]
&
\,+\frac1{|\k+\q|^2}\bigg[
-\frac{5 (\k\cdot\q)^3}{63 k^2 q^2}-\frac{11 (\k\cdot\q)^3}{63 q^4}-\frac{(\k\cdot\q)^2}{36 k^2}-\frac{43 (\k\cdot\q)^2}{252 q^2} \nonumber\\[2pt]
&\phantom{\quad+\frac1{|\k+\q|^2}\,}
-\frac{k^2 (\k\cdot\q)^2}{9 q^4}-\frac{\k\cdot\q}{252}-\frac{19 k^2 \k\cdot\q}{252 q^2}-\frac{k^2}{42}
\bigg].
\label{eq:G3}
\eea
Note that the general expression for $\Gs^{(3)}(\q_1,\q_2,\q_3)$ is rather lengthy so we have written it
in the only configuration needed for the power spectrum calculation. (In fact, the
the only part of $\Gs^{(3)}$ needed to calculate the one-loop spectrum is the angle-averaged
part, which depends on $\mu'\equiv\hat\k\cdot\hat\q$. And while integrating out $\mu'$ is
straightforward, the above form allows for fast numerical
evaluation by exploiting the Fast Fourier Transform; see Appendix \ref{app:FFT}.)

\begin{figure}[!t]
\makebox[\linewidth][c]{%
\begin{subfigure}[b]{.49\textwidth}
\centering
\includegraphics[width=1.\textwidth]{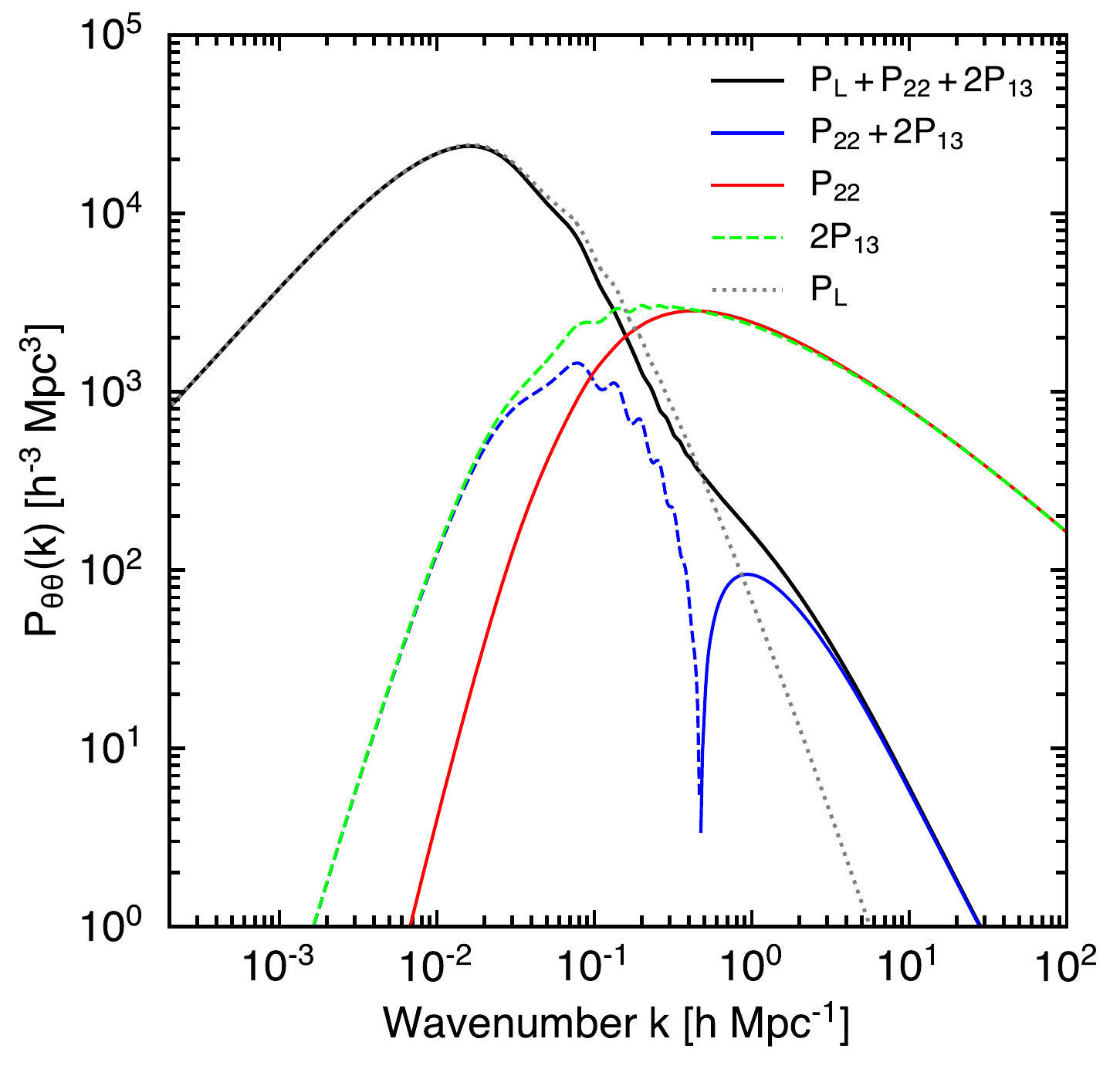}
\end{subfigure}%
\begin{subfigure}[b]{.49\textwidth}
\centering
\includegraphics[width=0.963\textwidth]{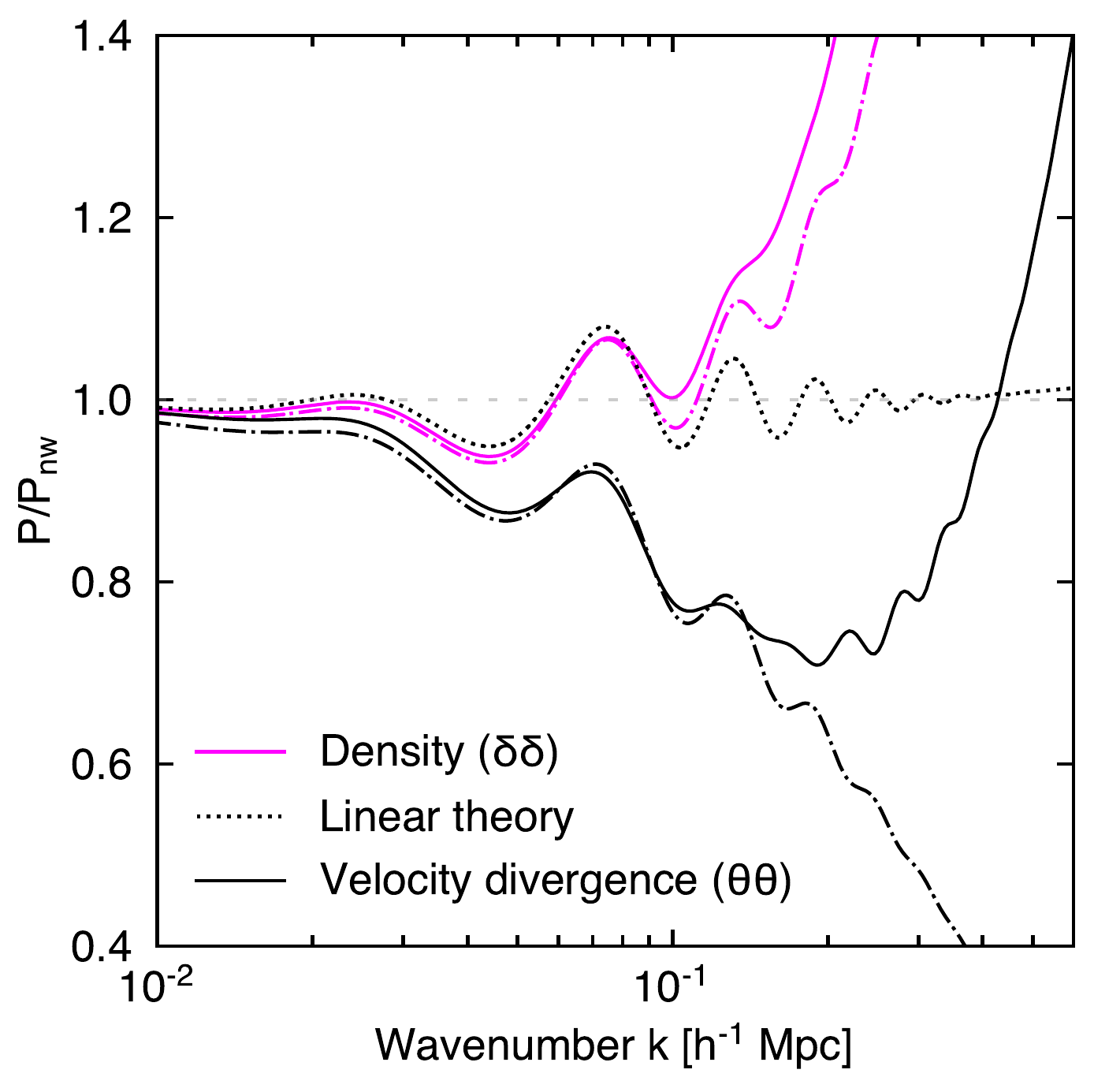}
\end{subfigure}%
}
\caption{Real-space power spectra at $z=0$.
        {\it Left panel}: The one-loop velocity divergence power spectrum (solid black line) is
        comprised of the linear power spectrum $\PL$ (dotted black line) and nonlinear
        corrections, $P_{22}$ (solid red line) and $P_{13}$ (dashed green line).
        Note that power spectrum contributions can be negative valued
        and in that case we show instead the absolute value (dashed lines).
        {\it Right panel}: Power spectra normalized to the linear ``no-wiggle'' power spectrum
        \cite{Eisenstein:1997ik}. The inclusion of one-loop corrections (solid lines) 
        is seen to break the degeneracy in linear theory (dotted black) 
        between the density and velocity divergence power spectra.
        The scales on which the one-loop power spectra is able to match simulations
        can be seen through comparison with the nonlinear power spectra (dash-dotted lines)
        indicated by the empirically-calibrated fitting formulae for density \cite{smith03,takahashi12} 
        (``Halofit'') and velocity divergence \cite{Hahn:2014lca} auto-power spectra.
        Notice in particular that for the velocity divergence the departure from linear theory
        (dotted black line) is seen to begin sooner (i.e.\ at smaller $k$) than for
        the density; that is, on scales where linear theory is adequate for density, nonlinear
        corrections are needed for the velocity divergence.}
  \label{fig:Poneloop}
\end{figure}

Turning to the power spectrum, the nonlinear contributions will come from a
nonvanishing bispectrum,
as well as the usual one-loop corrections to the real-space power spectrum:
\be\label{eq:one-loop}
P^\text{1-loop}_{\theta\theta}(k)
=   \PL(k) + P_{22}(k) + 2P_{13}(k),
\ee
where the power spectrum is defined by
$\langle\theta(\k)\theta(\k')\rangle=(2\pi)^3\delD(\k+\k')P_{\theta\theta}(k)$; the linear
power spectra is degenerate, $\PL\equiv P^\mrm{L}_{\delta\delta}=P^\mrm{L}_{\theta\theta}$;
and the one-loop contributions reads
\begin{subequations}\label{eq:Ploop-corr}
\bea
P_{22}(k)&= 2\fint{\q}\,\big[\Gs^{(2)}(\q,\k-\q)\big]^{2}\PL(q)\,\PL(|\k-\q|), \label{eq:22} \\
P_{13}(k)&= 3\PL(k)\!\fint{\q}\,\Gs^{(3)}(\k, \q,-\q) \,\PL(q). \label{eq:13}
\eea
\end{subequations}
The subscripts of these terms indicates the number of fields evaluated at each of the two points;
e.g.\ $P_{13}$ corresponds to three linear fields at the first location and a single linear field at
the second (the factor of two in eq.~\eqref{eq:one-loop} accounts for the reverse case). Note that the
factors of two and three in eqs.~\eqref{eq:22} and \eqref{eq:13} result from using the symmetrized
forms of the kernels.

Finally, the last ingredient we will need is the tree-level bispectrum,
\be\label{eq:Bttt}
B_{\theta\theta\theta}(k_1,k_2,k_3)=2\Gs^{(2)}(\k_1,\k_2)\PL(k_1)\PL(k_2) + \text{2 cyc.},
\ee
where we recall that the bispectrum is defined by ${\langle\theta(\k_1)\theta(\k_2)\theta(\k_3)\rangle
=(2\pi)^3\delD(\k_1+\k_2+\k_3) B_{\theta\theta\theta}(k_1,k_2,k_3)}$.\footnote{Note that depending
on the definition of $\theta$ a factor of $(-\calH f)^3$ may appear on the right-hand side of
eq.~\eqref{eq:Bttt}. With our definition \eqref{eq:theta} this factor has been absorbed into
the bispectrum.}
Since translation invariance implies the triangle condition, $\k_1+\k_2+\k_3=0$, the
bispectrum depends only on three numbers, either three lengths or two lengths and one angle.

In figure~\ref{fig:Poneloop} we show the one-loop prediction of the (real-space) power spectrum.
While in linear theory the (scaled) velocity divergence is equivalent to the matter density,
in nonlinear theory there is a decrease in power on scales $k\lesssim0.1\,h\Mpc^{-1}$. This has
been observed in simulations \cite{Scoccimarro:2004tg,Pueblas:2008uv,Carlson:2009it,Koda:2014}. By
contrast the linear density field remains valid up to $k\simeq 0.1\,h\Mpc^{-1}$.
Taking into account nonlinear effects therefore shows that the velocity field grows
more slowly than predicted by linear theory, leading to a decrease in power. This
would be the case for the density field as well if not for an accidental cancellation 
that occurs over the same range of scales. The difference is that the velocity divergence
field responds more readily to tidal forces, giving rise to non-radial motions that
tends to counter gravitational collapse \cite{Bernardeau:2001qr}.
Unlike the density field the onset of the nonlinear regime of the velocity field begins
at lower $k$ (larger scales).

\subsection{Redshift-space velocity divergence power spectrum}
So far we have presented the well-known real-space statistics by expanding
$\theta(\k)$ using third-order perturbation theory. We will now 
use these results to compute the one-loop redshift-space power
spectrum. To see what terms are relevant consider again
\be\label{eq:theta-s-3rd}
\theta^s(\k)=\theta^s_1(\k) +\theta^s_2(\k)+\theta^s_3(\k)+\cdots,
\ee
where the subscript denotes the number of (nonlinear) fields being coupled, so
$\theta^s_1(\k)\equiv\theta(\k)$ is just the real-space field given by eq.~\eqref{eq:theta-power},
$\theta^s_2(\k)$ denotes the convolution of two $\theta(\k)$ fields filtered by the kernel $\Ks^{(2)}$, etc.

Again, we have suppressed the dependence of $\theta^s_2,\theta^s_3,\ldots$ on the LOS $\n$.
[The subscripts in eq.~\eqref{eq:theta-s-3rd} are not to be confused with the superscripts in
eq.~\eqref{eq:theta-n} that are instead used to indicate the number of \emph{linearized} fields
being coupled.]
In this paper we will work under the plane-parallel
limit, allowing us to treat the LOS $\n$ as a constant across the imagined galaxy sample.%
\footnote{While we do not consider wide-angle effects in this work, given that realistic
peculiar velocity surveys are statistics limited (with $\sim20\%$ measurement uncertainties),
it might be expected that their impact on analysis is modest or negligible.}
Since we only have to consider a single LOS we will take this to be along the $z$-axis $\n=\hat\z$
and denote the LOS velocity component by $\vz$, rather than $v_\|$.

With the mapping and field specified we can now calculate the redshift-space power
spectrum. We have for the two-point function
\bea\label{eq:ts-ts}
\big\langle\theta^s(\k)\,\theta^s(\k')\big\rangle
=\big\langle\theta^s_1(\k)\,\theta^s_1(\k')\big\rangle
    +2\big\langle\theta(\k)\,\theta^s_2(\k')\big\rangle
    +2\big\langle\theta^s_1(\k)\,\theta^s_3(\k')\big\rangle
    +\big\langle\theta^s_2(\k)\,\theta^s_2(\k')\big\rangle
    +\cdots\, .
\eea
The power spectrum $P^s_{\theta\theta}(\k)$ is defined by
${\langle\theta^s(\k)\,\theta^s(\k')\rangle=(2\pi)^3\delta_\mrm{D}(\k+\k')P^s_{\theta\theta}(\k)}$.
The first term on the right-hand side of eq.~\eqref{eq:ts-ts} is the usual two-point function
in the absence of RSDs that we have already seen in Section \ref{sec:beyond}.
The remaining terms are then induced by RSD and will be calculated in this section.
Since we model the redshift-space power spectrum at one loop we truncate
$\theta^s$ at third order in $\delta^{(1)}$. Accordingly, eq.~\eqref{eq:ts-ts} reads
\be\label{eq:thetas-thetas}
\big\langle\theta^s(\k)\,\theta^s(\k')\big\rangle
\simeq\big\langle\theta(\k)\,\theta(\k')\big\rangle
    +2\big\langle\theta(\k)\,\theta^s_2(\k')\big\rangle
    +2\big\langle\theta^{(1)}(\k)\,\theta^s_3(\k')\big\rangle
    +\big\langle\theta^s_2(\k)\,\theta^s_2(\k')\big\rangle.
\ee
The first term on the right-hand side corresponds to the fact that the redshift-space velocity field is at
leading order equal to its real-space equivalent, as already discussed.
The other terms are therefore induced in redshift space and read, upon
substituting in eq.~\eqref{eq:theta-s-n} together with eq.~\eqref{eq:theta-power},
\begin{subequations}
\bea
\big\langle\theta(\k)\,\theta^s_2(\k')\big\rangle
&= f\!\fint{\q}\, \Ks^{(2)}(\q,\k'-\q) \,
    \big\langle\theta(\k)\,\theta(\q)\,\theta(\k'-\q)\big\rangle_\text{tree}, \label{eq:12s} \\[6pt]
\big\langle\theta^{(1)}(\k)\,\theta^s_3(\k')\big\rangle
&= f^2\!\fint{\q} \fint{\q'}\,
    \Ks^{(3)}(\q,\q',\k'-\q-\q')
    \big\langle\theta^{(1)}(\k)\,\theta^{(1)}(\q)\,
        \theta^{(1)}(\q')\,\theta^{(1)}(\k'-\q-\q')\big\rangle, \label{eq:13s} \\[6pt]
\big\langle\theta^s_2(\k)\,\theta^s_2(\k')\big\rangle
&=f^2\!\fint{\q} \fint{\q'}\,
    \Ks^{(2)}(\q,\k-\q)\Ks^{(2)}(\q',\k'-\q')
    \big\langle\theta^{(1)}(\q)\,\theta^{(1)}(\k-\q)\,
        \theta^{(1)}(\q')\,\theta^{(1)}(\k'-\q')\big\rangle. \label{eq:2s2s}
\eea
\end{subequations}
Equations \eqref{eq:13s} and \eqref{eq:2s2s} can be expressed in terms of two-point
functions using Wick's theorem; alternatively, if we notice that the loop integrals
are of the ``13'' and ``22'' forms (cf.~eqs.~\eqref{eq:22} and \eqref{eq:13}), we can write at once
\begin{subequations}
\bea
\big\langle\theta^{(1)}(\k)\,\theta^s_3(\k')\big\rangle
&= (2\pi)^3\delD(\k+\k') \, f^2 \PL(k)\!\fint{\q}\, 3\Ks^{(3)}(\k,\q,-\q) \, \PL(q), \\[5pt]
\big\langle\theta^s_2(\k)\,\theta^s_2(\k')\big\rangle
&= (2\pi)^3\delD(\k+\k') \, f^2 \fint{\q}\,
    2\big[\Ks^{(2)}(\q,\k-\q)\big]^2
    \PL(q) \, \PL(|\k-\q|).
\eea
\end{subequations}
This leaves eq.~\eqref{eq:12s}, which involves a three-point function;
thus, substituting in the tree-level bispectrum \eqref{eq:Bttt} we have
\begin{subequations}
\bea
\big\langle\theta(\k)\,\theta^s_2(\k')\big\rangle
&=(2\pi)^3\delD(\k+\k') \, f \fint{\q}\,\Ks^{(2)}(\q,\k-\q) \, B_{\theta\theta\theta}(k,q,|\k-\q|) \label{eq:iB1} \\
&= (2\pi)^3\delD(\k+\k') \, f \, \bigg[\,4\PL(k)\!\fint{\q}\,
	\Ks^{(2)}(\q,\k-\q) \, \Gs^{(2)}(\k,-\q) \, \PL(q)\nonumber \\
&\quad
      + 2\!\fint{\q}\,
	    \Ks^{(2)}(\q,\k-\q) \, \Gs^{(2)}(\q,\k-\q) \, \PL(q) \, \PL(|\k-\q|)\bigg].
\label{eq:iB}
\eea
\end{subequations}
In addition to substituting in the bispectrum, the first equality has been obtained by
taking $\q\to-\q$ in the integrand (since the integration is over all space), followed
by setting $\k'=-\k$ using the delta function;
the second equality is obtained by 
taking $\q\to\k-\q$;
in both equalities we have used that $\Ks^{(n)}$ and $\Gs^{(n)}$ are parity
symmetric, e.g.\ $\Ks^{(2)}(\q_1,\q_2)=\Ks^{(2)}(-\q_1,-\q_2)$.
The result of these simplifications shows that eq.~\eqref{eq:iB1} yields additional
``13'' and ``22'' loop contributions.

Now, gathering the previous results we can write the redshift-space power
spectrum succinctly as [cf.~eq.~\eqref{eq:one-loop}]
\be\label{eq:Ps-NL}
P_{\theta\theta}^s(\k) = \PL(k) + P^s_{22}(\k)+2P^s_{13}(\k),
\ee
where the 13- and 22-loop integrals are brought into the forms [cf.~eqs.~\eqref{eq:22} and \eqref{eq:13}]
\begin{subequations}\label{eq:P22-P13-s}
\bea
&P^s_{22}(\k)\equiv 2\!\fint{\q}\, \big[Z^{(2)}(\q,\k-\q)\big]^2 \PL(q) \, \PL(|\k-\q|), \\[4pt]
&P^s_{13}(\k)\equiv 3\PL(k)\!\fint{\q}\, Z^{(3)}(\k,\q,-\q) \, \PL(q),
\eea
\end{subequations}
with the \emph{redshift-space velocity divergence kernels},
\begin{subequations}\label{eq:Z-kernels}
\bea
Z^{(1)}(\q)&\equiv 1, \label{eq:Z1}\\[4pt]
Z^{(2)}(\q_1,\q_2) &\equiv \Gs^{(2)}(\q_1,\q_2) + f\Ks^{(2)}(\q_1,\q_2), \label{eq:Z2}\\[4pt]
Z^{(3)}(\q_1,\q_2,\q_3)
&\equiv \Gs^{(3)}(\q_1,\q_2,\q_3)
    +  \frac43 f\Ks^{(2)}(\q_2,\q_1+\q_3)\, \Gs^{(2)}(\q_1,\q_3)
    + f^2\Ks^{(3)}(\q_1,\q_2,\q_3). \label{eq:Z3}
\eea
\end{subequations}
[Although not needed here, $Z^{(1)}$ is defined in accordance with the
lowest-order redshift map being the identity map.]
These kernels are akin to the redshift-space {density} kernels given in ref.~\cite{Scoccimarro:1999ed}.
There is one obvious point of difference, however, which is that here linear theory is
inadequate to reveal the RSD effect on the velocity field: the first-order kernel $Z^{(1)}$
is unity, whereas for the density field it is equal to the Kaiser factor, $1+f\mu^2$
[$\mu=\hat\k\cdot\hat\z$].
We can notice that these kernels carry additional information about the growth rate
(and potentially be used to break degeneracies with other parameters).
And as with $\Ks^{(n)}$, these kernels are functions that depend on $\n$ through the scalars
$\n\cdot\q_i$ (though we have suppressed their dependence here). Finally, we remark that it is
straightforward enough to calculate the bispectrum of $\theta^s$ using these kernels, though
we point out that $Z^{(2)}$ is already in symmetrized form, but $Z^{(3)}$ is not (because of the
cross-term).

With these expressions we can thus compute $P^s_{\theta\theta}(\k)$ in an analogous way
to the one-loop power spectrum \eqref{eq:one-loop}, but instead with direction-dependent kernels.
However, for now we will keep separate the RSD terms (terms that depend on $\Ks$)
as doing so will help to simplify the extraction of multipoles later on.

\begin{figure}[!t]
\makebox[\linewidth][c]{%
\begin{subfigure}[b]{.48\textwidth}
\centering
\includegraphics[width=1.\textwidth]{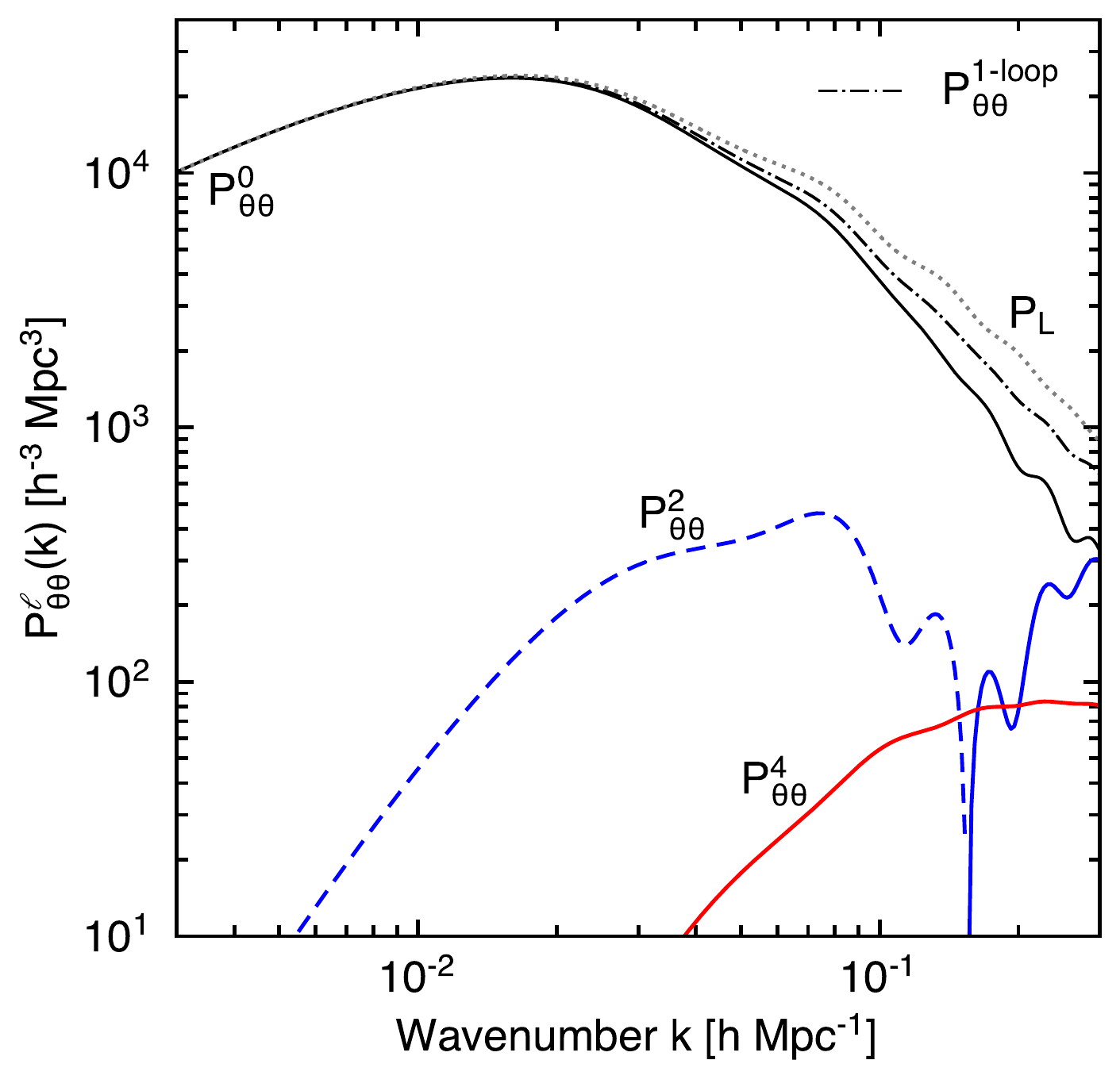}
\end{subfigure}%
\begin{subfigure}[b]{.48\textwidth}
\centering
\includegraphics[width=1.\textwidth]{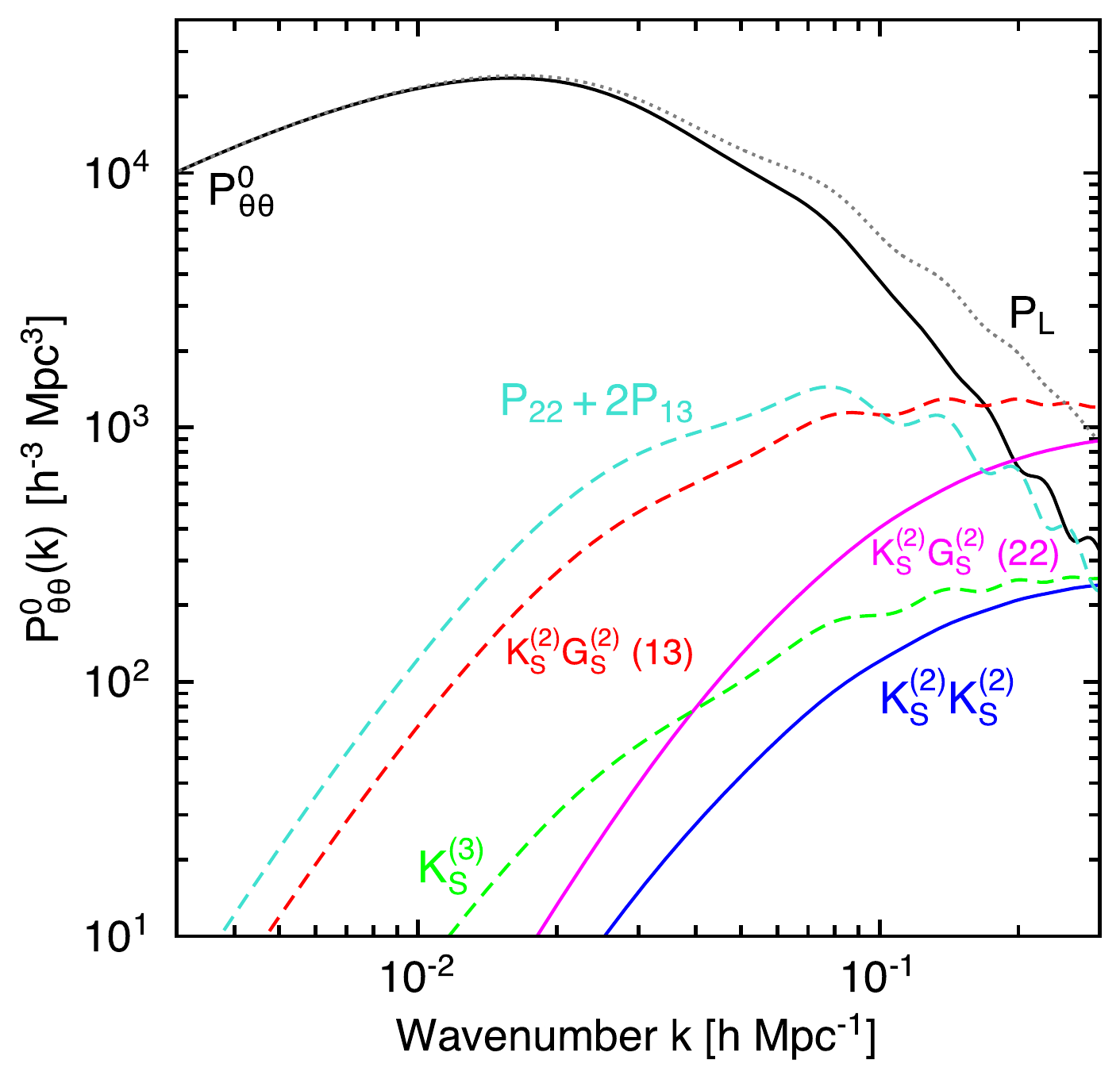}
\end{subfigure}%
}
\caption{Multipoles of the redshift-space velocity divergence power spectrum.
        {\it Left panel}: Comparison between multipoles in real and redshift space.
        In redshift space, there is in addition to the monopole (solid black)
        a quadrupole (solid blue) and hexadecapole (solid red).
        {\it Right panel}: Contributions to the redshift-space monopole from each
        (composite) kernel in eq.~\eqref{eq:Ps-NL}. As in figure~\ref{fig:Poneloop}
        dashed lines indicate negative values.
        In this work, one-loop integrals are evaluated using the FFTLog parameters
        $\nu=-0.6$, $N=256$, $k_\mrm{min}=10^{-4}\,h\Mpc^{-1}$
        and $k_\mrm{max}=100\,h\Mpc^{-1}$.}
  \label{fig:Pell_all}
\end{figure}

Like $\Gs^{(3)}$, the redshift-space kernel $\Ks^{(3)}$ also results in a $k$-dependent
correction to $\PL$. The contribution of this term to $P^s_{\theta\theta}(\k)$ is found to be
\bea
2 f^2 \PL(k)\fint{\q}\,3\Ks^{(3)}(\k,\q,-\q)\PL(q)
&= - f^2 \mu^2 k^2 \PL(k) \sigma_u^2,
\eea
where $\sigma_u^2$ is the one-dimensional (scaled) velocity dispersion,
\be\label{eq:sigu}
\sigma_u^2\equiv\frac13\fint{\q}\frac{\PL(q)}{q^2}
=\sigma_v^2/(\calH f)^2
\ee
(and can be interpreted as the one-dimensional displacement dispersion in
the Lagrangian description of the fluid).
As figure~\ref{fig:Pell_all} shows, this contribution is subdominant to the other
``13'' term given by $\Ks^{(2)}\Gs^{(2)}$.
The $\Ks^{(2)}\Ks^{(2)}$ contribution (as labelled in figure~\ref{fig:Pell_all})
is similarly small in magnitude. This term corresponds to a second-order redshift
mapping which, at one loop, amounts to a mode-coupling of linearized fields only
(i.e.\ the linear theory prediction is not sufficient to see a sizable redshift effect).

\subsubsection{Multipole moments}
The distortions that occur in redshift space are directed along the LOS and implies the loss
of statistical isotropy; the power spectrum is no longer invariant under rotations and will
in general depend on the wavevector $\k$. However, the power spectrum retains azimuthal
symmetry about the LOS $\n$, so depends not on $\k$ but two
components---the wavenumber $k$ and $\mu\equiv\hat\k\cdot\n$, the cosine of the angle formed
between the LOS and the wavevector.
These distortions are conveniently characterized in terms of multipole moments about $\n$
by expanding $P^s_{\theta\theta}(\k)$ in Legendre polynomials:
\be\label{eq:Ps-tt-multi}
P^s_{\theta\theta}(k,\mu)
=\sum_\ell P^\ell_{\theta\theta}(k) \calL_\ell(\mu),
\ee
where the multipole moments are
\be\label{eq:Ptt-ell-1}
P^\ell_{\theta\theta}(k)
\equiv(2\ell+1)\int^1_{-1}\!\frac{\dif\mu}{2}\,\calL_\ell(\mu) P^s_{\theta\theta}(k,\mu),
\ee
and $\calL_\ell(\mu)$ is the Legendre polynomial of order $\ell$.
The anisotropic contributions to $P^s_{\theta\theta}(\k)$ correspond to the $\ell>0$
multipoles and arise from all terms in eq.~\eqref{eq:Ps-NL}, except the first three, which
contribute to the $\ell=0$ multipole moment.

\begin{figure}[!t]
\makebox[\linewidth][c]{%
\begin{subfigure}[b]{.48\textwidth}
\centering
\includegraphics[width=1.\textwidth]{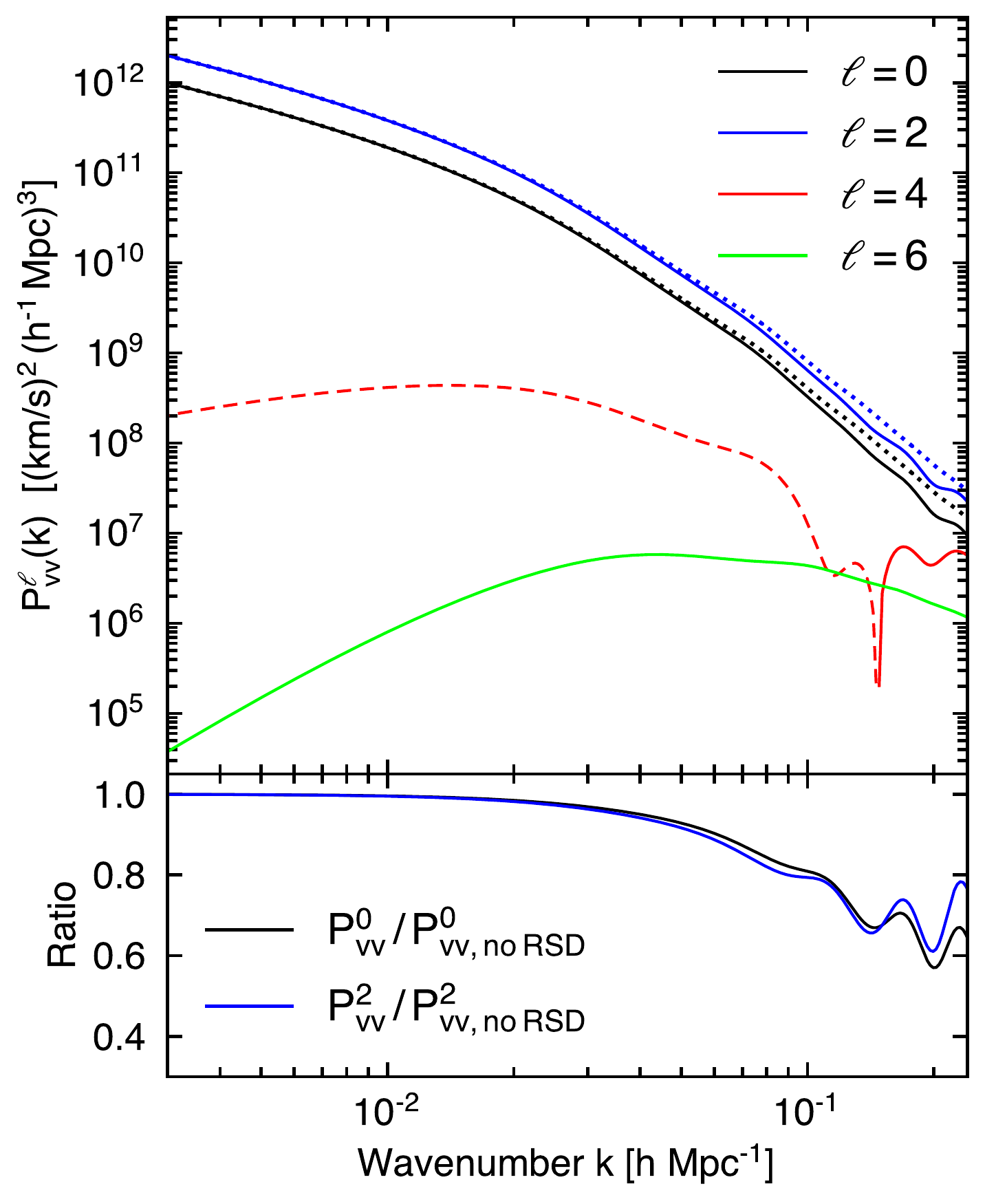}
\end{subfigure}%
\begin{subfigure}[b]{.48\textwidth}
\centering
\includegraphics[width=1.\textwidth]{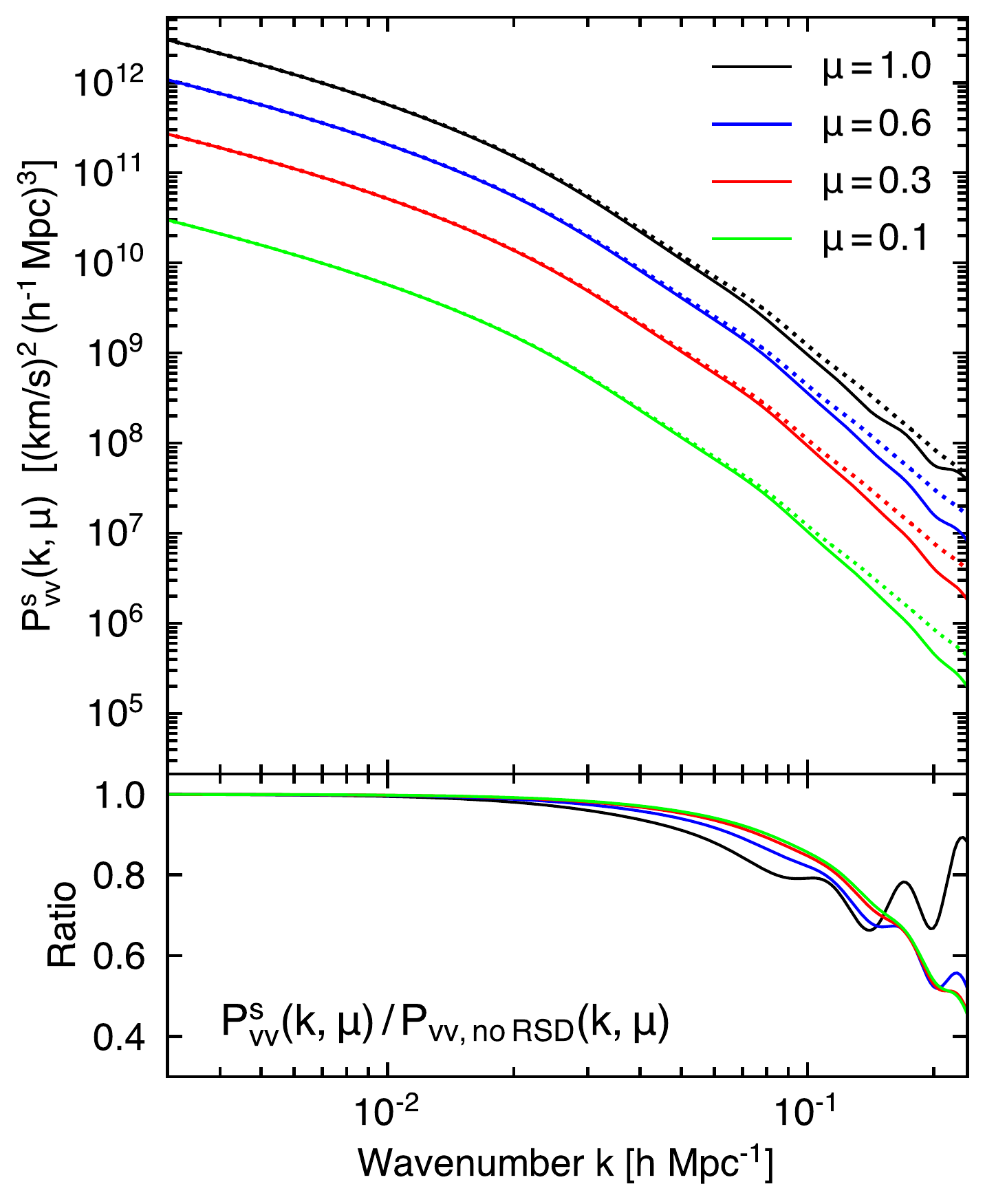}
\end{subfigure}%
}
\caption{Multipoles of the redshift-space velocity power spectrum (left panel)
        and the angular dependence of the total power spectrum (right panel).
        Note that in a typical survey the maximum separation between a pair of galaxies
        is $r_\mrm{max}\simeq 300\,h^{-1}\Mpc$, which roughly indicates the
        largest measurable mode is $k_\mrm{min}=2\pi/r_\mrm{max}\approx0.02\,h\Mpc^{-1}$.
        Notice that the dominant contribution comes in the form of a quadrupole moment,
        not the monopole moment as might be expected.}
  \label{fig:Pvv}
\end{figure}

\begin{figure}[!t]
\makebox[\linewidth][c]{%
\begin{subfigure}[b]{.48\textwidth}
\centering
\includegraphics[width=1.\textwidth]{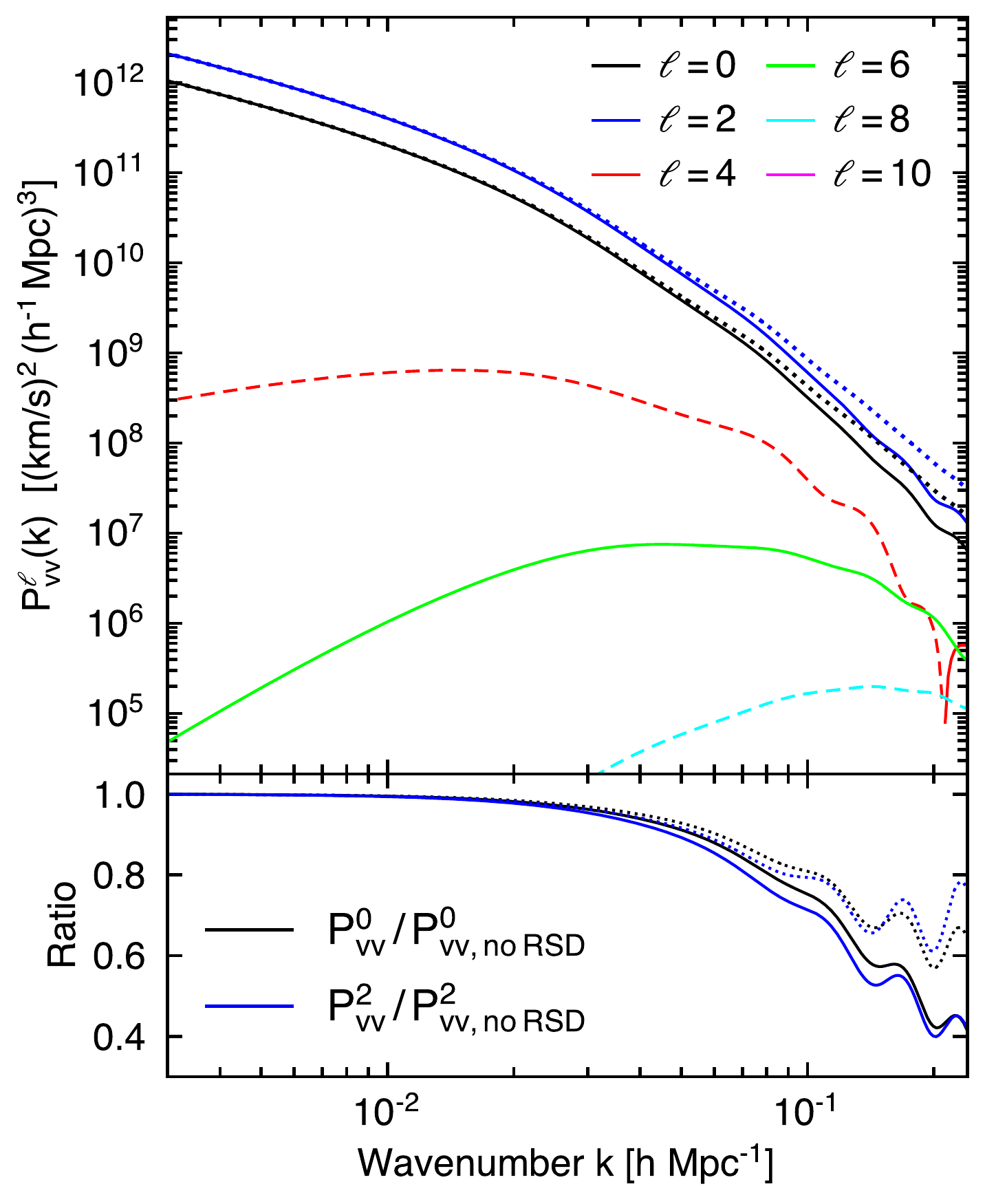}
\end{subfigure}%
\begin{subfigure}[b]{.48\textwidth}
\centering
\includegraphics[width=1.\textwidth]{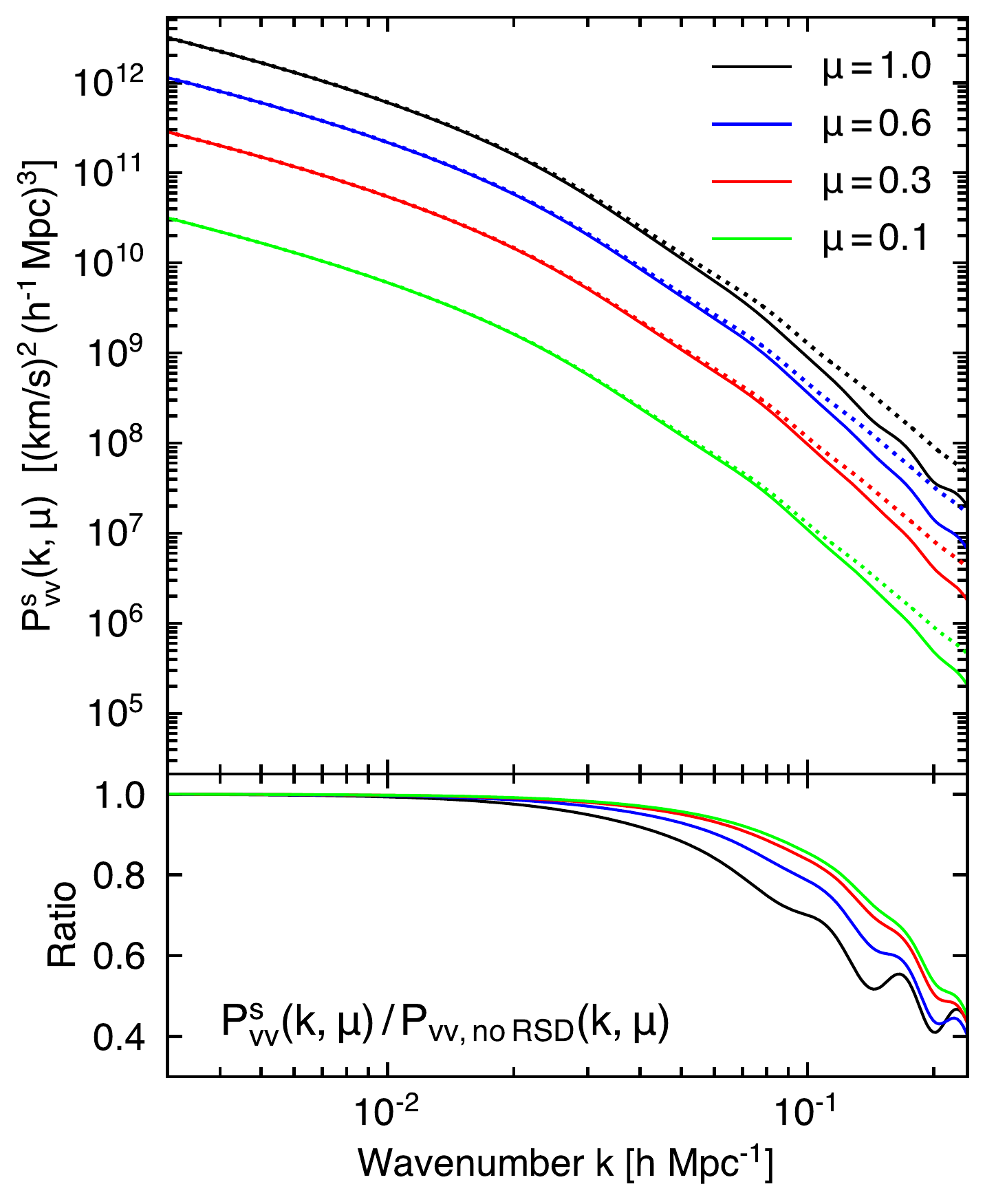}
\end{subfigure}%
}
\caption{Power spectrum damped by virial velocities with dispersion
        $\sigma_v=350\,{\rm km/s}$, and using the Gaussian $\exp(-x^2)$ damping model.
        The multipole moments of the power spectrum are shown in the left panel, while
        in the right panel we show the two-dimensional power spectrum $\Pv^s(k,\mu)$.}
  \label{fig:Pvv_disp}
\end{figure}

The real-space loop integrals are typically performed by choosing spherical coordinates with
the pole aligned with $\k$. In such coordinates the integral over $\phi$ is trivial and
reduces the loop integral to an integral over $q$ and $\mu'=\hat\k\cdot\hat\q$.
Here the presence of $\n$ in the kernels complicates this because there are now
$\q\cdot\n$ terms, and there is no longer rotational symmetry in $\phi$.
It can still, however, be carried out analytically by choosing coordinates
such that the $z$-axis is aligned with $\k$ (i.e.\ $\k=k\hat\z$).
In components parallel and perpendicular to $\hat\z$, we have
$\q=(q_\perp\cos\phi,\, q_\perp\sin\phi,\, q_z)$ and
$\n=(\hat{n}_\perp\cos\phi_o,\, \hat{n}_\perp\sin\phi_o,\, \hat{n}_z)$;
defining the separation angles, $\mu\equiv\hat\k\cdot\n$ and
$\mu'\equiv\hat\k\cdot\hat\q$, so that $q_z=q\mu'$, $\hat{n}_z=\mu$,
$q_\perp=q\sqrt{1-\mu'^2}$, and $\hat{n}_\perp=\sqrt{1-\mu^2}$,
we have the parametrization
\begin{subequations}
\bea
\q&=q\big(\sqrt{1-\mu'^2}\cos\phi, \, \sqrt{1-\mu'^2}\sin\phi, \, \mu'\big),  \\[3pt]
\n&=\big(\sqrt{1-\mu^2}, \, 0, \, \mu\big).
\eea
\end{subequations}
(Note that here we have chosen $\phi_o=0$ as allowed by the rotational freedom to orient the
$xy$-plane.)
The multipole moments then read
\bea
P^\ell_{\theta\theta}(k)
=\PL(k) \, \delta^\mrm{K}_{\ell0}
    + \frac{k^3}{2\pi^2}\bigg[&\int r^2\dif r\,\PL(kr)
        \!\int\!\frac{\dif\mu'}{2}\,I^\ell_{22}(r,\mu')
        \PL\big(k\psi(r,\mu')\big) \nonumber\\[4pt]
&
    + 2\PL(k)\!\int r^2\dif r\,\PL(kr)
        \!\int\!\frac{\dif\mu'}{2}\,I^\ell_{13}(r,\mu')\bigg],
\label{eq:Ptt-ell}
\eea
where $\delta_{\ell0}^\mrm{K}$ is the Kronecker delta, $r\equiv q/k$,
$\psi(r,\mu')\equiv\sqrt{1+r^2-2r\mu'}$, and
\begin{subequations}
\label{eq:Iell}
\bea
I^\ell_{22}(r,\mu')
&\equiv (2\ell+1)\int\!\frac{\dif\mu}{2}\,\calL_\ell(\mu)
    \!\int\!\frac{\dif\phi}{2\pi}\:2\big[Z_2(r,\mu,\mu',\phi)\big]^2, \\[4pt]
I^\ell_{13}(r,\mu')
&\equiv (2\ell+1)\int\!\frac{\dif\mu}{2}\,\calL_\ell(\mu)
    \!\int\!\frac{\dif\phi}{2\pi}\:3Z_3(r,\mu,\mu',\phi).
\eea
\end{subequations}
Closed-form expressions of these kernels can be found in Appendix \ref{app:Ikernels}.
Note that the $\mu'$ integral in the last term of eq.~\eqref{eq:Ptt-ell} is over
polynomials of $\mu'$ and is analytic.
As with the redshift-space density power spectrum (in the plane-parallel limit), only
the monopole ($\ell=0$), quadrupole ($\ell=2$), and hexadecapole ($\ell=4$) are non-vanishing.
While the above expression is presented in a form readily evaluated numerically by standard
quadrature methods, these computations are intensive. In Appendix
\ref{app:FFT} we show how these can be efficiently evaluated using the FFTLog approach
\cite{Talman:1978,Hamilton:1999uv}.

\subsection{Redshift-space velocity}
We now return to $\vz^s$, the observable in a peculiar velocity survey. It is easy to write down
its power spectrum since in Fourier space $\vz^s(\k)=\calH f(\ii\kz/k^2)\theta^s(\k)$; that is, we
can write the power spectrum of $\vz^s$ in terms of that for $\theta^s$ just presented,
\be\label{eq:Psvv}
\Pv^s(k,\mu)
=\bigg(\frac{\calH f\mu}{k}\bigg)^2 P^s_{\theta\theta}(k,\mu)
\equiv \sum_\ell \Pv^\ell(k)\mathcal{L}_\ell(\mu),
\ee
where $\mu=\kz/k$. The multipole moments are defined in an analogous way to
$P^\ell_{\theta\theta}$ \eqref{eq:Ptt-ell-1}, and are
\be\label{eq:Psvv-ell}
\Pv^\ell(k)
=\bigg(\frac{\calH f}{k}\bigg)^2\sum_{\ell'} A_{\ell\ell'}\, P^{\ell'}_{\theta\theta}(k),
\qquad
(A_{\ell\ell'})
=\begin{pmatrix}
\frac13 & \frac2{15} & 0 \\[2pt]
\frac23 & \frac{11}{21} & \frac4{21} \\[2pt]
0 & \frac{12}{35} & \frac{39}{77} \\[2pt]
0 & 0 & \frac{10}{33}
\end{pmatrix}.
\ee
The nonzero multipole moments are given by $\ell=0,2,4,6$ and are shown in figure~\ref{fig:Pvv}.
In particular, the multipole structure
exhibits a $\ell=6$ moment (``tetrahexacontapole''---or 64-pole), which we note is not present
in the power spectrum of the velocity divergence.
However, this merely arises from the geometric factor $\propto\mu^2$ in eq.~\eqref{eq:Psvv-ell}; it also
has the further effect of coupling different $P^\ell_{\theta\theta}$ to a given $\Pv^\ell$
with coefficients given through the matrix $(A_{\ell\ell'})$.
In $(A_{\ell\ell'})$ we see that the first column represents the
usual monopole--quadrupole split ($1/3$--$2/3$) of the real-space velocity power spectrum.
(Notice also that columns add to unity.) Evidently, we also have additional
contributions not found in real space as can be seen in eq.~\eqref{eq:Ptt-ell}.
Thus the second and third columns are new contributions to the anisotropy;
specifically these are contributions arising from nonzero $P^2_{\theta\theta}$
and $P^4_{\theta\theta}$.

\subsection{Further damping from velocity dispersion\label{sec:virial}}
The power spectrum we have so far presented describes the coherence of the galaxy motions
on sufficiently large scales where the fluid description is valid. In this regime the dark matter is
single-streaming and the velocity field is well defined. As such this model necessarily does not 
account for effects resulting from multi-streaming, i.e.\ the virial motions of galaxies associated
with clusters. Such effects, arising from various baryonic processes, are important but beyond PT.
Qualitatively though, in the context of galaxy clustering, the effect of viral motions is simply
to cause a damping of the small-scale power due to the elongation of nonlinear structures along the LOS
(the FoG effect). This is a smearing of the matter distribution, implying a shallower gravitational
potential in redshift space, and thus a decrease in the velocity power.

To account for this effect and complete the model we will simply adopt a
phenomenological damping (similar to that used to model the standard FoG effect).
The complete model is given by augmenting the previous model by taking
\be
P^s_{vv}(\k)\to D_v^2(k\mu) P^s_{vv}(\k),
\ee
where $P^s_{vv}(\k)$ is given by eq.~\eqref{eq:Psvv} and $D_v^2(x)\equiv\exp(-x^2)$ is the
damping factor, with $x\equiv k\mu\sigma_u$.

For the purpose of keeping the phenomenological model separate from the PT predictions, we
express the multipoles of the complete model in terms of the previous ones~\eqref{eq:Psvv} as
\be
\Pv^\ell(k)
\to(2\ell+1)\sum_{\ell',\ell''}
\bigg(\begin{matrix} \ell & \ell' & \ell'' \\ 0 & 0 & 0 \end{matrix}\bigg)^2
A^{\ell'}\!(k)\,\Pv^{\ell''}(k),
\ee
where $A^{\ell}(k)$ is the $\ell$th multipole moment of $D_v^2(k\mu)$,
$\Pv^{\ell''}(k)$ is given by eq.~\eqref{eq:Psvv-ell}, and the $2\times3$ array is the
Wigner 3j-symbol~\cite{Varshalovich_book} (which results from the integral over the product of
three Legendre polynomials). For typical damping models, a low-order truncation of the multipole
expansion (e.g.\ up to $\ell=4$) is accurate enough for our needs.
Figure~\ref{fig:Pvv_disp} shows the power spectrum with the inclusion of nonlinear damping.
As can be seen, the PT predictions on large scales are maintained but are increasingly modified
on small scales.
At $k\simeq0.1\,h\Mpc^{-1}$, the power is suppressed by $30\%$, about $10\%$ of which coming
from nonlinear damping described in this section. This is lower that the damping predicted by
Kaiser and Hudson~\cite{Kaiser:2014jca}, as discussed in Section~\ref{sec:intro}, but is not
surprising given that we have shown PT predicts an additional suppression of power.
As such the damping observed in the simulations of ref.~\cite{Koda:2014}
is only partially explained by the virial motions.

\section{Power spectrum model II: cumulant-expansion approach\label{sec:cum-exp-model}}
Many of the techniques developed for redshift-space statistics of galaxy clustering
(e.g.~\cite{Scoccimarro:2004tg,Matsubara:2008,Taruya:2010,Reid:2011,Carlson:2013,Zheng:2016,Vlah:2019})
carry over to the related velocity statistics. This will allow us to motivate FoG-like effects
that was missed in the straightforward moment expansion of the previous section.
To do this we need to return to the exact expression for $\vz^s(\s)$ given by eq.~\eqref{eq:v-s-G1}.
In Fourier space, $\vz^s(\k)$ is easily obtained from eq.~\eqref{eq:v-s-G1} by inserting
into it the Fourier representation of the Green's function \eqref{eq:G-Fourier};
after changing the order of integration, we read off [$\uz^s\equiv-\vz^s/(\calH f)$]
\be\label{eq:v-s-exact2}
\uz^s(\k)
=\frac{1}{-\ii\kz}\int\dif^3\x\: \rme^{\ii\k\cdot\x}\,
    \rme^{-\ii f\kz\uz(\x)}\, \nablaz \uz(\x),
\ee
with $\kz=\k\cdot\hat\z=k\mu$.
In Fourier space we see that redshift-space distortions arise
from a velocity-induced phase factor $\rme^{-\ii f\kz\uz(\x)}$; in its absence we recover $\uz^s(\k)=\uz(\k)$.
Now, using that by translation invariance the two-point function can only depend on $\r=\x-\x'$,
we have the power spectrum
\be\label{eq:P-s-exact}
\Pu^s(\k)
=\frac{1}{\kz^2} \int\dif^3\r\: \rme^{\ii\k\cdot\r}\,
    \Big\langle \rme^{-\ii f\kz\Delta\uz}\, \nablaz\uz(\x)\, \nablaz\uz(\x')\Big\rangle,
\ee
in which $\Delta\uz\equiv\uz(\x)-\uz(\x')$ is the pairwise relative velocity along the LOS.
Indeed, the perturbative model \eqref{eq:Ps-NL} can be recovered from the exact model \eqref{eq:P-s-exact};
that is, by Taylor expanding the velocity-dependent exponential in eq.~\eqref{eq:v-s-exact2} to
third order in $\uz$, then evaluating the resulting moments using one-loop PT.
(Equivalently, the same result can be obtained if in the power spectrum \eqref{eq:P-s-exact} the
second exponential is expanded to second order in $\Delta\uz$.)

Given the novelty of the redshift-space velocity power spectrum \eqref{eq:P-s-exact} it is instructive
to compare it to the analogous density power spectrum given by eq.~\eqref{eq:P-sco04} (but derived
assuming number conservation). It is not hard to see that the basic form of the two power spectra are
quite similar. The main difference is the weighting in the pairwise LOS
\emph{velocity-moment generating function} given by $\langle\cdots\rangle$; that is,
in eq.~\eqref{eq:P-sco04} the
moments are density weighted, while in eq.~\eqref{eq:P-s-exact} they are velocity-gradient weighted;
in both cases they share the same phase factor. This phase factor can be brought out by rewriting
eq.~\eqref{eq:P-sco04} in terms of connected $n$-point functions using that, by the cumulant
expansion theorem, we have in general
\be\label{eq:cum-thm-2}
\big\langle\rme^{\,j_1A_1}A_2A_3\big\rangle
=\exp\big\langle \rme^{\,j_1A_1}\big\rangle_\mrm{c}\,
	\Big[\big\langle \rme^{\,j_1A_1}A_2A_3\big\rangle_\mrm{c}
	+ \big\langle \rme^{\,j_1A_1}A_2\big\rangle_\mrm{c}
	    \big\langle \rme^{\,j_1A_1}A_3\big\rangle_\mrm{c} \Big],
\ee
where $A_1$, $A_2$, $A_3$ are random variables, and $j_1$ is a constant.
In the case of the density power spectrum [eq.~\eqref{eq:P-sco04}] we set
$A_2=1+\delta(\x)$, $A_3=1+\delta(\x')$,
whereas for the velocity power spectrum [eq.~\eqref{eq:P-s-exact}] we set
$A_2=\nablaz\uz(\x)$, $A_3=\nablaz\uz(\x')$;
in both cases $j_1=-\ii f\kz$ and $A_1=\Delta\uz$.%
\footnote{Another expression for $P^s_{\delta\delta}$, derived from eq.~\eqref{eq:P-sco04}, has
instead $A_2=\delta(\x)+f\nablaz\uz(\x)$,
$A_3=\delta(\x')+f\nablaz\uz(\x')$, with $j_1$ and $A_1$ the same as in eq.~\eqref{eq:P-sco04}.
This expression makes explicit use of the Jacobian of the redshift mapping \eqref{eq:mapping}.
The related Taruya, Nishimichi and Saito (TNS) model~\cite{Taruya:2010} is obtained by using
the cumulant expansion theorem on the resulting expression.
}
(Note that eq.~\eqref{eq:cum-thm-2} follows from the relation between the moment and cumulant
generating function,
$\langle\rme^{\,\mbf{j}\cdot\mbf{A}}\rangle=\exp\langle\rme^{\,\mbf{j}\cdot\mbf{A}}\rangle_\mrm{c}$,
with $\mbf{j}$ some constant vector.)

The factor $\exp\langle\rme^{\,j_1A_1}\rangle_\mrm{c}$ is typically understood as the putative FoG
damping prefactor, which has in the past been modelled phenomenologically in the ``dispersion models'' 
\cite{Ballinger:1996cd}.
Clearly we can see that a FoG-like effect is generic to both power spectra. This suggests that
we can treat ``FoG'' damping in the velocity power spectrum in much the same way (if not exactly
the same way) as for its density counterpart.

\subsection{Analytic model}
To specify the model we will need to evaluate the connected $n$-point functions in
eq.~\eqref{eq:P-s-exact}.
As in eq.~\eqref{eq:Ps-NL} we use one-loop PT, but only for terms within square brackets in
eq.~\eqref{eq:P-s-exact}; the FoG factor $\exp\langle\rme^{j_1A_1}\rangle$ will be treated
separately using more empirical arguments.

First, we take the large-scale limit $j_1\to 0$ (or $k\to0$). Thus treating $j_1$ as
an expansion parameter, and expanding the exponentials in eq.~\eqref{eq:cum-thm-2}, we get
\be\label{eq:cum-1loop}
\big\langle \rme^{\,j_1 A_1} A_2 A_3\big\rangle
\simeq \Da \, \Dab
\Big[
    \langle A_2 A_3\rangle_\mrm{c}
    +j_1\langle A_1A_2A_3\rangle_\mrm{c}
    +\frac12 \, j_1^2\langle A_1^2A_2A_3\rangle_\mrm{c}
    +j_1^2\langle A_1A_2\rangle_\mrm{c} \langle A_1A_3\rangle_\mrm{c}
\Big].
\ee
Here we have expanded to second order in $j_1$, as is sufficient for a one-loop 
calculation. In particular, we have only expanded the terms inside square brackets.
Following ref.~\cite{Zheng:2016}, the putative FoG damping factor $\exp\big\langle \rme^{\,j_1A_1}\big\rangle_\mrm{c}$
has been decomposed into two kinds of FoG factors: $\Da=\Da(\kz)$, which consists purely of one-point
contributions and so can be taken out of the spatial integral; and $\Dab=\Dab(\kz,\r)$, which consists
of both one- and two-point correlations.

A simple procedure to bring out an overall FoG damping term from the integral in eq.~\eqref{eq:P-s-exact}
is then to simply ignore spatial correlations---drop $\Dab$ from eq.~\eqref{eq:cum-1loop}.
This step might be justified by analogy with the density power spectrum, where it has been shown to
produce a model---the widely-used ``TNS model''~\cite{Taruya:2010}---that provides a good fit to
simulations~\cite{Kwan:2012}.
Thus, substituting eq.~\eqref{eq:cum-1loop} into eq.~\eqref{eq:P-s-exact}, and hence dropping
$\Dab$, we have for the one-loop power spectrum model,
\bea\label{eq:P-s-cumexp}
\Pu^s(\k)
= \Da\!&\int\dif^3\r\:\rme^{\ii\k\cdot\r} \frac{1}{\kz^2}
\Big[
    \langle A_2 A_3\rangle
    +j_1\langle A_1A_2A_3\rangle
    +j_1^2\langle A_1A_2\rangle \langle A_1A_3\rangle
    +\mathcal{O}(j_1^3)
\Big],
\nonumber\\[5pt]
\text{with}\:\:\:
&j_1=-\ii f\kz,\quad
A_1=\uz(\x)-\uz(\x'),\quad
    A_2=\nablaz\uz(\x),\quad
    \text{and}\quad
    A_3=\nablaz\uz(\x').
\eea
Note that $\langle A_2 A_3\rangle_\mrm{c}=\langle A_2 A_3\rangle$,
$\langle A_1A_2 A_3\rangle_\mrm{c}=\langle A_1A_2 A_3\rangle$, but that in general
$\langle A_1^2A_2A_3\rangle_\mrm{c}\neq\langle A_1^2A_2A_3\rangle$; in addition, we drop
$\langle A_1^2A_2A_3\rangle_\mrm{c}$ as it is a two-loop correction.%
\footnote{The connected moment $\langle A_1^2A_2A_3\rangle_\mrm{c}$, in Fourier space, involves
the trispectrum at the first nontrivial order. This term is therefore $\mathcal{O}(\PL^3)$,
i.e.\ a two-loop correction.}

\begin{figure}[!t]
	\centering
	\includegraphics[width=\linewidth]{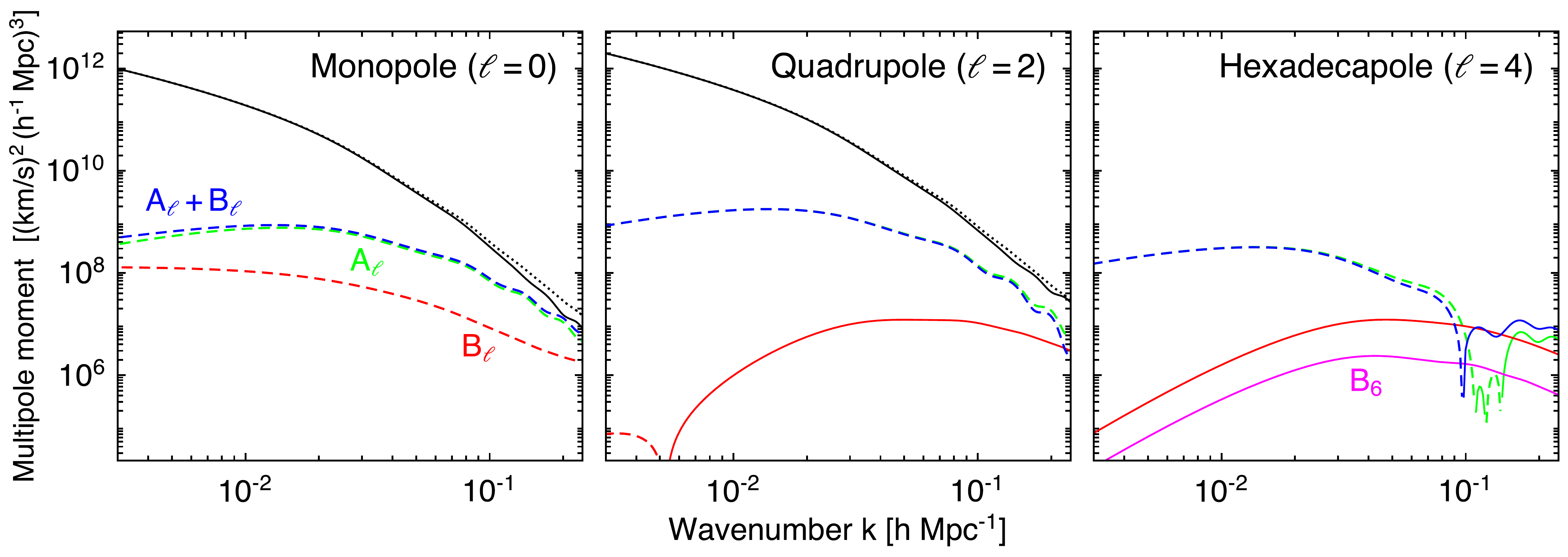}
\caption{Multipoles of distortion terms $A$ and $B$ in the cumulant expansion model.
        The effect on the total power spectrum  $\Pu^\ell+A_\ell+B_\ell$ (solid black)
        is to be compared to the corresponding undistorted, real-space result (dotted black).
        Note that in real space there is no $\ell=4$ multipole moment;
        all lines have been scaled by $(\calH f)^2$; and as before we display the
        absolute value of the moments (dashed lines indicate negative values).}
  \label{fig:cumexp_no_disp}
\end{figure}

The first term in eq.~\eqref{eq:P-s-cumexp} corresponds to the
leading-order real-to-redshift-space map, which as we have mentioned is the
identity map (no distortions). The second term in eq.~\eqref{eq:P-s-cumexp} contains
a mixture of effects arising from the next-to-leading-order redshift mapping and nonlinear
dynamics (the ``$\Ks^{(2)}\Gs^{(2)}$'' contributions in figure~\ref{fig:Pell_all}).%
\footnote{The other terms---labelled ``$\Ks^{(3)}$'' and ``$\Ks^{(2)}\Ks^{(2)}$'' in
figure~\ref{fig:Pell_all}---originate from $\langle A_1^2A_2A_3\rangle$ but do not appear in the
cumulant expansion.} In the limit $j_1\to0$, eq.~\eqref{eq:P-s-cumexp} reduces to
\be\label{eq:P-s-limit}
\Pu^s(\k) = \Da(\kz) \Pu(\k).
\ee
(Recall that taking the same limit but for the density field recovers the Scoccimarro model,
{$P^s_{\delta\delta}=\Da(P_{\delta\delta}+2f\mu^2P_{\delta\theta}+f^2\mu^4P_{\theta\theta})$}
\cite{Scoccimarro:2004tg}.)
Provided the velocity dispersion is nonzero, eq.~\eqref{eq:P-s-limit} shows that there
is an ``FoG'' effect, even for a velocity field described exactly by linear theory.
This is to be contrasted with the moment expansion of Section \ref{sec:vel-Pk}, where damping
effects are entirely absent in linear theory (since real- and redshift-space velocity
fields coincide).

Going beyond linear theory to next-to-leading order then shows there
are two additional contributions to consider. Let us write the
redshift-space power spectrum as%
\be\label{eq:Ps-cumexp-model}
\Pu^s(\k) = \Da(\kz)\Big[\Pu(\k) + A(\k) + B(\k)\Big].
\ee
This model can be compared with that of eq.~\eqref{eq:Ps-NL}. The obvious improvement is the
appearance now of damping from velocity dispersion (through resumming the one-point contributions).
Further, the $A$ and $B$ terms that we have treated perturbatively do not make the assumption
that the amplitude of the fields are small (which was assumed in the velocity-moment expansion
approach.) Rather, it is the correlations at large separations that are expected to be weak,
even if the amplitude of the field at each point is large.

\begin{figure}[!t]
\makebox[\linewidth][c]{%
\begin{subfigure}[b]{.48\textwidth}
\centering
\includegraphics[width=1.\textwidth]{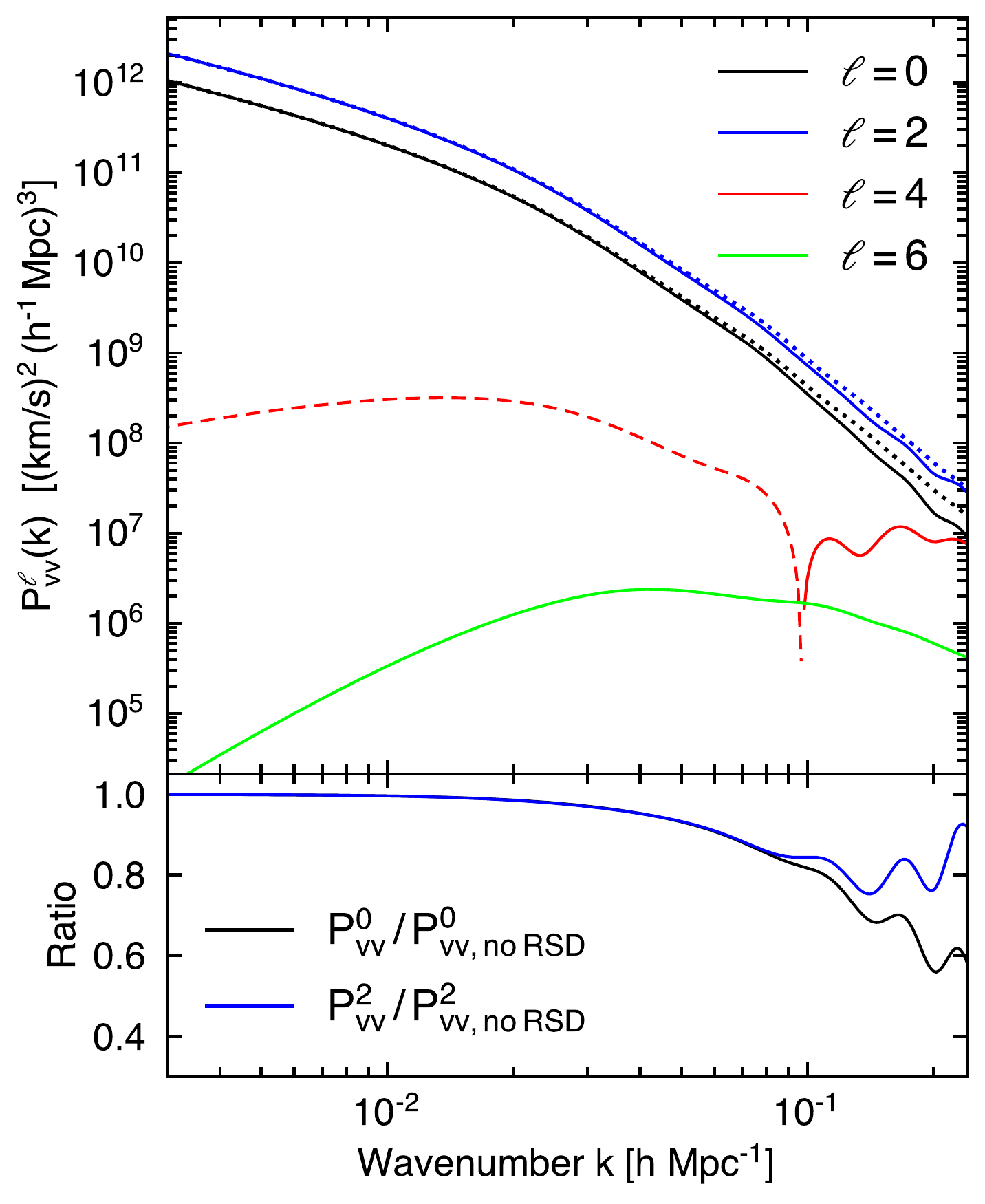}
\end{subfigure}%
\begin{subfigure}[b]{.48\textwidth}
\centering
\includegraphics[width=1.\textwidth]{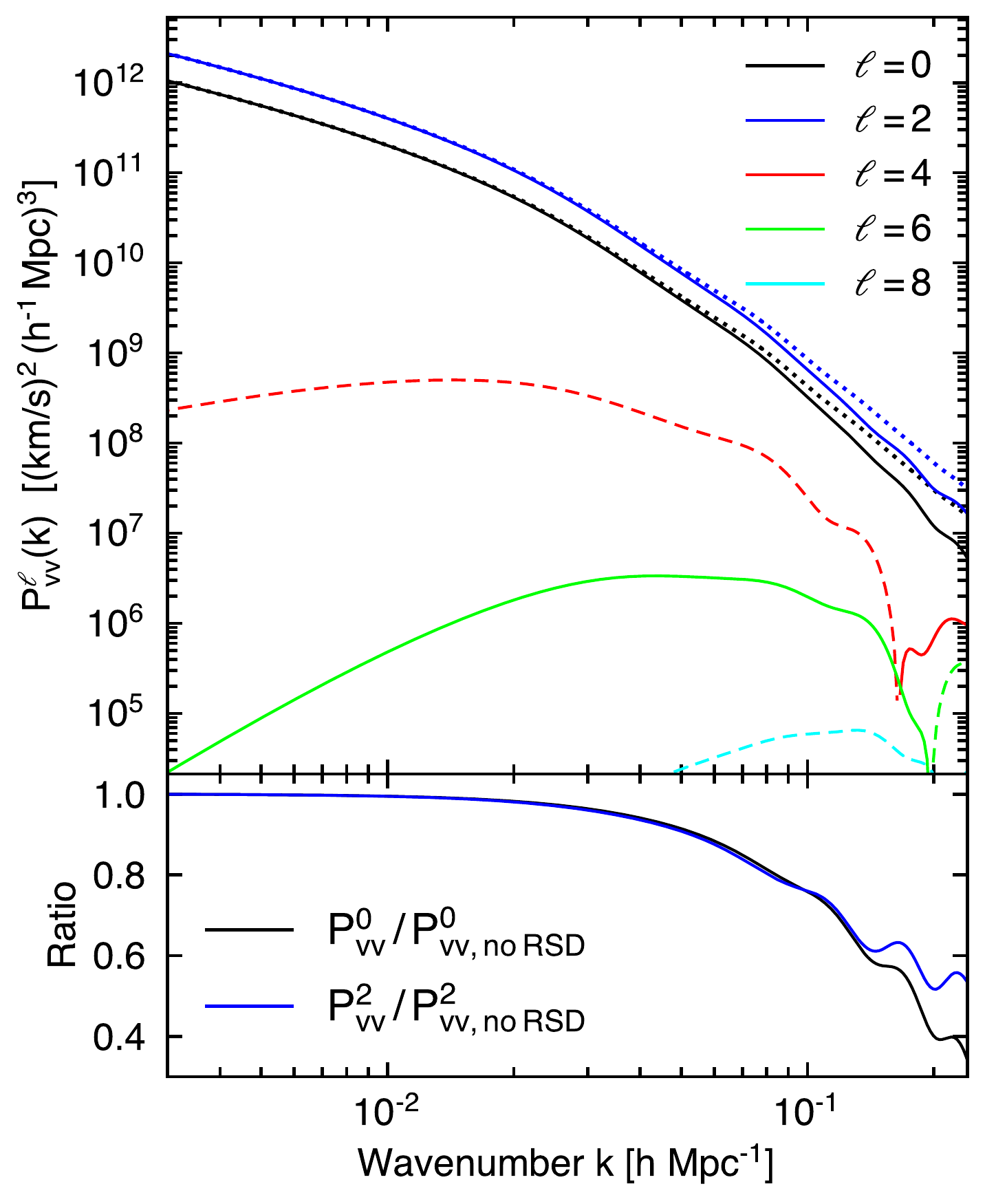}
\end{subfigure}%
}
\caption{Power spectrum multipoles without dispersion damping (left panel) and with
        dispersion damping (right panel). Note that the full power spectrum model,
        with exponential dispersion damping, generates higher-multipole
        orders ($\ell=10,12,\ldots$) but are too small to be seen in frame.
        Here, as in figure~\ref{fig:Pvv_disp}, we use $\sigma_v=350\ {\rm km/s}$
        for the Gaussian damping model, $\Da(x)=\exp(-x^2)$.}
  \label{fig:Pvv_disp_cumexp}
\end{figure}

The term $A$ involves a three-point function, whereas $B$ amounts
to a convolution; both are order $(\PL)^2$ (up to $\k$-dependent factors).
After carrying out the necessary Fourier transforms, they read
\begin{subequations}\label{eq:A-B}
\bea
A(\k)
&\equiv j_1 \int\dif^3\r\:\rme^{\ii\k\cdot\r} \frac{1}{\kz^2}\langle A_1 A_2 A_3\rangle
=2f\!\fint{\q} \KA(\q,\k-\q)\,
        B_{\theta\theta\theta}(q,|\k-\q|,k), \label{eq:Au} \\[3pt]
B(\k)
&\equiv j_1^2 \int\dif^3\r\:\rme^{\ii\k\cdot\r} \frac{1}{\kz^2}
    \langle A_1 A_2\rangle \langle A_1 A_3\rangle
= f^2\!\fint{\q}
    \KB(\q,\k-\q) \,
    \PL(q) \, \PL(|\k-\q|), \label{eq:Bu}
\eea
\end{subequations}
Here the equalities have been obtained by evaluating the $n$-point functions using
one-loop PT; $B_{\theta\theta\theta}$ is the tree-level bispectrum \eqref{eq:Bttt};
and, in deriving eq.~\eqref{eq:Bu}, we have used 
that $\langle\uz(\x)\nablaz\uz(\x')\rangle=-\langle\uz(\x')\nablaz\uz(\x)\rangle$.
The kernels are ${\KA(\q,\q')\equiv\qz^2\qz'\kz/(qq'k)^2}$ and
{$\KB(\q,\q')\equiv{(\qz\qz')^3}/{(q q')^4}$}. (Note that unlike $\Ks^{(n)}$ they are
not dimensionless, but have units inverse length-squared.) As before, since the
kernels are multiplied by functions symmetric under permutations of their arguments,
we may replace them with their symmetrized versions, which are (with the LOS here made
explicit)
\begin{subequations}
\bea
\KsA(\q,\q')&=\frac{(\q\cdot\hat\z)(\q'\cdot\hat\z)\big[(\q+\q')\cdot\hat\z\big]^2}{2(qq')^2|\q+\q'|^2}, \\[5.5pt]
\KsB(\q,\q')&=\frac{\big[(\q\cdot\hat\z)(\q'\cdot\hat\z)\big]^3}{(qq')^4}.
\eea
\end{subequations}
Comparing $\KsA$ with
$\Ks^{(2)}$ \eqref{eq:K2} shows they differ by a geometric factor $(\kz/k^2)^2$,
which appears when converting between the velocity field and its divergence, 
cf.~eq.~\eqref{eq:v-theta}.
In any case, whatever the details of the kernel, $A(\k)$ has the same form
as eq.~\eqref{eq:iB1} and can thus be reduced to [cf.~eq.~\eqref{eq:iB}]
\be\label{eq:Au-2}
\begin{split}
A(\k)
=2f\bigg(\frac{k_z}{k^2}\bigg)^2 \, \bigg[
&\,4\PL(k)\!\fint{\q} \Ks^{(2)}(\q,\k-\q) \, \Gs^{(2)}(\k,-\q) \, \PL(q) \\
    &+2\!\fint{\q} \Ks^{(2)}(\q,\k-\q) \, \Gs^{(2)}(\q,\k-\q) \, \PL(q) \, \PL(|\k-\q|)\bigg],
\end{split}
\ee
where we have taken out the geometric factor in order to express in terms of $\Ks^{(2)}$. This term
corresponds to $2\langle \theta(\k)\,\theta^s_2(\k)^*\rangle$ in eq.~\eqref{eq:thetas-thetas} and
is in fact equal after converting to the velocity divergence, as mentioned.
However, the $B(\k)$ term is different to any contribution we have considered so far.
Figure~\ref{fig:cumexp_no_disp} shows the multipole contributions of $A$ and $B$;
in particular, we see that $B$ is generally subdominant to $A$.

Finally, in order to fully specify the model a particular form for $\Da$ needs to be given.
There are a few possibilities. Here, as we are patterning our model after the TNS model for the density
power spectrum, we will simply take $\Da(x)=\exp(-x^2)$, with $x\equiv f\kz\sigma_u$, i.e.\ the Gaussian damping
model.\footnote{For numerical work, we will depart from the TNS model and use the linear-theory prediction for
$\sigma_u$ (where in the TNS model it is treated as a free parameter).} The complete model is shown in the
right panel of figure~\ref{fig:Pvv_disp_cumexp} (the left panel shows the model before damping).
In contrast to the moment-expansion model of Section~\ref{sec:vel-Pk}, this damping does not need to be
put in by hand; it can be fashioned entirely from $\exp\langle \rme^{j_1 A_1}\rangle_c$ in eq.~\eqref{eq:cum-thm-2}
by discarding all but the first and second one-point cumulants. Doing so amounts to considering a
scale-independent Gaussian probability density function for the pairwise velocities.

As the model stands, it should be noted that it does not yet capture any multi-streaming effects.
Taking into account such effects is of course important for realistic modelling of the usual FoG effect
due to galaxy virial motions. We thus emphasize that the exact model
\eqref{eq:P-s-exact} assumes the single-streaming approximation for the mapping \eqref{eq:mapping} (and
in spite of the appearance of $\Da$, this is the case even if the dynamics could be evaluated exactly).
As such, the usual shortcomings of models of redshift-space clustering can also be found in our model.
For instance, the velocity dispersion $\sigma_u$ should be treated as an empirical parameter.

\begin{figure}[t]
\centering
\includegraphics[width=0.66\textwidth]{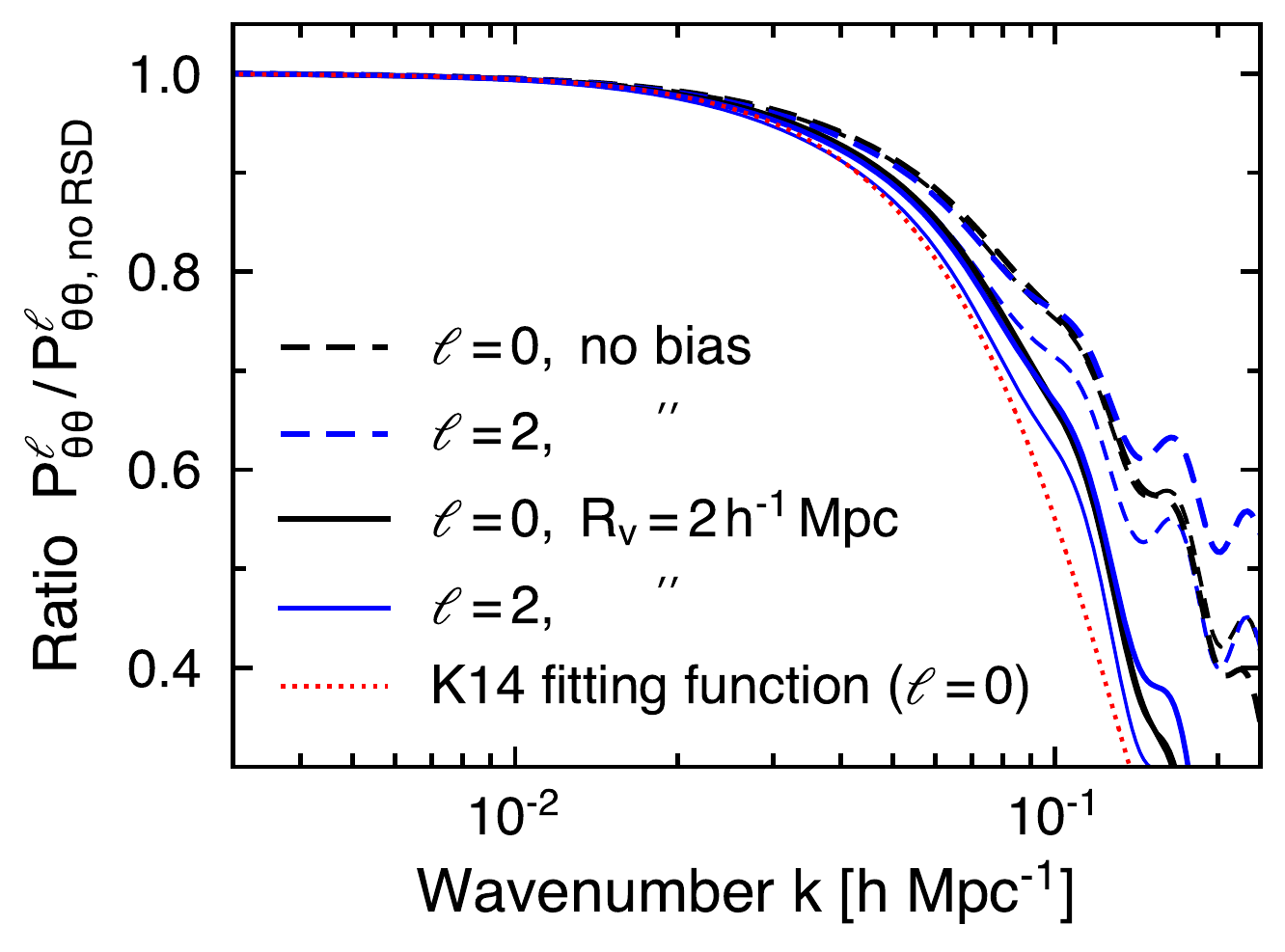}
\caption{Comparison between the cumulant-expansion model (thick lines) and the moment-expansion
        model (thin lines). As in figure~\ref{fig:Pvv_disp_cumexp}, black lines indicate the
        monopole power, while blue lines indicate the quadrupole power.
        We also plot for the monopole power the fitting function (dotted red line) given in
        ref.~\cite{Koda:2014}; here the shape of the suppression is described using a sinc function
        $D(k\sigma_u)=\sin(k\sigma_u)/(k\sigma_u)$, with the parameter $\sigma_u$
        calibrated on simulations. Note that the velocities measured in these simulations 
        correspond to subhalos, not dark matter, and because of this we need to allow an additional
        higher-derivative velocity bias term~\cite{bias_review} in the expansion of $\v^s$. At lowest order
        in derivatives, this term must be of the form $R_v^2\nabla^2\v$~\cite{bias_review}, where $R_v$ is of the
        order of the Lagrangian radius for halos.
        Here we use $\sigma_u=13\,h^{-1}\Mpc$, which corresponds to that found by ref.~\cite{Koda:2014}
        for subhalos with mass $M\simeq10^{12}\,h^{-1}\Mpc$ and Lagrangian radius
        $R_L=(3M/(4\pi\bar\rho))^{1/3}\simeq1.5\,h^{-1}\Mpc$. For this $R_L$ the simulations of
        ref.~\cite{Baldauf:2014fza} indicate $R_v=2\,h^{-1}\Mpc$.
        The solid lines include the velocity bias, whereas the dashed lines do not include velocity bias.}
  \label{fig:model-AB}
\end{figure}

\section{Configuration space\label{sec:config}}
We now pass from Fourier space to configuration space where the object of study is the
anisotropic correlation function. Of course, the correlation function is just the Fourier
transform of the power spectrum. But the presence of a LOS and the fact that we are
dealing with a vector field are complicating factors.
Without specifying the particular form of the power spectrum, this section derives exact expressions
for the correlation functions, working again in the plane-parallel limit.
We begin with a brief review of the well-known real-space theory to help us illustrate
later on the differences arising in redshift space.

\subsection{Real space}
The velocity divergence correlation function for any two points separated by $\r$ is
$\xi_{\theta\theta}(r)\equiv\langle \theta(\x)\theta(\x+\r)\rangle$, and reads in terms of
its power spectrum $P_{\theta\theta}(k)$,
\be
\xi_{\theta\theta}(r)
=\int\!\dk\, P_{\theta\theta}(k)\, \rme^{-\ii\k\cdot\r}.
\ee
Notice that by statistical isotropy and homogeneity $\xi_{\theta\theta}$ can only depend on
the separation distance $r=|\r|$ (and likewise for the power spectrum, but for wavenumber).

Recalling eq.~\eqref{eq:v-theta}, the velocity correlation tensor
$\Psi_{ij}(\r)\equiv\langle v_i(\x)v_j(\x+\r)\rangle$ is given in terms of $P_{\theta\theta}$ by
\be\label{eq:Psi-ij}
\Psi_{ij}(\r)
=\int\!\dk\,\frac{k_i}{k}\frac{k_j}{k} \Big(\frac{\calH f}{k}\Big)^2  P_{\theta\theta}(k)\, \rme^{-\ii\k\cdot\r}.
\ee
This symmetric tensor when decomposed into perpendicular and parallel parts (relative
to the separation $\r$) reads in general
\be\label{eq:perfect-corr}
\Psi_{ij}(\r)
=\Psi_\perp(r)\big(\delta^\mrm{K}_{ij}-\hat{r}_i\hat{r}_j\big)
+ \Psi_\|(r)\hat{r}_i\hat{r}_j,
\ee
for some radial functions $\Psi_\perp(r)$ and $\Psi_\|(r)$.
(Alternatively, this form can be deduced from symmetry considerations,
reasoning that any homogeneous and isotropic rank-2 tensor takes the
form given by eq.~\eqref{eq:perfect-corr} \cite{Monin:1975}.) As indicated by the subscripts these
correlation functions can be interpreted as those of the perpendicular and parallel
components of $\v$ relative to the pairwise separation $\r$.%
\footnote{The parallel component of the velocity relative to the separation is
$\hat{r}_i v^i$ so $\Psi_\|(r)=\hat{r}^i \hat{r}^j\langle v_i(\x) v_j(\x')\rangle$. The
perpendicular component $\mathcal{P}_{ij} v^j$ lies in a plane orthogonal to $\r$. With 
$\mathcal{P}^\perp_{ij}\equiv\delta^\mrm{K}_{ij}-\hat{r}_i\hat{r}_j$, we have
\be
\Psi_\perp(r)\mathcal{P}^\perp_{ij}
=\mathcal{P}^\perp_{ik}\mathcal{P}^\perp_{jl}\big\langle v_k(\x) v_l(\x')\big\rangle.
\ee
}
Note that the dependence of $\Psi_{ij}$ is on $\r$ and not $r$, as
eq.~\eqref{eq:perfect-corr} makes clear; intrinsic correlations only depend on $r$
through $\Psi_\perp$ and $\Psi_\|$. Integrating over directions $\hat\k$
in eq.~\eqref{eq:Psi-ij} the correlation functions are given by \cite{Gorski:1988}
\begin{subequations}
\begin{align}
\Psi_\perp(r)
&=\int^\infty_0\!\frac{k^2\dif k}{2\pi^2}\,
    K_\perp(kr) \bigg[\Big(\frac{\calH f}{k}\Big)^2 P_{\theta\theta}(k)\bigg], \label{eq:Psi-perp} \\[4pt]
\Psi_\|(r)
&=\int^\infty_0\!\frac{k^2\dif k}{2\pi^2}\,
    K_\|(kr) \bigg[\Big(\frac{\calH f}{k}\Big)^2 P_{\theta\theta}(k)\bigg], \label{eq:Psi-para}
\end{align}
\end{subequations}
where we have the kernels $K_\perp(x)\equiv j_1(x)/x$ and $K_\|(x)\equiv j_0(x)-2j_1(x)/x$. Note that
$\Psi_\|=\dif(r\Psi_\perp)/\dif r$ since $\v$ is solely determined by its velocity
divergence $\theta$ (so there can only be one independent correlation function).
Observations of the velocity field are limited to the LOS component $\v\cdot\n$.
For the purpose of comparison with our redshift-space results we assume
$\x/|\x|\approx(\x+\r)/|\x+\r|\approx\hat\z$, i.e.\ the plane-parallel limit.
The LOS correlation function is then
\be
\xiv(r,\mu)
\equiv\big\langle \vz(\x) \, \vz(\x+\r) \big\rangle
=(1-\mu^2)\Psi_\perp(r) + \mu^2\Psi_\|(r),
\ee
where $\mu=\hat\r\cdot\hat\z$. We see that $\xiv$ interpolates between 
$\Psi_\perp$ (when $\mu=0$) and $\Psi_\|$ (when $\mu=1$).
However, in general, for two distinct lines-of-sight, the correlation function depends on
three variables, not just two.

Evidently, the LOS identifies the observer's position as a preferred
location. Thus the orientation of the pairwise vector relative to the observer is now
relevant to the computation of the two-point correlations. That is, we need both the separation 
and orientation of the pair relative to the LOS; the correlation function is no longer
statistically isotropic. As with before, we separate out the angular dependence by yet
another multipole decomposition:
\bea
\xiv(r,\mu)
= \sum_\ell \xiv^\ell(r)\mathcal{L}_\ell(\mu).
\eea
The multipoles are given by
\begin{subequations}
\bea
\xiv^0(r)
&=+\frac13\int^\infty_0\!\frac{k^2\dif k}{2\pi^2}\,
    j_0(kr) \bigg[\Big(\frac{\calH f}{k}\Big)^2 P_{\theta\theta}(k)\bigg] \\[4pt]
\xiv^2(r)
&=-\frac23\int^\infty_0\!\frac{k^2\dif k}{2\pi^2}\,
    j_2(kr) \bigg[\Big(\frac{\calH f}{k}\Big)^2 P_{\theta\theta}(k)\bigg]
\eea
\end{subequations}
or $\xiv^0=1/3(\Psi_\|+2\Psi_\perp)$ and $\xiv^2=2/3(\Psi_\|-\Psi_\perp)$,
using that $j_0=K_\|+2K_\perp$ and $j_2=K_\perp-K_\|$.
We emphasize that the correlations are anisotropic; the quadrupole moment is present
($\xiv^2\neq0$) even in real space.
This is however nothing more than a consequence that we are computing two-point
correlations of a vector field and so must depend on the orientation of the galaxy pair.


\subsection{Redshift space\label{sec:config-space-redshift}}
Turn now to $\xi^s_{\theta\theta}(\s)$, the two-point function of the
\emph{redshift-space} velocity divergence field. This is related to the
redshift-space power spectrum $P^s_{\theta\theta}(\k)$ via Fourier transform:
\be\label{eq:xi-s-tt}
\xi^s_{\theta\theta}(\s)
=\big\langle \theta^s(0) \, \theta^s(\s)\big\rangle
=\int\!\frac{\dif^3\k}{(2\pi)^3}\, P^s_{\theta\theta}(\k)\, \rme^{-\ii\k\cdot\s}.
\ee
Substituting eq.~\eqref{eq:Ps-tt-multi} into eq.~\eqref{eq:xi-s-tt}, and using that the
angular integral evaluates to eq.~\eqref{eq:int-bess-id}, we can write
\be
\xi^s_{\theta\theta}(s,\mu)
=\sum_\ell  \xi^\ell_{\theta\theta}(s)\calL_\ell(\mu),
\ee
with the multipoles given by
\be\label{eq:xi-ell-tt}
\xi^\ell_{\theta\theta}(s)
\equiv \ii^\ell\int^\infty_0\frac{k^2\dif k}{2\pi^2}\:j_\ell(ks) P^\ell_{\theta\theta}(k),
\ee
which we note is the spherical Bessel transform of $P^\ell_{\theta\theta}$.
We recall only multipoles with even $\ell$ can contribute to the overall anisotropy, reflecting
the fact that we have symmetry under galaxy pair exchange ($\s\to-\s$).
In the case of the plane-parallel limit only the $\ell=0,2,4$ moments are nonvanishing
(as with the multipoles of density two-point correlations).

The velocity correlation tensor in terms of $P^s_{\theta\theta}$ is
\be\label{eq:Psi-s-ij}
\Psi^s_{ij}(\s)
= \int\!\frac{\dif^3\k}{(2\pi)^3}\,
    \Big(\frac{\calH f}{k}\Big)^2
    \frac{k_i}{k}\frac{k_j}{k}
    P_{\theta \theta}^{s}(\k)\, \rme^{-\ii\k\cdot\s}.
\ee
Notice that the difference between this expression and eq.~\eqref{eq:Psi-ij} is that the power spectrum
now depends on $\k$.
This slight difference however means that $\Psi^s_{ij}$ cannot be written in terms of parallel and
perpendicular correlation functions, as in eq.~\eqref{eq:perfect-corr}. In general, we find that the
parallel and perpendicular modes are not independent but become correlated. In addition, $\Psi^s_{ij}$
possesses a more complicated multipole structure; see Appendix \ref{app:tensor} for more details.

To obtain the LOS correlations from eq.~\eqref{eq:Psi-s-ij}, we can contract with $\hat{n}_i\hat{n}_j$,
insert eq.~\eqref{eq:Ps-tt-multi} then perform the necessary integrations. But a simpler way
is to instead begin with
\be\label{eq:xivv-ani}
\xiv^s(\s)
=\big\langle \vz^s(0) \, \vz^s(\s)\big\rangle
=\int\!\frac{\dif^3\k}{(2\pi)^3}\, \Pv^s(\k)\, \rme^{-\ii\k\cdot\s},
\ee
where $\Pv^s$ is LOS velocity power spectrum.
To relate this to the velocity divergence power spectrum we recall that the
velocity field is irrotational, so $\vz^s(\k)=-\calH f (\ii\kz/k^2){\theta}^s(\k)$,
which yields the exact relations \eqref{eq:Psvv} and \eqref{eq:Psvv-ell}
(i.e.\ independent of how $\theta$ is modelled).
Since we have symmetry about the LOS we can write again as a sum of multipoles,
\be\label{eq:xiv-s}
\xiv^s(s,\mu)
= \sum_{\ell} 
    \xi^\ell_{vv}(s) \calL_\ell(\mu).
\ee
Substituting eq.~\eqref{eq:Psvv-ell} into eq.~\eqref{eq:xivv-ani} we find that the
multipoles are given by
[cf.~eq.~\eqref{eq:xi-ell-tt}]
\be\label{eq:xiell-vv}
\xiv^\ell(s)
=\ii^\ell\!\int^\infty_0 \frac{k^2\dif k}{2\pi^2}\: j_\ell(ks)
    \bigg[\Big(\frac{\calH f}{k}\Big)^2\sum_{\ell'}A_{\ell\ell'}\,P^{\ell'}_{\theta\theta}(k)\bigg],
\ee
i.e.\ in terms of a combination of the velocity divergence multipole moments, with $A_{\ell\ell'}$
given by eq.~\eqref{eq:Psvv-ell}.
As we have mentioned in the preceding section the $\mu$ dependence is not an indication of
RSD as it is in galaxy clustering (though statistical homogeneity is still preserved in the
plane-parallel limit).

We have already obtained a relation between the power spectrum multipoles of $\theta^s(\k)$ and $\vz^s(\k)$;
see eq.~\eqref{eq:Psvv-ell}. We now derive an analogous relation between the {configuration}
space multipoles, i.e.\ between $\xi^\ell_{\theta\theta}(s)$, from $\xiv^\ell(s)$.
This is perhaps most easily done if we use that spherical Bessel functions satisfy the second-order
differential equation,
\be\label{eq:SB-ODE}
\bigg[\frac{\partial^2}{\partial s^2}+\frac2s\frac{\partial}{\partial s}-\frac{\ell(\ell+1)}{s^2}\bigg] j_\ell(ks)
=-k^2 j_\ell(ks).
\ee
Substituting this into eq.~\eqref{eq:xiell-vv} we obtain the following relation between multipoles:
\be\label{eq:xi-diff-rel}
\mathcal{D}_\ell\,\xiv^\ell + (\calH f)^2\sum_{\ell'} A_{\ell\ell'}\, \xi^{\ell'}_{\theta\theta}=0.
\ee
[Here $\mathcal{D}_\ell$ is defined by the contents of the square brackets on the left-hand side of 
eq.~\eqref{eq:SB-ODE}.]
Unsurprisingly, the algebraic relation in Fourier space \eqref{eq:Psvv-ell} becomes a differential
relation in configuration space. This expression cannot however be inverted to obtain
an explicit formula for the velocity divergence multipoles since $A_{\ell\ell'}$ is
overdetermined (the number of rows exceeds the number of columns). In practice, it is likely that only
the monopole $\xiv^0$ and quadrupole $\xiv^2$ can be measured at a satisfactory level, given that
$P^4_{vv}$ and $P^6_{vv}$ are subdominant (see figure~\ref{fig:Pvv_disp}). Still, this leaves
$\xi^0_{\theta\theta}$ overdetermined.
In any case, the practical utility of estimating $\xi^\ell_{\theta\theta}$ from eq.~\eqref{eq:xi-diff-rel}
will crucially depend on the ability to differentiate what is in practice a noisy estimate of $\xiv^\ell$.
Provided this can be done eq.~\eqref{eq:xi-diff-rel} shows in principle how one can estimate the divergence
field's two-point function directly from the measured LOS velocity multipoles in configuration space.

Finally, we stress that the expressions derived above are valid in the plane-parallel limit, but aside
from this assumption they are exact and independent of any input cosmology or PT modelling.
The $\xi^\ell_{\theta\theta}$'s can therefore be taken to be arbitrary functions (though we have assumed
that the multipole moments are only nonzero for $\ell=0,2,4,6$).
Alternatively, by taking the inverse Fourier transform of eq.~\eqref{eq:P-s-exact}, we can obtain
the following exact relation for the LOS velocity two-point function:
\be\label{eq:xiv-stream}
\xiv^s(s_\|,s_\perp)
=\int^\infty_{-\infty}\dif{r_\|}\:\int^\infty_{-\infty} \frac{\dif\kappa}{2\pi}\:
    \rme^{\ii\kappa(r_\|-s_\|)}
    \Big(\frac{\calH f}{\kappa}\Big)^2
    \Big\langle \rme^{-\ii f\kappa \Delta\uz}\nablaz\uz(\x) \nablaz\uz(\x') \Big\rangle,
\ee
where $s_\|=s\mu$, $s_\perp=r_\perp=(s^2-s_\|^2)^{1/2}$, and $\Delta\uz=\uz(\x)-\uz(\x')$.

The inner integral in eq.~\eqref{eq:xiv-stream} defines a probability distribution function,
expressed as the Fourier transform of the pairwise velocity generating function.
Equation~\eqref{eq:xiv-stream} can thus be considered the velocity analogue of the streaming
model given by eq.~\eqref{eq:xi-stream}. The object of interest here is the generating function and
so the challenge thus lies in modelling the pairwise velocity statistics. An interesting question we
can ask is whether a Gaussian distribution is valid, for then one requires only a model of the
(real-space) power spectrum. We note that such an assumption has been validated in the case
of clustering in the Gaussian streaming model~\cite{Reid:2011,Wang:2014} (but see also
ref.~\cite{Cuesta-Lazaro:2020ihk} for a non-Gaussian extension).

\section{Discussion and conclusions\label{sec:discussion}}

\subsection{RSD and the gradient expansion}
Using the distribution-function approach to RSD \cite{Seljak:2011tx}, we obtained an
expression, eq.~\eqref{eq:v-s-series}, describing the velocity field as would be seen in redshift space.
The expression is given as a series expansion and shows that at leading order the velocity field in
real space and redshift space coincide. This means that distortions are a second-order effect and
are therefore absent in linear theory.
Whether considered in real space or redshift space, the series expansion explicitly
shows that the velocity field is, as we expect, a volume-weighted quantity, despite the apparent
density weighting in eq.~\eqref{eq:v-s-frac} (i.e.\ there is no coupling to the real-space density field).

We have observed that the series expansion is organized as a hydrodynamic gradient expansion:
zeroth-order terms are derivative-free and correspond
to the perfect fluid that is usually considered;  higher-derivative terms (which counts
products of lower-order derivatives) represent short-wavelength, dissipative corrections to the
perfect-fluid description (among other higher-order hydrodynamic effects).
Understood in this way,  distortions to the real-space motions of a perfect fluid
can be likened to the dissipative effects of an \emph{imperfect} fluid (e.g.\ one with heat
conduction, shear viscosity, etc). However, the kind of dissipation is different to any that
might be encountered in nature. This is because it depends on who is observing
it. That is, the mapping \eqref{eq:mapping} singles out the observer's
LOS as a preferred direction; gradients appearing in the derivative expansion are really LOS derivatives
($\nabla_\|=\n\cdot\nabla$), and it is these terms that give rise to apparent observer-dependent dynamics
in the fluid.

The gradient expansion was shown to follow from a simple integral formula 
given by eq.~\eqref{eq:v-s-G1} [see eqs.~\eqref{eq:1pdelta-s} and \eqref{eq:m-s} for the case of the density
and momentum counterparts, respectively]. This is perhaps not surprising given that the redshift-space distribution 
function~\eqref{eq:fs} is itself given by a convolution.
While formally equivalent to the gradient expansion when considered to all orders, this nonperturbative
form provides heuristic way to understand how the mapping gives rise to distortions---namely, as
a certain convolution which only operates on the LOS modes. Indeed, this agrees with the idea that the correlation
function of galaxies in redshift space takes on characteristics of the (LOS) velocity field through the
pairwise-velocity probability density function \cite{Peebles:1980,Fisher:1995,Carlson:2013}.
Technically speaking, because the shift term in the mapping---the peculiar velocity---depends
on space, we have more correctly a convolution of the real-space field with a spectrum of plane waves, each
with a different velocity-induced phase; in the case of the velocity field, which is not density weighted,
the plane waves have an amplitude $1/(\k\cdot\n)$.
(Note that despite the appearance of the Dirac delta function, the convolution cannot be carried out
in general, for the shift carries $\x$ dependence.)

That we have been able to write redshift-space fields as an integral transformation of their real-space
counterpart is made possible by asserting the PPF assumption at the level of the (real-space)
distribution function; see eq.~\eqref{eq:f-PPF}. (Recall this is usually assumed {after} taking moments
of the Boltzmann equation.) We are thus working in the regime of single-streaming
in which phase-space particle trajectories do not cross in real space. However, in the more realistic
case of multi-streaming the reassignment of mass tracers implied by the mapping \eqref{eq:mapping} allows
for multiple tracers at the same position $\x$: Even if single-streaming is valid in real space it
does not preclude the possibility of multi-streaming in redshift space.
The validity of our formulae should be understood with these caveats in mind.

\subsection{Power spectrum}
Two models for the redshift-space LOS velocity power spectrum have been presented.
The first model we presented, eq.~\eqref{eq:Ps-NL}, is based on the gradient expansion, and largely
follows the approach taken in refs.~\cite{Seljak:2011tx,DF2,DF3,DF4} for the density field, and
ref.~\cite{Okumura:2014} for the momentum field.
The second model~\eqref{eq:Ps-cumexp-model} is based on the integral formula, and derives from
a nonperturbative expression for the power spectrum given in terms of the pairwise-velocity generating
function [eq.~\eqref{eq:P-s-exact}].
This model is constructed in a similar way to the well-known TNS clustering model \cite{Taruya:2010};
that is, it is based on the cumulant expansion of the generating function \cite{Scoccimarro:2004tg},
upon which the connected moments are evaluated using PT.

Both models show a damping of the power spectrum. Quantitatively, the damping begins on quasilinear
scales ($k\gtrsim 0.01\,h\Mpc^{-1}$), and reaches about $20\%$ at $k\simeq 0.1\,h\Mpc^{-1}$.
There is also an ``FoG'' effect in the velocity field, much as seen in redshift-space clustering.
In the first model, we have included FoG damping empirically, invoking the ``dispersion models''
of the galaxy power spectrum.
The second model has the virtue that FoG damping naturally arises from the cumulant expansion of the generating
function and appears in the exactly the same way as for clustering models [cf.~eq.~\eqref{eq:cum-thm-2}].
The FoG effect in the velocity field---being related to virial motions of galaxies---is no less difficult to model
from first principles, and we have thus adopted a Gaussian model, $D^2_\mrm{1pt}(x)=\exp(-x^2)$, $x=k\mu\sigma_u$.
Using the PT prediction for the velocity dispersion at $z=0$, $\sigma_v\equiv(H_0 f)\sigma_u\simeq 300\,{\rm km\,s}^{-1}$,
leads to an additional damping of about $10\%$.

The damping can be explained qualitatively in terms of the gradient expansion.
Dissipation here has the effect of erasing density gradients in the fluid, and with it the correlated
motions. In practice, this is the familiar FoG effect in action whereby galaxies are scattered out from
clusters. As time goes on, and the velocity dispersion grows larger, galaxies are scattered
further and further away from the centers of their host halos. From the dynamical point of view, there is an outflow of
material (galaxies) from dense regions to less dense regions and this is akin to a heat conduction (movement from hotter
to cooler regions). Of course, this is the reverse of what actually occurs when viewed in real space, though, which is that
(on large scales) matter is acted on solely by gravity causing it to be drawn towards higher-density regions and away
from lower-density regions.

\subsection{Comparison with previous work}
Our framework provides a physical model for the damping observed in $N$-body simulations by
Koda et al.~\cite{Koda:2014} (hereafter K14).
These simulations showed two regimes of behaviour: (\emph{i}) a damping in the measured monopole moment
of the power spectrum beginning on scales $k\simeq0.01\,h\Mpc^{-1}$ and lasting to $k\simeq0.1\,h\Mpc^{-1}$;
(\emph{ii}) an enhancement in power at $k\gtrsim0.2\,h\Mpc^{-1}$ over the damping in (\emph{i}) (when
extrapolated to larger $k$).  K14 explained (\emph{ii}) as arising from a (largely) scale-independent
random component in the velocity with assumed Gaussian statistics (zero-centered with variance $\sigma_*^2$);
on the other hand, K14 explained (\emph{i}) as being due entirely to the familiar FoG effect. The latter
was modelled phenomenologically using an angle-independent damping function $D^2_\mrm{K14}(k\sigma_\mrm{K14})=\sin^2(k\sigma_\mrm{K14})/(k\sigma_\mrm{K14})^2$.
Assuming (\emph{i}) and (\emph{ii}) arise from independent effects, K14 found that
$\sigma_\mrm{K14}\simeq13\,h^{-1}\Mpc$ was required for concordance with simulation.
(Note that the empirical parameter $\sigma_\mrm{K14}$ was found to depend somewhat on the subhalo mass bin chosen.)

A comparison between the monopole moment of our power spectrum model \eqref{eq:Ps-cumexp-model} and the K14
fitting function $D^2_\mrm{K14}\Pv$ shows that our model predicts about $10\%$ of excess power at $k=0.1\,h\Mpc^{-1}$
(see dashed lines in figure~\ref{fig:model-AB}). This is however not surprising given that the halo velocities
are known to be biased tracers of the dark matter velocity field~\cite{Baldauf:2014fza,bias_review}.
As figure~\ref{fig:model-AB} also shows, a reasonable fit to simulations is obtained by allowing a halo velocity bias.
For simplicity, we have taken into account the bias by considering only the effect at lowest order in
perturbations, i.e.~on the linear power spectrum.
Note that the velocity bias results in a further suppression of power---e.g.~at $k=0.1\,h\Mpc^{-1}$
the subhalo velocity (in real space) is lower by about $4\%$ compared to dark matter,
and the velocity lower power by about $8\%$. 

Separately, we reiterate that our model uses the linear-theory prediction for the velocity dispersion
that controls the amount of FoG damping. Given the theoretical uncertainty around this parameter it is more
appropriate to treat it as a nuisance parameter. Of course, this introduces an extra fitting parameter,
which ensures concordance with simulation.

But it is perhaps unsurprising that we need an empirical FoG parameter. After all, the model given by eq.~\eqref{eq:Ps-cumexp-model}
is styled on the TNS model, which itself requires that the FoG velocity dispersion be treated as an empirical parameter.
It is likely that a more careful treatment of the FoG damping model (as in refs.~\cite{Zheng:2016}) will be needed
than the simple one we have given here.
Nevertheless, our results show that the damping cannot be entirely blamed on FoG effects: about half of the
observed damping should be attributed to the coherent streaming motions (i.e.\ not related to the internal motions
of clusters).

A more comprehensive comparison of our models with simulations may also need to consider the possibility of a velocity bias.
In this work we have assumed no velocity bias between galaxies and matter, $\v_g=\v$. The issue of bias is a complicated
subject (see~ref.~\cite{bias_review} and references therein) and well beyond the scope of this work. However, we note
that on scales $k\lesssim 0.2\,h\Mpc^{-1}$ halos do not appear to biased velocity tracers~\cite{Chen:2018,Zheng:2015}.
In the case of subhalos, considered in K14, the situation is different. On small scales a subhalo velocity bias
is fairly well established from simulations~\cite{Carlberg:1989,Carlberg:1990,Carlberg:1994,Colin:2000,Jennings:2015}.
And this bias does not necessarily need to arise from baryonic effects; owing to dynamical friction, dark-matter-only
simulations have also observed such a bias~\cite{Carlberg:1990}.

In addition, while we have computed the leading-order effect on the power spectrum,
a higher-order calculation may be required. Firstly, the velocity power spectrum
is more sensitive to nonlinear effects than is the case for density; see figure~\ref{fig:Poneloop}.
Secondly, PT breaks down on larger scales in redshift space than in real space \cite{Scoccimarro:1999ed}, though
we note that this applies more to the first model, which is based on treating both the redshift mapping and
dynamics perturbatively.
An obvious first step in this direction, however, is to check whether the connected four-point moment in
eq.~\eqref{eq:2s2s} (which does not appear in our one-loop calculation) is sizable. We leave this to future
work. In the end, however, such a calculation may not be necessary, given that how accurate the model needs
to be specified will depend on the quality of the data at hand. Since measurement errors on the peculiar
velocities directly propagate to the power spectrum's shot noise error, any systematic bias present may not
be significant enough to warrant the higher-order calculation.

Finally, we note that the second model (based on the cumulant expansion) should in principle capture more
of the nonlinearity missed in the first model (based on moment expansion). This is because the
second model treats the redshift mapping exactly, whereas the first model treats it perturbatively.
Furthermore, while both models treat the dynamics perturbatively, the second model does not explicitly
assume the smallness of the field's amplitude. Rather, it is the correlations that are expected to be weak,
and this was the logic in performing an expansion in powers of $j_1\propto k$ in eq.~\eqref{eq:cum-1loop}.

\section{Summary and outlook\label{sec:conclusions}}
We have studied the effect of RSD on the motions of tracers as inferred from their
redshift-space positions. Beginning with the distribution-function approach to RSD~\cite{Seljak:2011tx},
we derived two expressions for the \emph{redshift-space} velocity field---the derivative
expansion~\eqref{eq:v-s-series} and the convolution formula~\eqref{eq:v-s-G1}. These expressions are
formally equivalent at all orders, but permit different perturbative treatments. Using one-loop PT,
we computed the leading-order effect of RSD on the velocity power. Working in the plane-parallel limit,
two models for the redshift-space velocity power spectrum were presented, each based on a different
perturbative approach:
\begin{enumerate}[I.]
\item Power spectrum model~\eqref{eq:Ps-NL} is obtained from the derivative expansion~\eqref{eq:v-s-series}.
The effect of RSD is captured in a set of LOS-dependent mode-coupling kernels~\eqref{eq:Z-kernels};
these kernels are modified from the standard one-loop kernels, and are akin to the redshift-space
density kernels given in ref.~\cite{Scoccimarro:1999ed}. This model is closely related to those~\cite{Seljak:2011tx,Okumura:2014}
derived from the (density-weighted) velocity-moment expansion.
\item Power spectrum model~\eqref{eq:Ps-cumexp-model} is obtained from the convolution formula~\eqref{eq:v-s-G1}.
It follows from using the cumulant-expansion theorem on the pairwise-velocity generating
function appearing in the exact expression \eqref{eq:P-s-exact}.
This model may be considered the velocity analogue of the TNS model~\cite{Taruya:2010} for the
galaxy power spectrum.
\end{enumerate}

Our main findings are as follows.
Both models I and II predict a damping of the power spectrum beginning on
quasilinear scales $k\gtrsim 0.01\,h\Mpc^{-1}$; at large-scales $k\to0$ the effect
is suppressed by higher-derivative terms.
Heuristically, the damping may be understood as a RSD-induced dissipation: in redshift space,
we have an apparent outflow of galaxies directed along the LOS towards lower-density regions, behaviour
which is not described by a purely gravitating perfect fluid.
From the gradient-expansion perspective of hydrodynamics this implies an apparent nonvanishing heat conductivity.
The overall effect is to suppress the tendency for galaxies moving under gravity to fall towards
regions of higher density.
This is a long-range FoG effect; it is present in addition to the usual FoG effect due to
the virial motions of galaxies, which also exists for the velocity field.
(There is no analogous Kaiser effect, however). In the case of model I an FoG-type damping
is entirely absent from the model and needs to be put in by hand; in the case of model II the damping
arises from one-point moments within the pairwise velocity generating function. The damping is qualitatively
consistent with behaviour observed in $N$-body simulations \cite{Koda:2014}, and a quantitative fit to simulations
thus requires  treating the velocity dispersion parameter empirically (as with galaxy clustering models).

Our broader motivation for this study has been to supply in part the theoretical predictions needed for an
eventual multi-tracer analysis of galaxy density and peculiar velocities~\cite{Koda:2014}. In the past, a proper
comparison between data and theory has not been possible, with workers
(e.g.~\cite{Burkey:2003rk,Adams:2020,Amendola:2021}) having relied on a phenomenological model.
In this regard the framework we have presented provides much of the needed analytic modelling (and numerical 
implementation).
While it is clear that more detailed modelling is needed before confronting with data
(e.g.\ of the dynamics, in relaxing the plane-parallel assumption, etc), we hope that the
framework we have developed can nevertheless provide a template for future efforts on this front.
Suffice to say this first study has largely been devoted to theory.
In future work we will investigate the advantages of performing analysis in
redshift space using mock data. Questions of particular interest include:
quantifying the information gain in constraining the growth rate and
breaking of parameter degeneracies in redshift space; the extent to which systematic
biases arise from using the phenomenological damping model; and assessing the trade-off
between smaller errors in galaxy redshift-space positions versus the loss of cosmological signal
from the power suppression.

\section*{Acknowledgements}
We thank Chris Blake for helpful comments and suggestions,
and for his comments on this manuscript. We also thank Jun Koda for making us
aware of the heat-conduction analogy and bringing to our attention ref.~\cite{Kaiser:2014jca}.
LD is supported by the Australian government Research Training Program.
The code used to obtain the numerical results in this work is publicly available and can be found
at \url{https://github.com/lhd23/RSDPT-FFTLog/}.
We acknowledge use of the software libraries
NumPy~\cite{numpy}, SciPy~\cite{scipy}, and Matplotlib~\cite{matplotlib}.

\clearpage
\appendix
\section{Power-law FFTLog numerical method\label{app:FFT}}
In this appendix we give details on the numerical evaluation method for the
power spectrum model. In this model, and in PT more generally, we frequently encounter convolutions,
such as
\be\label{eq:conv-int}
P_{22}(k)= 2\int\!\frac{\dif^3\q}{(2\pi)^3}\,\big[\Gs^{(2)}(\q,\k-\q)\big]^{2}\PL(q)\,\PL(|\k-\q|).
\ee
These loop integrals are generally unpleasant to evaluate efficiently and precisely.
Firstly, the mode coupling is over a large dynamic range of the power spectrum; secondly,
many such integrals need to be performed if $P_{22}(k)$ is to be returned at all wavenumbers $k$
of interest---as well as with different values of the cosmological parameters (e.g.\ for
Markov chain Monte Carlo sampling).

To evaluate these integrals (and others) we use a recent method \cite{Simonovic:2017mhp}
based on the FFTLog algorithm \cite{Talman:1978} (see also ref.~\cite{Hamilton:1999uv}).
The key idea is to exploit the fact that by representing the power spectrum as a discrete
Fourier Transform in $\ln k$, rather than $k$, we can express the linear power spectrum
$\PL(k)$ over some finite range of scales of interest as
\be\label{eq:P-FFT}
\PL(k)=\sum_{m=-N/2}^{N/2} c_m \, k^{\nu+\ii\eta_m}\,;
\ee
i.e. as the sum of (complex) power laws [the symbols are defined below]. This last
fact is important as many integrals, including those of the form \eqref{eq:conv-int},
are analytic in the case of a power-law power spectrum. Because no numerical integration
is involved this method is significantly faster than standard quadrature or
Monte Carlo integration. In particular, the FFTLog approach requires $\mathcal{O}(N\log N)$
steps, outputting all $k$ at once. (This is in contrast to the $\mathcal{O}(N^3)$ steps needed
for quadrature integration.)

Strictly speaking, the equality in eq.~\eqref{eq:P-FFT} is only approximate for finite
$N$, and the periodicity of the right-hand side of eq.~\eqref{eq:P-FFT} means we need to restrict
attention to some finite range. The chosen $N$ will depend on how featureful
the function to be approximated is; in the case of the standard $\Lambda$CDM power spectrum
an $N$ of only about $200$ is sufficient to accurately represent the linear power spectrum
down to the BAO wiggles.
In this method the coefficients of the discrete Fourier transform encode the
cosmological information, allowing the cosmology to be separated out from the integrals.
These integrals can then be performed analytically (and only once), then stored in
look-up tables. Below we discuss in more detail the FFTLog approach to convolution
integrals. Other difficult integrals, many involving spherical Bessel functions, can also
be evaluated using another variant of the FFTLog method.

For a logarithmic sampling of points in $k$-space, the Fourier coefficients in
eq.~\eqref{eq:P-FFT} are given by
\be\label{eq:cm}
c_m=\frac1N\sum_{j=0}^{N-1} \PL(k_l) \,
    k_j^{-\nu} \, k_\mrm{min}^{-\ii\eta_m} \, \rme^{-\ii2\pi m j/N},
\qquad
\eta_m = \frac{2\pi m}{\ln(k_\mrm{max}/k_\mrm{min})},
\ee
with the understanding that $c_{\pm N/2}$
is multiplied by a factor of $1/2$ to get the correct endpoint weighting.
Here $c_m^*=c_{-m}$ by the reality of the power spectrum; $\nu$ is a real
number called the \emph{bias}, which is to be chosen to avoid spurious divergences (see
below); and $k_j=k_\mrm{min}\, (k_\mrm{max}/k_\mrm{min})^{j/N}$, since the points are
uniformly spaced in $\ln k$. In practice, when we compute the FFT, the input signal is
the ``biased'' power spectrum $P(k) k^{-\nu}$.

Now, convolutions of the form \eqref{eq:conv-int} reduce to a linear combination
of irreducible integrals, which can be carried out analytically \cite{Scoccimarro:1996se}:
\be\label{eq:int-Ifunc}
\int\!\frac{d^3\q}{(2\pi)^3}\frac{1}{q^{2\nu_1}|\k-\q|^{2\nu_2}}
=k^{3-2(\nu_1+\nu_2)}\,\mathsf{I}(\nu_1,\nu_2),
\ee
where
\be\label{eq:Ifunc}
\mathsf{I}(\nu_1,\nu_2)
\equiv\frac{1}{8\pi^{3/2}}
\frac{\Gamma(\frac32-\nu_1)\Gamma(\frac32-\nu_2)\Gamma(\nu_1+\nu_2-\frac32)}
    {\Gamma(\nu_1)\Gamma(\nu_2)\Gamma(3-\nu_1-\nu_2)}.
\ee

\paragraph*{Implementation.} The discrete Fourier transform suffers two edge effects---ringing
and aliasing. \emph{Aliasing}, in which small-scale features leak into large scales,
can be mitigated by padding the input signal array with zeros on both ends.
Rapid oscillations at the ends of the input signal, or \emph{ringing}, can be suppressed
by passing the signal through a low-pass filter to the Fourier coefficients.

\subsection{Integrals of type 22}
Terms of the type $P_{22}$ are convolutions and may be written in the FFTLog approach as
\bea
\fint{\q}\, I_\ell(\q,\k-\q) \PL(q) \PL(|\k-\q|)
&=\sum_{m_1,m_2} c_{m_1} c_{m_2} \sum_{n_1, n_2} f^\ell_{n_1n_2} 
    k^{-2(n_1+n_2)} \! \fint{\q}\,
    \frac{1}{q^{2(\nu_{m_1}-n_1)}|\k-\q|^{2(\nu_{m_2}-n_2)}} \nonumber\\[4pt]
&=k^3\!\!\sum_{m_1,m_2} 
    \bigg[\sum_{n_1, n_2} f^\ell_{n_1n_2} \,
    \mathsf{I}(\nu_{m_1}-n_1, \nu_{m_2}-n_2)\bigg]
    c_{m_1}\,k^{-2\nu_{m_1}}\, c_{m_2}\, k^{-2\nu_{m_2}},
    \label{eq:22-int}
\eea
with $\nu_{m_1}\equiv-\frac12(\nu+\ii\eta_{m_1})$ and $\nu_{m_2}\equiv-\frac12(\nu+\ii\eta_{m_2})$.
We thus see that for a given $k$ the convolution reduces to matrix multiplication involving two
copies of the vector $(c_{-N/2} k^{-2\nu_{-N/2}},\ldots,c_{N/2} k^{-2\nu_{N/2}})$, and a matrix
with components given by the contents of the square brackets above.\footnote{The functions
$\mathsf{I}(\cdot,\cdot)$ have several useful properties and  satisfy a set of recursion relations
that allow the linear combination in the square brackets of eq.~\eqref{eq:22-int} to be rewritten in
terms of a single $\mathsf{I}$ \cite{Simonovic:2017mhp}. As these functions are undemanding
to evaluate we will not simplify further.} As mentioned above, the matrix is independent
of the wavenumbers and cosmology, and  can be precomputed for a given $N$, $k_\mrm{min}$, $k_\mrm{max}$,
and $\nu$.

\subsubsection*{Kernel expansion}
To take advantage of the FFTLog method we first need to put the $I_\ell(r,\mu')$ kernels
\eqref{eq:Iell} in the form \eqref{eq:int-Ifunc}. This is done by replacing $r$ with $q/k$, and
rewriting $\mu'\equiv\hat\k\cdot\hat\q$ in terms of $k$, $q$, and $|\k-\q|$, with the help of
the identity $|\k-\q|^2=k^2+q^2-2kq\mu'$.
All resulting terms contain integer powers of $k^2$, $q^2$, and $|\k-\q|^2$ and thus
have the general form $k^{-2(n_1+n_2)} q^{2n_1}|\k-\q|^{2n_2}$. The kernels can then be
encoded in a set of indices ($n_1,n_2$) and coefficients $f^\ell_{n_1n_2}$.
A summary of all kernels using the FFTLog expansion is given in table~\ref{fft-ind}.

For example, consider the kernel $2[\Ks^{(2)}(\q,\k-\q)]^2$. This depends on $\n$ but does not
pose a real problem, as we can simply factor out its dependence by expanding into multipoles as
\bea
2[\Ks^{(2)}(\q,\k-\q)]^2&= \sum_\ell I_\ell(\q,\k-\q)\calL_\ell(\hat\k\cdot\n), \\
I_\ell(\q,\k-\q)&\equiv(2\ell+1)\int^1_{-1}\frac{\dif\mu}{2}\mathcal{L}_\ell(\mu)
	\int^{2\pi}_0\frac{\dif\phi}{2\pi}\: 2[\Ks^{(2)}(\q,\k-\q)]^2.
\eea
(In the case of the SPT kernels only the $\ell=0$ multipole is nonzero since there is
no explicit $\mu$ dependence.) We can then apply the method for each $\ell$; e.g.\
for $\ell=0$, we can write in FFTLog form
\bea
I_0(\q,\k-\q)
&=\sum_{n_1,n_2} f^0_{n_1n_2}\, \frac{k^{-2(n_1+n_2)}}{q^{-2n_1}|\k-\q|^{-2n_2}} \nonumber\\[4pt]
&=\frac{k^8}{60 |\k-\q|^4 q^4}-\frac{k^6}{30 |\k-\q|^4 q^2}-\frac{k^6}{30 |\k-\q|^2 q^4} \nonumber\\[4pt]
&\quad + \frac{k^4}{60 |\k-\q|^4}+\frac{k^4}{15 |\k-\q|^2 q^2}+\frac{k^4}{60 q^4}.
\eea
From here it is straightforward to read off the coefficients and indices of each term,
then input them into eq.~\eqref{eq:22-int}.

The bias parameter $\nu$ is the one tuning parameter in the FFTLog approach.
This parameter is needed to ensure the convergence of each
integral \eqref{eq:Ifunc} corresponding to $\mathsf{I}(\nu_{m_1}-n_1,\nu_{m_2}-n_2)$,
which is not guaranteed if we naively set $\nu=0$. Though \eqref{eq:Ifunc} will always
give a finite answer \cite{Simonovic:2017mhp}, even for values of $\nu_1$ and $\nu_2$
that would yield a divergent integral, we nevertheless need to choose the bias
so that the integral is free of divergences.
The range of $\nu$ for which the integral is convergent can be determined by
inspecting the asymptotic behaviour of the kernels. For each $\ell$ we have
\begin{subequations}
\bea
I_\ell(\q,\k-\q)&\to \frac{k^2}{q^2}, \qquad q\to0, \\[3pt]
I_\ell(\q,\k-\q)&\to \frac{k^4}{q^4}, \qquad q\to\infty,
\eea
\end{subequations}
where we have ignored $\mathcal{O}(1)$ multiplicative factors.
In the asymptotic limit the integrands of type $P_{22}$ read
\begin{subequations}
\begin{alignat}{2}
\text{(IR)}\qquad &q^2 I_\ell(\q,\k-\q)\PL(q)\PL(|\k-\q|)\to q^2 q^{-2} q^\nu = q^{\nu}, \qquad && q\to0, \\[4pt]
\text{(UV)}\qquad &q^2 I_\ell(\q,\k-\q)\PL(q)\PL(|\k-\q|)\to q^2 q^{-4} q^{2\nu} = q^{-2+2\nu}, \qquad && q\to\infty,
\end{alignat}
\end{subequations}
where the factor of $q^2$ comes from using volume element in spherical coordinates.
From this we see that to avoid divergences for any $\ell$ we require a bias in the range of $-1<\nu<1/2$.
In practice we find $\nu=-0.6$ gives good results.

\subsection{Integrals of type 13}
Terms of the type $P_{13}$ can also be evaluated using the FFTLog approach. For example
\bea
2P_{\theta\theta,13}(k)
&=6\PL(k)\fint{\q}\,\Gs^{(3)}(\k,\q,-\q)\PL(q) \nonumber\\[4pt]
&=6\PL(k)\sum_{m_1} c_{m_1}\sum_{n_1,n_2} f_{n_1n_2}^\ell k^{-2(n_1+n_2)} \!
    \fint{\q}\,\frac{1}{q^{2(\nu_{m_1}-n_1)}|\k-\q|^{-2n_2}} \nonumber\\[4pt]
&=6\PL(k)k^3\sum_{m_1} c_{m_1}k^{-2\nu_{m_1}}\bigg[\sum_{n_1,n_2} f_{n_1n_2}^\ell\,\mathsf{I}(\nu_{m_1}-n_1,-n_2)\bigg]
    \label{eq:13-fftlog}
\eea
These integrals are in fact more straightforward to evaluate than the convolutions
above, since the angular dependence only enters through the kernel and thus
allowing us integrate out $\mu$ analytically \cite{Suto:1990wf,Makino:1991rp}:
\be\label{eq:Ptt13}
2P_{\theta\theta,13}(k)
=\frac{1}{84}\frac{k^3}{4\pi^2}\PL(k)\int^\infty_0\dif r\:\PL(kr)
\left[\frac{12}{r^2} - 82 + 4r^2 - 6r^4 
	+ \frac{3}{r^3}(r^2-1)^3 (r^2+2) \ln\Big|\frac{r+1}{r-1}\Big|\right],
\ee
where $r=q/k$.

\subsubsection*{Asymptotic limits in the IR and UV}
In the large-$k$ asymptotic limit (or IR limit) we find
\be
2P_{\theta\theta,13}(k)
=-\frac13 k^2 \PL(k)\int \frac{q^2\dif q}{2\pi^2}\, \frac{\PL(q)}{q^2}
	\Big(1+\frac{4 q^2}{35 k^2}+\frac{12 q^4}{49 k^4}+\cdots\Big)
\ee
where $\sigma_u^2$ is given by eq.~\eqref{eq:sigu}.
In the $k\to0$ asymptotic limit (UV limit) we find
\be
2P_{\theta\theta,13}(k)
=-\frac13 k^2 \PL(k)\int \frac{q^2\dif q}{2\pi^2}\, \frac{\PL(q)}{q^2}
	\Big(\,\frac95-\frac{156 k^2}{245q^2}+\frac{76 k^4}{735q^4}+\cdots\Big)
\ee
It is easy now to read off the limits in the IR and UV:
\be
P^\mrm{IR}_{\theta\theta,13}(k) = -\frac12 k^2 \PL(k) \sigma_u^2, \qquad
P^\mrm{UV}_{\theta\theta,13}(k) = -\frac{9}{10} k^2 \PL(k) \sigma_u^2.
\ee
Either $P^\mrm{IR}_{\theta\theta,13}(k)$ or $P^\mrm{UV}_{\theta\theta,13}(k)$ needs to be
added on to eq.~\eqref{eq:13-fftlog}, depending on the choice of bias parameter. For example
choosing $\nu=-0.6$ we add $P^\mrm{UV}_{\theta\theta,13}(k)$.

For $\Ks^{(2)}(\q,\k-\q)\Gs^{(2)}(\k,-\q)$ we have
\begin{alignat}{2}
&P^{0,\mrm{IR}}_{\theta\theta,13}(k)  = -\frac{1}{12}k^2 \PL(k)\sigma_u^2,\qquad
&&P^{0,\mrm{UV}}_{\theta\theta,13}(k) = -\frac{19}{84}k^2 \PL(k)\sigma_u^2, \\[4pt]
&P^{2,\mrm{IR}}_{\theta\theta,13}(k)  = -\frac{1}{6}k^2 \PL(k)\sigma_u^2, \qquad
&&P^{2,\mrm{UV}}_{\theta\theta,13}(k) = -\frac{23}{210}k^2 \PL(k)\sigma_u^2.
\end{alignat}

\begin{figure}[t]
\centering
\includegraphics[width=0.95\textwidth]{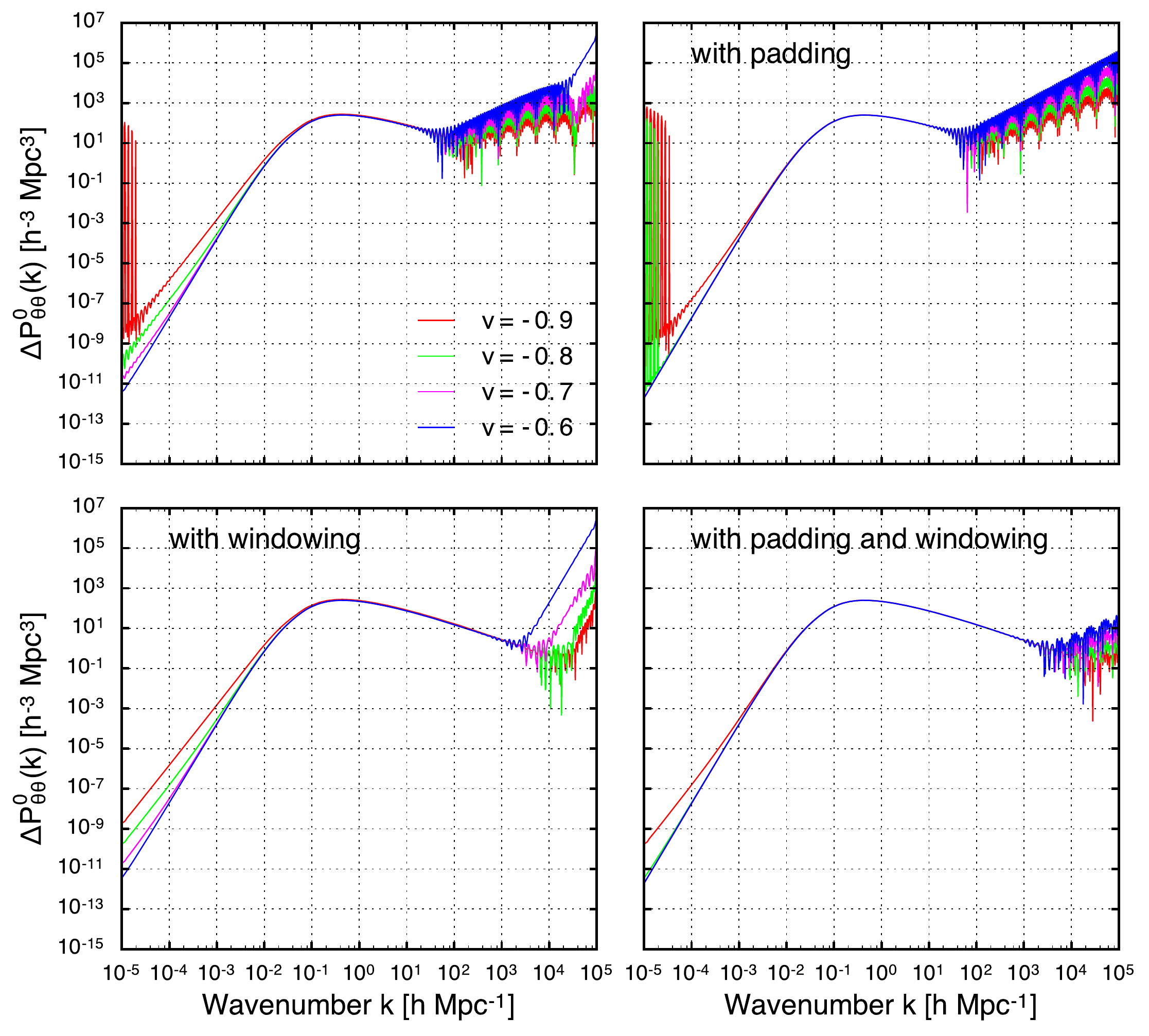}
\caption{Edge effects of power spectra when represented using the FFTLog
        approach (upper left panel). The signal shown is for
        $\Delta P^0_{\theta\theta}\equiv P^0_{\theta\theta}-\PL$, where
        $P^0_{\theta\theta}$ is given by eq.~\eqref{eq:Ptt-ell} and $\PL$ is the
        linear power spectrum.
        The rest of the panels show how edge effects can be mitigated with the
        use of padding and windowing (where we use the window function given by
        equation (C1) in ref.~\cite{McEwen:2016fjn}).}
  \label{fig:windowing}
\end{figure}

\begin{table}[t]
\begin{tabular}{ |r|r||r|r|r|r|r|r|r|r|r||r|r|r|r| }
\hline
\multirow{2}{*}{$n_1$} & \multirow{2}{*}{$n_2$} & \multicolumn{3}{c|}{$[2\Ks^{(2)}\Ks^{(2)}]_{22}$} & \multicolumn{2}{c|}{$[4\Ks^{(2)}\Gs^{(2)}]_{22}$} & \multicolumn{2}{c|}{$[\Ks^{(2)}\Gs^{(2)}]_{13}$} & \multicolumn{1}{c|}{$[2\Gs^{(2)}\Gs^{(2)}]_{22}$}
& \multicolumn{1}{c||}{$[\Gs^{(3)}]_{13}$} & \multicolumn{4}{c|}{$k^2\times[\KsB]_{22}$} \\
\cline{3-15}
{} & {} & \multicolumn{1}{c|}{$\ell=0$} & \multicolumn{1}{c|}{$\ell=2$} & \multicolumn{1}{c|}{$\ell=4$} 
& \multicolumn{1}{c|}{$\ell=0$} & \multicolumn{1}{c|}{$\ell=2$} & \multicolumn{1}{c|}{$\ell=0$} & \multicolumn{1}{c|}{$\ell=2$} & \multicolumn{1}{c|}{$\ell=0$} & \multicolumn{1}{c||}{$\ell=0$} & \multicolumn{1}{c|}{$\ell=0$} & \multicolumn{1}{c|}{$\ell=2$} & \multicolumn{1}{c|}{$\ell=4$} & \multicolumn{1}{c|}{$\ell=6$}\\
\hline
$-2$  &  $-2$  &  $ 1/60$  &  $   1/84$  &  $ 3/1120$  &  $   1/21$  &  $   1/42$  &  $      0$  &  $      0$  &  $   2/49$  &  $      0$ &  $ 1/140$ & $ 1/168$ & $ 27/12320$ & $5/14784$ \\
$-1$  &  $-2$  &  $-1/30$  &  $  5/168$  &  $  1/280$  &  $  -5/84$  &  $   1/24$  &  $      0$  &  $      0$  &  $  -1/49$  &  $      0$ &  $-3/140$ & $  1/56$ & $ 39/12320$ & $1/2464$ \\
$-2$  &  $-1$  &  $-1/30$  &  $  5/168$  &  $  1/280$  &  $  -5/84$  &  $   1/24$  &  $ -1/112$  &  $ -1/224$  &  $  -1/49$  &  $  -1/84$ &  $-3/140$ & $  1/56$ & $ 39/12320$ & $1/2464$ \\
$ 0$  &  $-2$  &  $ 1/60$  &  $  -2/21$  &  $  9/560$  &  $  -1/42$  &  $-17/168$  &  $      0$  &  $      0$  &  $-23/392$  &  $      0$ &  $ 3/140$ & $ -5/56$ & $ 123/6160$ & $5/4928$ \\
$-1$  &  $-1$  &  $ 1/15$  &  $   1/21$  &  $  3/280$  &  $   2/21$  &  $  13/84$  &  $  3/112$  &  $      0$  &  $ 25/196$  &  $  5/168$ &  $  3/35$ & $     0$ & $    3/440$ & $3/2464$ \\
$-2$  &  $ 0$  &  $ 1/60$  &  $  -2/21$  &  $  9/560$  &  $  -1/42$  &  $-17/168$  &  $  1/168$  &  $  -1/96$  &  $-23/392$  &  $ -1/168$ &  $ 3/140$ & $ -5/56$ & $ 123/6160$ & $5/4928$ \\
$ 1$  &  $-2$  &  $    0$  &  $   3/56$  &  $  -3/56$  &  $   1/28$  &  $  -1/56$  &  $      0$  &  $      0$  &  $  3/196$  &  $      0$ &  $-1/140$ & $17/168$ & $-621/6160$ & $25/3696$ \\
$-2$  &  $ 1$  &  $    0$  &  $   3/56$  &  $  -3/56$  &  $   1/28$  &  $  -1/56$  &  $  5/336$  &  $ 11/672$  &  $  3/196$  &  $   1/21$ &  $-1/140$ & $17/168$ & $-621/6160$ & $25/3696$ \\
$ 0$  &  $-1$  &  $    0$  &  $  -3/56$  &  $   3/56$  &  $  -1/28$  &  $   1/56$  &  $ -3/112$  &  $  3/112$  &  $ -3/196$  &  $  -1/56$ &  $-9/140$ & $  1/56$ & $ 261/6160$ & $5/1232$  \\
$-1$  &  $ 0$  &  $    0$  &  $  -3/56$  &  $   3/56$  &  $  -1/28$  &  $   1/56$  &  $  -1/56$  &  $ -3/224$  &  $ -3/196$  &  $-37/504$ &  $-9/140$ & $  1/56$ & $ 261/6160$ & $5/1232$  \\
$ 2$  &  $-2$  &  $    0$  &  $      0$  &  $   1/32$  &  $      0$  &  $   3/56$  &  $      0$  &  $      0$  &  $  9/392$  &  $      0$ &  $     0$ & $ -1/28$ & $ 291/2464$ & $-25/704$ \\
$-2$  &  $ 2$  &  $    0$  &  $      0$  &  $   1/32$  &  $      0$  &  $   3/56$  &  $  -1/84$  &  $ 11/672$  &  $  9/392$  &  $ -5/168$ &  $     0$ & $ -1/28$ & $ 291/2464$ & $-25/704$ \\
$ 1$  &  $-1$  &  $    0$  &  $      0$  &  $   -1/8$  &  $      0$  &  $  -3/14$  &  $  1/112$  &  $  -1/28$  &  $  -9/98$  &  $ -1/168$ &  $     0$ & $ -1/28$ & $ -111/616$ & $5/176$ \\
$-1$  &  $ 1$  &  $    0$  &  $      0$  &  $   -1/8$  &  $      0$  &  $  -3/14$  &  $ -1/112$  &  $ -3/112$  &  $  -9/98$  &  $ 25/504$ &  $     0$ & $ -1/28$ & $ -111/616$ & $5/176$ \\
$ 0$  &  $ 0$  &  $    0$  &  $      0$  &  $   3/16$  &  $      0$  &  $   9/28$  &  $   1/84$  &  $ 31/672$  &  $ 27/196$  &  $  -1/72$ &  $     0$ & $   1/7$ & $ 153/1232$ & $5/352$ \\
$-2$  &  $ 3$  &  $    0$  &  $      0$  &  $      0$  &  $      0$  &  $      0$  &  $      0$  &  $  -1/56$  &  $      0$  &  $      0$ &  $     0$ & $     0$ & $ -15/352$ & $15/352$ \\
$ 3$  &  $-2$  &  $    0$  &  $      0$  &  $      0$  &  $      0$  &  $      0$  &  $      0$  &  $      0$  &  $      0$  &  $      0$ &  $     0$ & $     0$ & $ -15/352$ & $15/352$ \\
$-1$  &  $ 2$  &  $    0$  &  $      0$  &  $      0$  &  $      0$  &  $      0$  &  $      0$  &  $  9/224$  &  $      0$  &  $ -1/168$ &  $     0$ & $     0$ & $  45/352$ & $-45/352$ \\
$ 2$  &  $-1$  &  $    0$  &  $      0$  &  $      0$  &  $      0$  &  $      0$  &  $      0$  &  $  3/224$  &  $      0$  &  $  1/168$ &  $     0$ & $     0$ & $  45/352$ & $-45/352$ \\
$ 0$  &  $ 1$  &  $    0$  &  $      0$  &  $      0$  &  $      0$  &  $      0$  &  $      0$  &  $ -3/224$  &  $      0$  &  $   1/56$ &  $     0$ & $     0$ & $ -15/176$ & $15/176$ \\
$ 1$  &  $ 0$  &  $    0$  &  $      0$  &  $      0$  &  $      0$  &  $      0$  &  $      0$  &  $ -5/224$  &  $      0$  &  $  -1/56$ &  $     0$ & $     0$ & $ -15/176$ & $15/176$ \\
$ 1$  &  $ 1$  &  $    0$  &  $      0$  &  $      0$  &  $      0$  &  $      0$  &  $      0$  &  $      0$  &  $      0$  &  $      0$ &  $     0$ & $     0$ & $       0$ & $  5/16$ \\
$ 0$  &  $ 2$  &  $    0$  &  $      0$  &  $      0$  &  $      0$  &  $      0$  &  $      0$  &  $      0$  &  $      0$  &  $      0$ &  $     0$ & $     0$ & $       0$ & $-15/64$ \\
$ 2$  &  $ 0$  &  $    0$  &  $      0$  &  $      0$  &  $      0$  &  $      0$  &  $      0$  &  $      0$  &  $      0$  &  $      0$ &  $     0$ & $     0$ & $       0$ & $-15/64$ \\
$-1$  &  $ 3$  &  $    0$  &  $      0$  &  $      0$  &  $      0$  &  $      0$  &  $      0$  &  $      0$  &  $      0$  &  $      0$ &  $     0$ & $     0$ & $       0$ & $  3/32$ \\
$ 3$  &  $-1$  &  $    0$  &  $      0$  &  $      0$  &  $      0$  &  $      0$  &  $      0$  &  $      0$  &  $      0$  &  $      0$ &  $     0$ & $     0$ & $       0$ & $  3/32$ \\
$-2$  &  $ 4$  &  $    0$  &  $      0$  &  $      0$  &  $      0$  &  $      0$  &  $      0$  &  $      0$  &  $      0$  &  $      0$ &  $     0$ & $     0$ & $       0$ & $ -1/64$ \\
$ 4$  &  $-2$  &  $    0$  &  $      0$  &  $      0$  &  $      0$  &  $      0$  &  $      0$  &  $      0$  &  $      0$  &  $      0$ &  $     0$ & $     0$ & $       0$ & $ -1/64$ \\
\hline
\end{tabular}
\caption{\label{fft-ind}Summary of FFTLog expansion indices $(n_1,n_2)$ and coefficients
            $f^\ell_{n_1n_2}$ characterizing each kernel (top row). 
            Note that the kernels of type $P_{13}$ generally contain terms with
            $|\k+\q|$ but can be replaced (as we have done here) with
            $|\k-\q|$, since the integration is over all $\q$. Also note that for kernels of type $P_{22}$
            the coefficients are symmetric in $n_1$ and $n_2$, i.e.\ $f^\ell_{n_1n_2}=f^\ell_{n_2n_1}$.
            The last four columns give the coefficients for the LOS velocity power spectrum model
            described in Section \ref{sec:cum-exp-model}.
            (Here $\KsB$ has been multiplied by $k^2$ to obtain a dimensionless kernel.)}
\end{table}

\subsection{Configuration-space multipoles}
We have noted in Section \ref{sec:config-space-redshift} that eq.~\eqref{eq:xiell-vv} is a
spherical Bessel (or Hankel) transform. Given the highly oscillatory nature of the spherical
Bessel functions, evaluating such transforms efficiently and precisely requires special
integration methods.
There is however an efficient way to evaluate $\xiv^\ell(s)$ using a slightly
different FFTLog implementation \cite{Talman:1978,Hamilton:1999uv}. Here the FFTLog algorithm is based on
the idea that the one-dimensional radial integral can be recast as a convolution under the transformation
$k\to\ln k$ and $s\to\ln s$. In the (one-dimensional) Fourier dual space this is the
product of the Fourier transforms of the power spectrum and the spherical Bessel function.
The calculation of $\xiv^\ell$ is then
equivalent to taking the inverse Fourier transform of this product. In practice, in the
discrete case, one takes fast Fourier transforms, with the Fourier transform of the
spherical Bessel function computed exactly using the known analytic form.
In the case of the matter power spectrum only about 200 logarithmically-spaced points
in the range of $k_\mrm{min}=10^{-4}\,h\Mpc^{-1}$ and 
$k_\mrm{max}=10^{2}\,h\Mpc^{-1}$ of $\PL(k)$ are required to capture all features.%
\footnote{We use the {\tt mcfit} implementation available at
\url{https://github.com/eelregit/mcfit}.}
In fact given the convolution contained in eq.~\eqref{eq:xiell-vv}, $\xiv^\ell$ can be
further developed into another one-dimensional spherical Bessel transform~\cite{Schmittfull:2016jsw}
involving the generalized correlation function
\be
\zeta^\ell_n(r)\equiv\int\frac{k^2\dif k}{2\pi^2}\,k^{n} j_\ell(kr) \PL(k),
\ee
which forms a transform pair with $k^n \PL(k)$. Note $\zeta^0_0(r)=\xi(r)$,
the autocorrelation function of $\delta$. Because the RSD
enters at nonlinear order this means that $\xiv^\ell(s)$ can
be expressed as integrals over the products of different $\zeta^\ell_n$.


\section{Consistency check of derivative expansion}\label{app:v-details}
In the main text we gave an expression \eqref{eq:v-s-series} for the redshift-space velocity field 
$v^s_\|$, which we obtained by formally expanding in the real-space fields. Since the apparent dependence on $\delta$
vanishes, this expression depends on terms involving only $v_\|$. In this appendix we will verify this
result by checking that when we multiply $v^s_\|$ by the overdensity $1+\delta^s$ we indeed get the
momentum  $\pi^s_\|$. Precisely, we will show that the product of the series expansions,
eqs.~\eqref{eq:1pdel-s-sum} and \eqref{eq:v-s-series}, is equal to that of eq.~\eqref{eq:pi-s-sum} at
all orders. We specialize to the LOS component but note that the expressions given here do not assume
the plane-parallel limit.

First, it will be convenient to reorganize each expansion so that the summands gather all terms
consisting of a given number $n$ of the real-space fields (loosely the ``perturbation'' order,
though note that there is no requirement for the fields to be small fluctuations).
Under this rearrangement the product to be shown reads [$\rho^s\equiv1+\delta^s$, $u^s\equiv v^s_\|/(-\calH)$, etc]
\be\label{eq:cauchy}
\sum_{n=0}^\infty\rho^s_{(n)}\sum_{m=0}^\infty u^s_{(m)}=\sum_{m=0}^\infty\pi^s_{(N)},
\qquad\text{where}\qquad
\pi^s_{(N)}=\sum_{n+m=N}\rho^s_{(n)}u^s_{(m)},
\ee
where the constrained sum is over all non-negative integers $n$ and $m$ with $n+m=N$;
the subscripts denote the field order, so
$\rho^s_{(0)}=1$, $u^s_{(0)}=0$, $\pi^s_{(0)}=0$, $\pi^s_{(1)}=u$, and
\bea
\rho^s_{(n)}&\equiv\frac{1}{(n-1)!}\,\partial^{n-1}(u^{n-1}\delta)+\frac{1}{n!}\,\partial^n u^n, && n\geq1, \label{eq:rho-s-n} \\[2pt]
u^s_{(m)}&\equiv\frac{1}{m!}\,\partial^{m-1}u^m, && m\geq1, \label{eq:u-s-n} \\[2pt]
\pi^s_{(N)}&\equiv\frac{1}{(N-2)!}\,\partial^{N-2}(u^{N-1}\delta)+\frac{1}{(N-1)!}\,\partial^{N-1}u^N, && N\geq2, \label{eq:pi-s-n}
\eea
where $\partial$ is a shorthand for $\nabla_\|\equiv\n\cdot\nabla=\partial/\partial r$.
Since eq.~\eqref{eq:cauchy} defines a \emph{Cauchy product}~\cite{Riordan},
showing that eq.~\eqref{eq:pi-s-sum} for $\pi^s$ is the product of 
eqs.~\eqref{eq:1pdel-s-sum} and \eqref{eq:v-s-series}, is equivalent to showing that,
order-by-order, $\pi^S_{(N)}$ is of the form given in eq.~\eqref{eq:cauchy}.
The first two cases, $N=0,1$, are easily checked. For $N\geq2$, substituting eqs.~\eqref{eq:rho-s-n}, \eqref{eq:u-s-n},
and \eqref{eq:pi-s-n}, into eq.~\eqref{eq:cauchy}, we have
\bea
\frac{1}{(N-2)!}\,&\partial^{N-2}(u^{N-1}\delta)+\frac{1}{(N-1)!}\,\partial^{N-1}u^N \nonumber\\[3pt]
&=\rho^s_{(0)}u^s_{(N)}+\rho^s_{(N)}u^s_{(0)}
+\sum_{n=1}^\infty\sum_{m=1}^\infty\delta^\mrm{K}_{n+m,N}
\bigg[\Big(\,\frac{1}{(n-1)!}\,\partial^{n-1}(u^{n-1}\delta)+\frac{1}{n!}\,\partial^n u^n\Big)
    \frac{1}{m!}\,\partial^{m-1}u^m\bigg] \nonumber\\[3pt]
&=\frac{1}{N!}\,\partial^{N-1}u^N
+\sum_{n+m=N-2}
\bigg[\,\frac{1}{n!\,(m+1)!}\,\partial^{n}(u^{n}\delta)\,\partial^{m}u^{m+1}
    +\frac{1}{(n+1)!\,(m+1)!}\,\partial^{n+1}u^{n+1}\,\partial^{m}u^{m+1}\bigg],     \label{eq:pi-s-N-all}
\eea
where in the second equality we have redefined the labels such that $n\to n+1$ and $m\to m+1$.
Notice that there are terms that depend on $\delta$, and terms that do not depend on $\delta$.
Identifying on the left- and right-hand sides the $\delta$-dependent terms, taking $N\to N+2$,
and rearranging slightly, we then have
\be\label{eq:del-dpt}
\partial^{N}(u^{N+1}\delta)
=\sum_{n+m=N}\frac{N!}{n!\,(m+1)!}\,\partial^{n}(u^{n}\delta)\,\partial^{m}u^{m+1}.
\ee
For the $\delta$-independent terms, by similar steps, we have
\be\label{eq:del-indpt}
\frac{1}{N+2}\,\partial^{N+1}u^{N+2}
=\sum_{n+m=N}\frac{N!}{(n+1)!\,(m+1)!}\,\partial^{n+1}u^{n+1}\,\partial^{m}u^{m+1}.
\ee
Note that these two relations are general; they hold for arbitrary functions. This implies
that eqs.~\eqref{eq:del-indpt} and \eqref{eq:del-dpt} are not independent relations but
that the latter follows from the former.
To see this take $\delta=u$, symmetrize over $n$ and $m$ the summand in eq.~\eqref{eq:del-dpt},
then differentiate; the end result is equivalent to eq.~\eqref{eq:del-indpt}, upon symmetrizing,
then writing
$\partial^{n+1}u^{n+1}\,\partial^{m}u^{m+1}+\partial^{n}u^{n+1}\,\partial^{m+1}u^{m+1}
    =\partial(\partial^{n}u^{n+1}\,\partial^{m}u^{m+1})$.
Thus we need only show that eq.~\eqref{eq:del-dpt} holds. To show that the right-hand
side of eq.~\eqref{eq:del-dpt} simplifies to the left-hand side, we find it convenient to use
the Fourier representation, in which differentiation becomes algebra:
\bea
\partial^n (u^{n}\delta)\partial^{m}u^{m+1}
&=\int\dif r\,\rme^{-\ii kr}\!\! \int_{q_1,\ldots,q_{N+2}}\!\!\!\!\!\!\!
	\delD(k-k_1-k_2)\, (\ii k_1)^{n}\,(\ii k_2)^{m} \,
    \tilde{u}(q_1)\cdots\tilde\delta(q_{n+1})\,\tilde{u}(q_{n+2})\cdots\tilde{u}(q_{N+2}),
\eea
where $k_1$ and $k_2$ are shorthands for $q_1+q_2+\cdots+q_{n+1}$ and
$q_{n+2}+q_{n+3}+\cdots+q_{N+2}$, respectively. Inserting back into eq.~\eqref{eq:del-dpt} and
symmetrizing, using the multinomial identity~\cite{Riordan}
\be
\big(x_1+\cdots+x_{N+2}\big)^{N}
=\sum_{n+m=N}
\frac{N!}{n!\,(m+1)!}\:\mrm{Sym}
	\Big[\big(x_1+\cdots+x_{n+1}\big)^n\big(x_{n+2}+\cdots+x_{N+2}\big)^{m}\Big],
\ee
where $\mrm{Sym}$ is an instruction to symmetrize over $\{x_1,x_2,\ldots,x_{N+2}\}$, we obtain
the left-hand side of eq.~\eqref{eq:del-dpt}.


\section{Explicit expressions for redshift-space kernels}\label{app:Ikernels}
In this appendix we present closed-form expressions for the multipole moments of the
power spectrum kernels $2[Z_2(\q,\k-\q)]^2$ and $3Z_3(\k,\q,-\q)$ appearing in
eq.~\eqref{eq:P22-P13-s}. In the parametrization of eq.~\eqref{eq:Ptt-ell} the multipole
moments read
\begin{subequations}
\bea
I^\ell_{22}(r,\mu')
&\equiv (2\ell+1)\int^1_{-1}\!\frac{\dif\mu}{2}\,\calL_\ell(\mu)
    \!\int^{2\pi}_0\!\frac{\dif\phi}{2\pi}\:2\big[Z_2(r,\mu,\mu',\phi)\big]^2, \\[4pt]
I^\ell_{13}(r,\mu')
&\equiv (2\ell+1)\int^1_{-1}\!\frac{\dif\mu}{2}\,\calL_\ell(\mu)
    \!\int^{2\pi}_0\!\frac{\dif\phi}{2\pi}\:3Z_3(r,\mu,\mu',\phi),
\eea
\end{subequations}
where $r=q/k$, $\mu=\hat\k\cdot\n$, and $\mu'=\hat\k\cdot\hat\q$.
The only non-zero kernels are for $\ell=0,2,4$, giving a total of six kernels.
The first three are found to be
\bea
I^0_{22}(r,\mu')
&=f^2\,\frac{3r^2 - 6r \mu'+ 2\mu'^2 + 1}{30r^2(1+r^2-2r\mu')^2} \nonumber\\[7pt]
&\quad
  + f\,\frac{6r^2\mu'^2 + r^2 - 6r\mu'^3 - 8r \mu'+ 7\mu'^2}{21r^2(1+r^2-2r\mu')^2} \nonumber\\[7pt]
&\quad
  + \frac{36r^2\mu'^4 + 12r^2\mu'^2 + r^2 - 84r\mu'^3 - 14r\mu' + 49\mu'^2}{98r^2(1+r^2-2r\mu')^2}\,, \\[12pt]
I^2_{22}(r,\mu')
&=f^2\,\frac{18 r^2 \mu'^2-6 r^2-18 r \mu'^3-6 r \mu'+11 \mu'^2+1}{42r^2(1+r^2-2r\mu')^2} \nonumber\\[7pt]
&\quad
    +f\,\frac{18r^2\mu'^4 - 3r^2\mu'^2 - r^2 - 33r\mu'^3 + 5r\mu' + 14\mu'^2}{21r^2(1+r^2-2r\mu')^2}\,, \\[12pt]
I^4_{22}(r,\mu')
&=f^2\,\frac{35 r^2 \mu'^4-30 r^2 \mu'^2+3 r^2-40 r \mu'^3+24 r \mu'+12 \mu'^2-4}{70r^2(1+r^2-2r\mu')^2}\,.
\eea
Note terms with the prefactor $f$ arise from the part in $Z_2$ linear in $\Ks^{(2)}$, while those with a
$f^2$ prefactor arise from the part quadratic in $\Ks^{(2)}$, namely $2[\Ks^{(2)}(\q,\k-\q)]^2$;  the monopole
term ($\ell=0$) without a prefactor is a mixture of all kernels comprising eq.~\eqref{eq:Z2}. The real-space
kernel $\Gs^{(2)}(\q,\k-\q)$ of course only contributes to the monopole.

We also have
\bea
I^0_{13}(r,\mu')
&=-f^2\,\frac{2r^2\mu'^2 + r^2 - 4r\mu'^3 - 2r\mu' + 2\mu'^2 + 1}{30r^2(1+r^2-2r\mu')} \nonumber\\[7pt]
&\quad
    - f\,\frac{7r^3\mu' - 15r^2\mu'^2 - 6r^2 + 8r\mu'^3 + 13r\mu' - 7\mu'^2}{21r^2(1+r^2-2r\mu')} \nonumber\\[7pt]
&\quad
    +\frac{6r^3\mu'^3 + r^3\mu' - 29r^2\mu'^2 - 6r^2 + 30r\mu'^3 + 19r\mu' - 21\mu'^2}{42r^2(1+r^2-2r\mu')}\,, \label{eq:I013}\\[12pt]
I^2_{13}(r,\mu')
&=-f^2\,\frac{11r^2\mu'^2 + r^2 - 22r\mu'^3 - 2r\mu' + 11\mu'^2 + 1}{42r^2(1+r^2-2r\mu')} \nonumber\\[7pt]
&\quad
    -f\,\frac{21r^3\mu'^3 - 7r^3\mu' - 24r^2\mu'^4 - 24r^2\mu'^2 + 6r^2 + 37r\mu'^3 + 5r\mu' - 14\mu'^2}{21r^2(1+r^2-2r\mu')}\,, \\[12pt]
I^4_{13}(r,\mu')
&=-f^2\,\frac{2(3 \mu'^2-1)}{35 r^2}\,.
\eea
(Note that in the last term in eq.~\eqref{eq:I013} we have taken $\q\to-\q$; these terms correspond to the second group of terms in eq.~\eqref{eq:G3},
which are integrated over the space of all $\q$.)

Finally, let us also note here the angular integrals in the ``13'' loop integrals yield
\bea
\int\!\frac{\dif\mu'}{2}\,I^0_{13}(r,\mu')
&=-\frac{f^2}{18 r^2}
    - \frac{f}{504r^2}\bigg[\frac{18}{r^2} - 104 - 18r^2
    + \frac{9}{r^3}(r^2-1)^3 \ln\Big|\frac{r+1}{r-1}\Big|\bigg] \nonumber\\[5pt]
&\quad
    +\frac{1}{336r^2}\bigg[\frac{12}{r^{2}}-82+4 r^{2}-6 r^{4}
        +\frac{3}{r^{3}}(r^{2}-1)^{3}(r^{2}+2) \ln\Big|\frac{r+1}{r-1}\Big|\bigg]\,,\\[5pt]
\int\!\frac{\dif\mu'}{2}\,I^2_{13}(r,\mu')
&=-\frac{f^2}{9 r^2} 
    - \frac{f}{1008r^2}\bigg[\frac{18}{r^2} - 218 + 126r^2 - 54r^4
    + \frac{9}{r^3}(r^2-1)^3 (3 r^2+1) \ln\Big|\frac{r+1}{r-1}\Big|\bigg]\,, \\[5pt]
\int\!\frac{\dif\mu'}{2}\,I^4_{13}(r,\mu')
&=0\,.
\eea

\section{Redshift-space correlation tensor\label{app:tensor}}
In this appendix we give details on the velocity correlation tensor
$\Psi^s_{ij}(\s)\equiv\langle v^s_i(0)\, v^s_j(\s)\rangle$, the redshift space
version of eq.~\eqref{eq:Psi-ij}.

As in real space, the redshift-space velocity field is sourced by scalar
perturbations, and reads in terms of the velocity divergence power spectrum,
\be
\Psi^s_{ij}(\s)
= \int\!\frac{\dif^3\k}{(2\pi)^3}\,
    \Big(\frac{\calH f}{k}\Big)^2
    \frac{k_i}{k}\frac{k_j}{k}
    P_{\theta \theta}^{s}(\k)\, \rme^{-\ii\k\cdot\s}.
\ee
Inserting the multipole expansion
$P^s_{\theta\theta}(\k)=\sum_\ell P^\ell_{\theta\theta}(k)\calL_\ell(\hat\k\cdot\n)$, and
using the identity
\be\label{eq:int-bess-id}
\int\frac{\dif^2\hat\k}{4\pi}\,\calL_\ell(\hat\k\cdot\n)\, \rme^{-\ii\k\cdot\s}
=(-\ii)^\ell j_\ell(ks) \calL_\ell(\hat\s\cdot\n),
\ee
we have
\bea
\Psi^s_{ij}(\s)
&= \int \frac{k^2\dif k}{2\pi^2} \sum_\ell \Big(\frac{\calH f}{k}\Big)^2 P_{\theta\theta}^\ell(k)
    \frac{1}{(-\ii)^2k^2}\frac{\partial}{\partial s^i} \frac{\partial}{\partial s^j}
    \int\frac{\dif^2\hat\k}{4\pi}\,
        \calL_{\ell}(\hat\k\cdot\n)\, \rme^{-\ii\k\cdot\s} \nonumber\\[3pt]
&=\sum_\ell(-\ii)^{\ell-2} \int\!\frac{k^2\dif k}{2\pi^{2}}\,
    \Big(\frac{\calH f}{k}\Big)^2 {P_{\theta\theta}^\ell(k)}
    \mathcal{T}^\ell_{ij} \label{eq:Psi-s},
\eea
where we have replaced $k_i$ with $\partial/\partial s^i$ (acting on the plane waves),
and in the second line we have the Hessian,
\be\label{eq:Tij}
\mathcal{T}^\ell_{ij} \equiv \partial_i\partial_j\Big[j_\ell(ks)\calL_\ell(\mu)\Big],
\ee
with $\mu\equiv\hat\s\cdot\n$ and $\partial_i\equiv k^{-1}\partial/\partial s^i$.
To compute the Hessian note that
$\partial_i (ks) = \hat{s}_i$,
$\partial_i\hat{s}_j = (\delta^\mrm{K}_{ij}-\hat{s}_i\hat{s}_j)/(ks)$, and
$\partial_i\mu = (\hat{n}_i-\mu\hat{s}_i)/(ks)$.
If we then write $\mathcal{T}^\ell_{ij}$ as a decomposition with respect to $\s$ as
\be\label{eq:Tij-2}
\mathcal{T}^\ell_{ij}
= \mathcal{T}^{\ell,\|}_{ij} 
    + 2\mathcal{T}^{\ell,\times}_{ij} + \mathcal{T}^{\ell,\perp}_{ij},
\ee
we have for each term,
\begin{subequations}
\bea
\mathcal{T}^{\ell,\|}_{ij}
&\equiv\mathcal{P}^\|_{ik}\mathcal{P}^\|_{jl} \mathcal{T}^\ell_{kl}
=\mathcal{P}^{\|}_{ij}\,j_\ell''(y)\,\calL_\ell(\mu), \label{eq:Tpara}\\[9pt]
\mathcal{T}^{\ell,\times}_{ij}
&\equiv\mathcal{P}^\perp_{k(i}\,\mathcal{P}^\|_{j)l}\mathcal{T}^\ell_{kl}
=\hat{s}_{(i}\hat{n}^\perp_{j)}
    \Big[\frac{j_\ell'(y)}{y}-\frac{j_\ell(y)}{y^2}\Big]\frac{\dif\calL_\ell}{\dif\mu}, \\[4pt]
\mathcal{T}^{\ell,\perp}_{ij}
&\equiv\mathcal{P}^\perp_{ik}\mathcal{P}^\perp_{jl}\mathcal{T}^\ell_{kl}
=\mathcal{P}^\perp_{ij}\,\frac{j_\ell'(y)}{y} \calL_\ell(\mu)
    - \mathcal{P}^\perp_{ij}\, \frac{j_\ell(y)}{y^2}\frac{\dif\calL_\ell}{\dif\mu}\mu
    + \hat{n}^\perp_i\hat{n}^\perp_j\,\frac{j_\ell(y)}{y^2}\frac{\dif^2\calL_\ell}{\dif\mu^2} \label{eq:Tperp},
\eea
\end{subequations}
where a prime denotes differentiation with respect to the function argument $y$,
and indices enclosed in parentheses denotes the symmetric part, e.g.\ 
$\hat{s}_{(i}\hat{n}^\perp_{j)}=(\hat{s}_i\hat{n}^\perp_j+\hat{s}_j\hat{n}^\perp_i)/2$;
also we have defined $y\equiv ks$, $\hat{n}^\perp_i\equiv\mathcal{P}^\perp_{ij}\,\hat{n}^j$, and
projection tensors, $\mathcal{P}^\|_{ij}(\hat\s)\equiv\hat{s}_i\hat{s}_j$
and $\mathcal{P}^\perp_{ij}(\hat\s)\equiv\delta^\mrm{K}_{ij}-\hat{s}_i\hat{s}_j$.
Note that the first term in eqs.~\eqref{eq:Tpara} and \eqref{eq:Tperp}
together generalizes the usual isotropic correlation functions.
In particular, if we neglect RSD then we need only consider $\ell=0$ and we may
write $\mathcal{T}^\ell_{ij}=\mathcal{T}^0_{ij}\delta^\mrm{K}_{\ell0}$,
\be
\mathcal{T}^0_{ij}
= -\mathcal{P}^{\|}_{ij}(\hat\s) K_\|(y) - \mathcal{P}^\perp_{ij}(\hat\s) K_\perp(y)
\ee
[recall $\calL_0(\mu)=1$].
Substituting the foregoing expression into eq.~\eqref{eq:Psi-s} [eq.~\eqref{eq:perfect-corr}] we
recover the usual (real-space) $\Psi_{ij}$ in terms of $\Psi_\perp$ [eq.~\eqref{eq:Psi-perp}], and
$\Psi_\|$ [eq.~\eqref{eq:Psi-para}].

It is easy to obtain the LOS part of eq.~\eqref{eq:Tij-2},
\bea
\hat{n}^i\hat{n}^j \mathcal{T}^\ell_{ij}
&=\Big[\mu^2 j_\ell'' + (1-\mu^2) \frac{j_\ell'}{y}\,\Big]\calL_\ell
    +\mu(1-\mu^2)
        \Big[\,2\frac{j_\ell'}{y} - 3\frac{j_\ell}{y^2}\Big]\frac{\dif\calL_\ell}{\dif\mu}
    +(1-\mu^2)^2\, \frac{j_\ell}{y^2} \frac{\dif^2\calL_\ell}{\dif\mu^2}.
\eea
When the foregoing expression is substituted back into eq.~\eqref{eq:Psi-s}, the resulting
expression should be consistent with eqs.~\eqref{eq:xiv-s} and \eqref{eq:xiell-vv}.
Indeed this can be shown by making use of the recursion relations for the Legendre polynomials.

\end{document}